\documentclass[conference]{IEEEtran}

\usepackage{paralist}
\usepackage{booktabs}
\usepackage{xcolor}
\usepackage{graphicx}
\usepackage{colortbl}
\usepackage{xcolor}
\usepackage{graphicx}
\usepackage{etoolbox}
\usepackage{hyperref}
\usepackage{cleveref}
\usepackage{orcidlink}

\graphicspath{{Fig/}}

\newcommand{\Paragraph}[1]{\vspace{3pt plus 2pt minus 1pt}\noindent{\bf #1}\ }
\newcommand\ph{\phantom{-}}

\newcommand{\Qone}{\textbf{Q1}}
\newcommand{\Qtwo}{\textbf{Q2}}
\newcommand{\Qthree}{\textbf{Q3}}
\newcommand{\Qfour}{\textbf{Q4}}
\newcommand{\Qfive}{\textbf{Q6}}
\newcommand{\Qsix}{\textbf{Q7}}
\newcommand{\Qseven}{\textbf{Q8}}
\newcommand{\Qeight}{\textbf{Q9}}
\newcommand{\Qnine}{\textbf{Q5}}
\newcommand{\Qten}{\textbf{Q10}}
\newcommand{\Qeleven}{\textbf{Q11}}
\newcommand{\Qtwelve}{\textbf{Q12}}

\definecolor{negative_dark}{HTML}{ECAA69}
\definecolor{negative_medium}{HTML}{F6D6B6}
\definecolor{negative_light}{HTML}{FBEBDB}
\definecolor{positive_light}{HTML}{E0F3F5}
\definecolor{positive_medium}{HTML}{A3DAE1}
\definecolor{positive_dark}{HTML}{63C0CC}
\definecolor{white}{HTML}{FFFFFF}

\newcommand{\opt}{\\$\circ\,$}

\newcommand{\new}[1]{{#1}}

\begin{document}

\title{``You do understand that people don't trust technology?'': Explaining Trusted Execution Environments to Non-Experts}

\author{
    \IEEEauthorblockN{McKenna McCall\IEEEauthorrefmark{1}\IEEEauthorrefmark{2}\orcidlink{0009-0007-5642-4717}}
    \IEEEauthorblockA{\textit{Colorado State University}}
\and
    \IEEEauthorblockN{Carolina Carreira\IEEEauthorrefmark{1}\orcidlink{0000-0002-4526-6510}}
    \IEEEauthorblockA{\textit{Carnegie Mellon University} and \\
    \textit{IST University of Lisbon, INESC-ID}}
\and
    \IEEEauthorblockN{Miguel Flores}
    \IEEEauthorblockA{\textit{Carnegie Mellon University} \\
    }
\and
    \IEEEauthorblockN{Lorrie Faith Cranor\orcidlink{0000-0003-2125-0124}}
    \IEEEauthorblockA{\textit{Carnegie Mellon University}}
}

\maketitle

\thispagestyle{plain}
\pagestyle{plain}

\renewcommand{\thefootnote}{\fnsymbol{footnote}}
\footnotetext[1]{Both authors contributed equally to this work.}
\footnotetext[2]{This work was completed while McKenna McCall was at Carnegie Mellon University.}
\renewcommand{\thefootnote}{\arabic{footnote}}

\begin{abstract}
Trusted Execution Environments (TEEs) protect confidentiality and integrity of trusted applications by creating an isolated environment for executing code. Prior work has shown that users may feel more comfortable sharing data when they know it will be protected by a TEE, especially if they understand what a TEE is. In this study, we evaluated text-based explanations introducing TEEs to non-experts. We analyzed existing TEE explanations to develop candidate explanations and evaluated them via vignette scenarios with 966 crowdworkers. The explanations that enhanced understanding most were \textit{non-technical} ones that highlighted specific threats that can be \textit{prevented} by a TEE. Surprisingly, even the explanations that enhanced understanding had little effect on willingness to use the TEE-enhanced technology. These results provide insights into ways to communicate technical security concepts more effectively but also suggest that explaining security technology might not be enough to address users’ privacy concerns.
\end{abstract}

\section{Introduction}
\label{sec:intro}
While the demand for users to share their data grows, consumers are expressing concern and confusion about how their information is used~\cite{pewresearch}. 
Confidential computing~\cite{confidentialcomputing} seeks to protect users by restricting computations on sensitive data to Trusted Execution Environments (TEEs). These environments guarantee the authenticity of the executed code, the integrity of the runtime states, and the confidentiality of the code and data~\cite{whataretees}. Confidential computing and TEEs have applications in AI and machine learning~\cite{confidentialcomputingandAI, chen2020training, mo2024machine}, IoT~\cite{ayoade2018decentralized}, and blockchain smart contracts~\cite{wang2020hybridchain}.

TEEs are not only being explored for their technical strengths. Prior work also suggests that when users are made aware of cloud-based TEEs in home IoT devices, they may feel more comfortable with their data being collected~\cite{pittTEE}, especially when they understand what a TEE is. 
However, this work did not investigate how technologists should explain TEEs to end-users, which is itself a challenge. Unlike some security concepts like passwords, most people are not familiar with TEEs, or even the technologies they rely on.
While fully explaining TEEs to a non-technical audience may not be feasible, understanding all of the technical details may not be necessary to understand enough of the security benefits they offer to help users feel secure.
Indeed, the original study investigating the impact of TEEs on comfort evaluated understanding based on just three high-level TEE concepts~\cite{pittTEE}.

In this study, we evaluate strategies for explaining TEEs to enhance both \emph{understanding} of the capabilities of TEEs and \emph{comfort} using TEE-enhanced technologies. Ideally, an explanation would be nuanced enough to communicate what guarantees TEEs offer, without over-promising or being overly pessimistic, which could discourage people from using technology. We based our explanations on common themes we found in existing TEE explanations from technical websites, forums, research papers, and popular media. We evaluated candidate explanations through a series of True/False questions via two online surveys of 966 Prolific crowd workers. We used vignettes in our surveys to evaluate explanations across different scenarios where TEEs might be used. 
We include home IoT scenarios, AI, and medical research applications, as they have been identified as potential use cases for TEEs~\cite{geppert2022trusted}.

Our first survey addresses two research questions:
\begin{itemize}
\item {\bf RQ1:} Which explanations improve TEE \textit{understandability} for non-experts? Is there a best overall explanation or do different scenarios benefit from different explanations?
\item {\bf RQ2:} Which explanations enhance \textit{willingness} to use the TEE-enhanced technology? Which ones promote the feeling that data will be \textit{safe}?
\end{itemize}
Based on the results of our first survey, we developed an FAQ to supplement our explanations by answering real questions asked by participants. We also asked follow-up questions to better understand what contributes to perceptions of safety. In the second survey, we answer the following research questions:
\begin{itemize}
\item {\bf RQ3:} Does an FAQ further improve understandability? Does it increase willingness to use TEE-enhanced technology or the feeling of safety?
\item {\bf RQ4:} Which aspects of TEE scenarios contribute to the belief that data would be safe/unsafe?
\end{itemize}

To the best of our knowledge, ours is the first study to investigate strategies for explaining TEEs to non-experts. 
While we found that many existing explanations use technical jargon and focus on broad security guarantees (e.g., attestation, confidentiality, and integrity), what performed best in our experiments were \textit{non-technical} explanations that highlighted specific attacks \textit{prevented} by a TEE. We also found that people generally answered comprehension questions correctly when we provided information directly in our explanations or FAQs, but struggled to answer questions that required them to make inferences based on our explanations. 

Surprisingly, in contrast with prior work~\cite{pittTEE}, we found that our explanations had little effect on willingness to use TEE-enhanced technology or feelings of safety. We believe that this is due to methodological differences between ours and the previous study (namely that we focus on high-level feelings of comfort and safety, while they focused on specific data-sharing conditions) and, importantly, our observation that TEEs cannot address all privacy concerns. These results provide insights into ways to communicate technical security concepts more effectively but also suggest that explaining security technology, while useful for improving transparency, might not be enough to address users' privacy concerns.

The paper is organized as follows: Section~\ref{sec:background} covers background and related work; 
Section~\ref{sec:explanations} explains our methods for collecting and analyzing existing TEE explanations to identify themes for testing via our surveys; Section~\ref{sec:methodology} describes our survey methods;  Sections~\ref{sec:results} and~\ref{sec:follow-up} present the results of our two surveys; Section~\ref{sec:discussion} includes additional discussion; finally, Section~\ref{sec:conclusion} concludes.

\section{Background and Related Work}
\label{sec:background}

In this section, we describe background on TEEs. 
Next, we outline related work on the importance of communicating with users about security. Finally, we summarize work that has attempted to explain technical security concepts to end-users.

\subsection{Trusted Execution Environments}
\label{sec:back-tee}

TEEs are combinations of several security processes, including hardware security extensions, cryptographic modules, secure distributed systems protocols, and more. 
According to Sabt et al.~\cite{whataretees}, a TEE can be defined as a tamper-resistant processing environment that guarantees the \textit{confidentiality} and \textit{integrity} of the executed code and data (preventing unauthorized reading and modification, respectively). A TEE also provides remote \textit{attestation}, a process where the TEE proves to a remote verifier (such as a server) that it is operating securely and that the integrity of its code and data has not been compromised.
TEEs can be found in many Android phones. Authentication in Android is typically handled by code residing in a TEE based on ARM TrustZone~\cite{authentication},  a set of security extensions that enable ARM processors to run in two distinct modes---secure and non-secure. Most Androids also use TEEs to process mobile payments, secure banking, device reset protection, and detect malware~\cite{trusty}. 

Another type of TEE, Intel SGX, allows applications to create protected areas in memory in some Intel CPUs~\cite{sgx} and also has applications in smart home devices.
Ayoade et al.~\cite{ayoade2018decentralized} propose using Intel SGX  for decentralized data management in smart home applications. There are several other TEE technologies in the realm of confidential computing that target cloud computing, such as AMD SEV-SNP~\cite{AMDSEV-SNP}, Intel TDX~\cite{tdx}, and ARM CCA~\cite{cca}.

Confidential computing is particularly relevant in the medical domain because patient data, such as data from clinical trials, has some of the strongest legal protections in the US~\cite{gostin2009beyond,hipaa1996}. Data aggregation enabled by confidential computing allows healthcare providers to improve patient or research outcomes while safeguarding patient privacy~\cite{medical_ccc}.     
TEEs could be used to protect machine learning for medical applications and ensure compliance with medical regulations~\cite{chen2020training}.

\subsection{Importance of Understanding Security Technology Basics}

While technical expertise may be required to understand the details of security technology, even a basic understanding can help users make informed decisions and better protect themselves, while misconceptions and poor usability can lead to worse outcomes.
An early example of opaque security technology hindering users can be found in the seminal paper by Whitten et al., which provides empirical evidence that users who lack understanding about how public key encryption works behave in ways that undermine the security and privacy of their encrypted email. The authors conclude that an unusable or incomprehensible security mechanism will not be used effectively and thus not provide security~\cite{whitten1999johnny}. 

Misunderstanding the use and limitations of security tools can lead to a false sense of security~\cite{abu2020evaluating,habib2018away}. This overconfidence may lead them to engage in riskier behaviors under the mistaken belief that they are protected. For instance, Bravo-Lillo et al.~\cite{bravo2010bridging} showed that misconceptions about web browser security warnings can give users an illusion of safety. 
In addition, interview~\cite {ur2015added} and survey studies~\cite{ur2016users} have investigated users' misconceptions about how attackers steal passwords, finding that misconceptions led users to believe vulnerable passwords were secure. 

Misconceptions about security tools and design choices can also hinder their adoption~\cite{pearman2019people}. 
Users may also face usability challenges with cookie banners due to their design~\cite{bouma2023us}. Similarly, users struggle to comprehend iOS privacy labels because of jargon and unfamiliar terminology~\cite{wu2020risk,zhang2022usable}.
On the other hand, informing users about security technology can have a positive impact. For example, Furnell et al.~\cite{furnell2018enhancing} find that more information can motivate users to choose strong passwords.
When security- and privacy-enhancing technologies 
are mentioned to users as part of the consent process, users need a basic understanding of what protections these tools can and cannot offer if they are to make an informed decision. 
The European Union's GDPR~\cite{GDPR} requires that organizations must provide clear and accessible information to ensure users understand how their data is used. Similarly, in the US, HIPAA~\cite{hipaa1996} mandates that healthcare providers give patients clear information about their privacy rights and how their medical information is shared.

While most of the work in this space focused on technologies users employ to protect themselves, explaining TEEs is a fundamentally different task since they are hidden from users. Our goal in explaining TEEs is more to improve transparency and comfort than to change user behavior. 

\subsection{Explaining Technical Concepts}

Our study builds upon prior work on short explanations to communicate technical concepts and evaluate them using online surveys~\cite{shen2021can,akgul2021evaluating,distler2020making}. Prior work on explaining security concepts mostly focused on perceived security~\cite{distler2020making,stransky2021limited} and did not address comprehension or focus on non-TEE related contexts~\cite{akgul2021evaluating,bravo2013your,shen2021can,akgul2021evaluating}. 
\label{sec:back-expln}
Several research studies have proposed and tested explanations of other technical security concepts with end-users. 
Research on formal verification has also emphasized the importance of communicating technical concepts to non-technical audiences and identified it as a priority and a challenge for future work~\cite{carreira2021}.

Xiong et al.~\cite{Xiong2020} attempted to explain differential privacy with experiments to investigate the effects of different communication approaches.
They found that, despite the positive effect of the explanations, participants struggled with understanding some of the more technical jargon. Karegar et al.~\cite{karegar2022exploring} studied a possible solution by addressing the impact of metaphor-based explanations of differential privacy. They found that metaphors can aid understanding but may lead to misconceptions.
Cummings et al.~\cite{cummings2021need} attempted to design better explanations about differential privacy but highlighted the difficulty of crafting explanations that satisfy user interest and preserve the integrity of the technical content.

Similar to our work, Akgul et al.~\cite{akgul2021evaluating} investigated whether text-based explanations improve users' mental models of encryption. 
They found that changing pre-existing mental models can be challenging, but educational interventions can work. Interestingly, they concluded that their explanations may have slightly oversold the capabilities of encryption. 
Shen et al. investigated users' understanding of smartphone permissions and observed that short explanations within user interfaces led to better comprehension. The authors found that adding information to permissions dialogues made it more clear to users how their choice affected the way that their location would be tracked~\cite{shen2021can}. 
When it comes to explaining encryption, not all authors agree. Distler et al. ~\cite{distler2020making} attempted to explain encryption and concluded that explaining encryption does not necessarily maximize perceived security. They focused primarily on the feeling of security and did not study users' comprehension of encryption. This is also the case for Stransky et al.~\cite{stransky2021limited}, but their results suggest that using text disclosures about encryption makes users feel more secure seem more effective than iconography.

OS and browser security warnings are designed to provide actionable security information to non-technical users~\cite{bravo2013your,schechter2007emperor}.
Wu et at.~\cite{wu2019something} show that warning notifications in Signal can improve comprehension of the purpose of security mechanisms and promote favorable privacy outcomes.
Well-designed password meters can be an effective communication tool to inform users about their password complexity and are a good way to provide actionable feedback about password strength~\cite{ur2017design}. 
Privacy and security ``nutrition'' labels are designed to provide succinct information to users that can inform their decision-making~\cite {iotlabel,nutritionlabel}.

To our knowledge, the only other attempt to communicate about TEEs to end-users is from Musale et al.~\cite{pittTEE}, who investigated the impact of TEEs on data-sharing preferences. They also looked at the impact of understanding TEEs, finding that people who understood TEEs were more likely to be comfortable sharing their data.
For example, they found that participants who understood TEEs were significantly more comfortable with their data being collected if they were ``notified'' of the data collection than those who did not understand TEEs.
To assess TEE comprehension, the authors asked three True/False questions about secure storage, secure computing, and remote attestation. 
In one question, they ask whether the statement ``non-authorized persons can modify/change the nature of the algorithm being used or gain access to the image database'' is true or false.
While their study focused on understanding the impact of TEEs on existing privacy norms, ours focuses on how to effectively explain TEEs. For this reason, we conduct a broader assessment of 10-12 questions that address different aspects of a TEE.
We also focus on high-level feelings of comfort and safety instead of specific data-sharing conditions. 
While their work did attempt to explain TEEs to their participants, the goals and methodologies of that study were fundamentally different from ours.

\section{Developing Candidate TEE Explanations}
\label{sec:explanations}
In this section, we describe our approach for developing candidate TEE explanations, which is based on a technique from prior work on differential privacy~\cite{cummings2021need} and other guidelines for writing effective explanations~\cite{distler2020making,schaewitz2021peeking}.
First, we describe how we analyzed existing TEE explanations from technical websites, forums, research papers, and popular media to identify common themes. Next, we explain how we used these themes to develop explanations to evaluate in our study.

\subsection{Identifying Existing TEE Explanations}
\label{sec:method-expln}
 
We conducted a Google search using the term ``Trusted Execution Environment'' and restricted results to the last five years.  The first five pages of results included 42 unique URLs that had 32 TEE explanations.
These results came from diverse sources, mostly aimed at an audience of technical experts. 
We obtained eight additional explanations through searches targeting well-known, general audience platforms like the New York Times, Medium, and Forbes. 

We removed 12 sources from the initial 50 that did not include substantive TEE explanations (i.e., TEEs were mentioned, but not explained). We removed two others because their explanations were incorrect or misleading. This filtering was done by one of the authors, and the removed sources were discussed with the team. We analyzed the explanations from the 36 remaining sources to identify themes to test in our experiments. 
19 sources came from technology-focused websites from companies that provide TEEs (e.g. Intel, NVIDIA, AWS). Media sources, including general audience magazines and news websites, accounted for 6 explanations. The remainder came from a mix of scientific publications, forums, social media websites, governmental websites, and Wikipedia (see supplementary materials for all explanations and sources). This diversity of sources ensured a broad spectrum of explanations to reflect the variety of information available to the public.

\begin{table}[t]
  \centering
  \resizebox{\columnwidth}{!}{%
  
\begin{tabular}{@{}lll@{}}
\toprule

\textbf{Code} & 
\textbf{Description} & 
\textbf{Frequency} \\ 
\midrule
Reputation    & 
  Leverages pre-existing trust/reputation of & 
  2           \\
  & recognizable companies \\
Verified      & 
Application running in the TEE is verified & 
  2           \\
Attestation   & 
Process to check that the software supporting the TEE & 
  4           \\
  & is the code we expect \\
Trust     & 
  Explanation mentions the word ``trust'' & 
  5           \\
Unsubstantial & 
  Generic/un-detailed description & 
  8           \\
Threat        & 
  TEE protects against untrusted OS/peripherals & 
  8           \\
Techniques    & Describes particular TEE & 
  10          \\
  & (e.g., Intel SGX, Arm TrustZone) \\
Cryptography  & 
  Mentions cryptographic concepts & 
  10          \\
Technical     & 
  Explanation uses technical terminology & 
  11          \\
  & (e.g., ``confidentiality,'' ``attestation'') \\
Integrity     & TEE prevents unauthorized modification                                     & 16          \\
Prevents      & TEE prevents some undesirable behavior                                     & 17          \\
Secrecy       & TEE prevents unauthorized access                                            & 21          \\
Isolation     & TEE ensures isolation from the rest of the system                           & 23          \\
Hardware      & Mentions that a TEE is \emph{hardware}-supported                                         & 23          \\ \bottomrule
\end{tabular}

  }
  \vspace{2mm}
  \caption{Codebook for explanations found in the wild and how frequently each code was identified in the explanations. Each explanation could have up to 7 different codes. }
  \label{tab:codebookexplanation}
\end{table}

Two authors independently reviewed the explanations to identify  themes and assigned a code for each theme.
The number of codes per explanation largely depended on the size and complexity of the text. We ended up assigning eight codes to the most complex explanation. The coding process began with a few initial codes based on our 
prior knowledge of TEEs. Codes were added based on themes that emerged during the analysis. After reviewing all explanations, the coders discussed the themes to develop a shared codebook. They repeated the process of reviewing explanations, coding, discussing all disagreements, and refining the codebook twice more until they reached 100\% agreement. The final codebook has 14 codes. Here we list the codes that appeared in at least five explanations with the number of explanations in which they appeared in parentheses:  \emph{Isolation} (23),  \emph{Hardware} (23), \emph{Confidentiality} (21), \emph{Prevents} (17),  \emph{Integrity} (16), \emph{Technical} (11), \emph{Cryptography} (10). \emph{Techniques} (10), \emph{Threat} (8), \emph{Unsubstantial} (8), and \emph{Trust} (5).  
The complete codebook, including a description of each code, can be found in Table~\ref{tab:codebookexplanation}.

\subsection{Designing Candidate TEE Explanations}
We developed new explanations that used key themes found in existing explanations and iterated on their wording through pilot testing. We designed our explanations to be composable so that we could separately test each component in controlled experiments.
We identified \emph{Confidentiality}, \emph{Isolation}, and \emph{Integrity} as themes that seemed fundamental to a TEE explanation and should be included in every candidate explanation using either \emph{Technical} or \emph{Non-technical} language.  In addition, we identified \emph{Hardware}, \emph{Trust}, and \emph{Prevents} as themes that might aid understanding. We decided to test explanations that included \emph{Hardware}, \emph{Trust}, or an \emph{Unsubstantial} explanation, as well as explanations that either explained what a TEE \emph{Prevents} or includes \emph{No Prevents} clause. In order to keep the number of treatments in our survey manageable, we did not evaluate the less common themes. 

\begin{figure*}
    \centering  
    \includegraphics[alt={A figure showing the structure of a TEE explanation as a mathematical formula: Explanation equals (begin orange highlighting) Hardware or Trust or Unsubstantial (end orange highlighting) plus (begin green highlighting) Techncial or Non-Technical (end green highlighting) plus (begin purple highlighting) Prevents or No Prevents (end purple highlighting). The candidate TEE explanation shows how these pieces are combined using highlighted text: A Trusted Execution Environment (TEE) is a technique for running programs and interacting with data securely (begin orange highlighting) using a protected area of the physical computer (end orange highlighting). A program running in a TEE is isolated from the rest of the computer to (begin green highlighting) protect the confidentiality and integrity of (end green highlighting) the program and data. (begin purple highlighting) The TEE protects the program and data even when other software on the computer is behaving maliciously. (end purple highlighting)},width=0.43\textwidth]{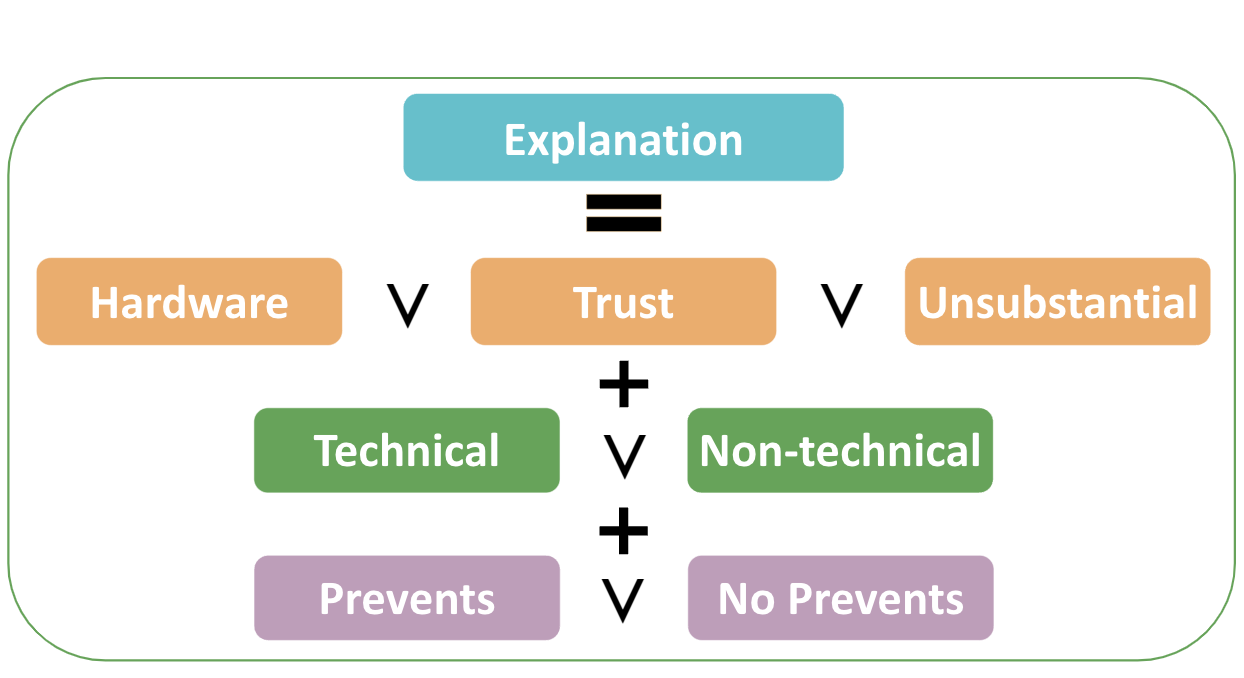}
    \hspace{0.04\textwidth}
    \includegraphics[width=0.4\textwidth]{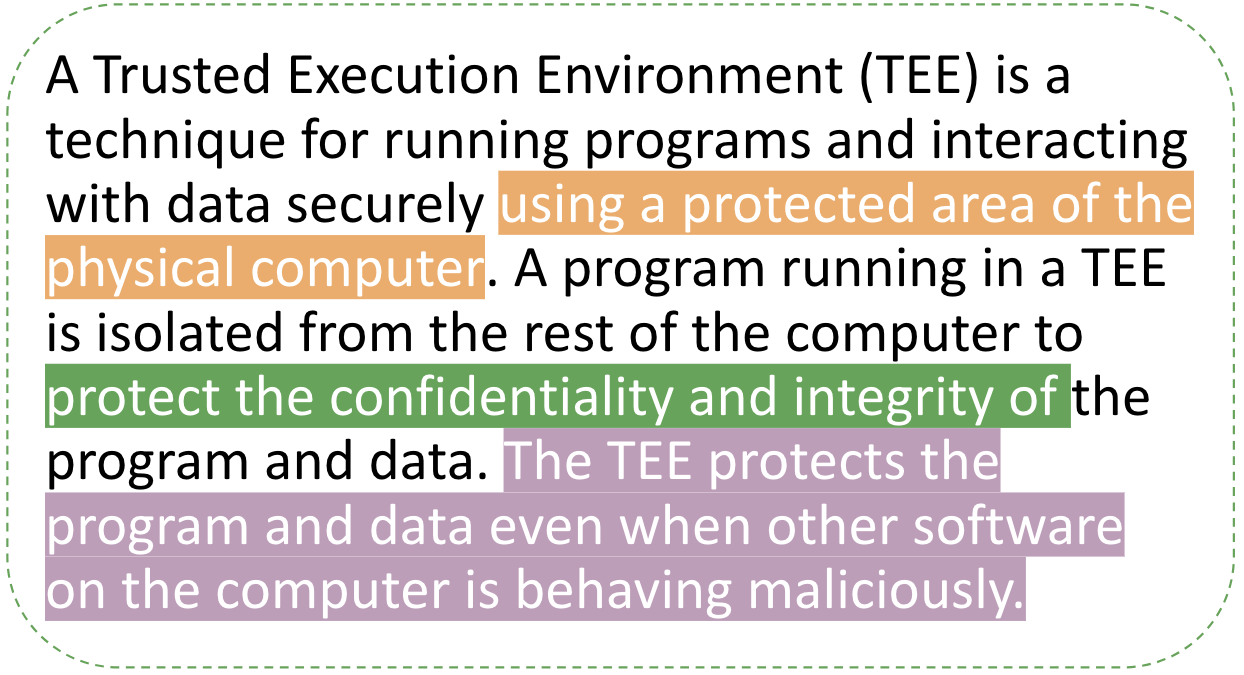}
    \caption{Diagram illustrating the design of the initial TEE explanations and an example candidate TEE explanation.}
    \label{fig:explanation-diagram}
\end{figure*}

As shown in Figure~\ref{fig:explanation-diagram}, the structure of each explanation is: (1) a high-level sentence introducing the concept of a TEE as a security mechanism (one of the following themes: \emph{Hardware}, \emph{Trust}, or \emph{Unsubstantial}), followed by (2) a sentence introducing the concepts of isolation, confidentiality, and integrity in either technical or non-technical language (one of the following themes: \emph{Technical} or \emph{Non-technical}), and, only for some explanations, (3) a third sentence introducing a specific threat that a TEE can prevent (theme: \emph{Prevents} or \emph{No Prevents}). Our candidate TEE explanations are the set of all 12 possible combinations of themes that follow the structure above. Complete TEE explanations are shown in Appendix~\ref{app:tee}.

\section{Survey Methods}
\label{sec:methodology}
 To evaluate our candidate TEE explanations, we conducted two 
 surveys. The Survey 1 focused on evaluating our explanations (\textbf{RQ1-2}), while Survey 2 tested some follow-up research questions (\textbf{RQ3-4}) based on the results of Survey 1. In this section, we describe the methods we used to conduct and analyze data from both surveys. 

\subsection{Survey 1: Evaluating TEE Explanations} 
\label{sec:method-survey} 
The purpose of Survey 1 was to evaluate our candidate TEE explanations to identify which themes, adopted from existing explanations, are best at enhancing understanding ({\bf RQ1}), willingness to use TEE-enhanced technology, and feelings of safety ({\bf RQ2}).
We constructed four scenarios to use in our surveys. Each scenario describes a situation where personally identifiable data is collected for some purpose. The data collected in each scenario is the same, but the \textit{setting} and \textit{purpose} of collection depends on the scenario. 

We choose medical research and smart home settings because they are both promising TEE use cases~\cite{medical_ccc,chen2020training,ayoade2018decentralized,AMDSEV-SNP,tdx,cca}. 
We chose not to use a smartphone scenario, despite it being another TEE use case~\cite{authentication,trusty} because we wanted to focus on emerging TEE applications in the healthcare setting and smart home scenarios to match prior work~\cite{pittTEE}. The fact that the vast majority of Americans already use smartphones~\cite{peoplehavesmartphones} could also have biased our results. 
In the medical research setting, we ask participants to imagine there is a medical research study that involves collecting personal information if they choose to participate. In the smart home setting, we ask them to imagine shopping for a smart device that will collect personal information about them if they choose to purchase it. 

We also have two variations of each scenario, one where the purpose of the data collection is to develop technology involving AI and one not involving AI. We included AI in our scenarios because the adoption of AI has been growing in both medical research~\cite{karpov2023analysis} and the smart home context (e.g., Google Home~\cite{aigooglehome} and Alexa~\cite{aialexa}) and there is evidence that people are wary of AI~\cite{pewresearch}, which could factor into their willingness to use the technology.

Each participant receives one medical research scenario and one smart home scenario in a randomized order to account for learning effects. They are also randomly assigned the technology with AI or without AI. 
For example, one participant may receive the ``medical research with AI'' scenario followed by the ``smart home with AI'' scenario while another may receive the ``smart home without AI'' scenario followed by the ``medical research with AI'' scenario. For each scenario, participants are told that the data is stored in the cloud and protected by a TEE. The complete scenario text for all four scenarios is shown in Appendix~\ref{app:scenarios}.
The scenario text is followed by a random candidate TEE explanation (from the set of 12 explanations), and they receive the same explanation for both scenarios. 

In the first part of the survey, we introduce the scenario and confirm participants are paying attention by asking them to select the purpose of the medical research study or what device they're shopping for. If they answer incorrectly, they are asked to re-read the scenario text and try again.\footnote{Eight participants answered incorrectly the first time, but four succeeded after we gave them a second chance at the attention check.}

Next, we asked participants to rate their willingness to participate in the medical research study (for the medical research scenarios) or willingness to purchase the smart home device (for the smart home scenarios), and how safe they believe their data would be, each on a 3-point scale. The willingness scale also included a "Not sure" free-response option that we manually categorized into the same 3-point scale for analysis. We then evaluated comprehension via 10 True/False questions and allowed participants to ask us any lingering questions they had about TEEs. We ended the survey by collecting demographic data. 
The complete survey instrument can be found in Appendix~\ref{app:survey}.
We solicited feedback from TEE experts outside of our team on the technical accuracy of our explanations, scenarios, and comprehension questions. We refined the survey questions through multiple pilots.

\subsection{Survey 2: FAQs and Understanding Aspects of Feeling Safe}

Survey 2 was similar to Survey 1 other than the introduction of an FAQ to answer some of the most frequently asked questions from Survey 1 ({\bf RQ3}) and asked additional questions to understand which \textit{aspects} of the scenarios led to the belief that data shared with the TEE-enhanced technology would be safe or unsafe ({\bf RQ4}). 
We developed this FAQ based on direct feedback from participants. However, given its considerably more technical nature, we kept it apart from the explanations.

Participants were randomly assigned to one of three FAQ conditions: one where they were \textit{Shown} the FAQ on its own page after the first scenario was introduced (which they could not click past for 60 seconds) \textit{and} as expandable text on subsequent pages; one where the FAQ was \textit{Hidden} by default and \textit{only} offered as expandable text; and one where they were not given an FAQ (\textit{None} condition). To keep the number of survey conditions reasonably small, we did not re-test all of the TEE explanations from Survey 1. Since the \textit{Technical} and \textit{No Prevents} themes generally led to worse comprehension scores, we used the \textit{Non-Technical} and \textit{Prevents} themes in all of the explanations. Thus, we had 3 explanation conditions (\textit{Hardware}, \textit{Trust}, and \textit{Unsubstantial}), plus we added a fourth no-explanation condition (\textit{None} condition) to serve as a baseline for the questions about aspects of safety.

Because some participants mentioned in Survey 1 that they do not believe data could ever be ``Completely safe,'' when we asked about safety in Survey 2, we used a 4-point scale, adding ``Mostly safe'' to the 3-point scale (Completely safe, Somewhat safe, Not at all safe) from Survey 1. 
We did not adjust the willingness scale and kept the "Not sure" free-response option.
We also asked participants to rate how much different aspects of the scenarios contributed to the belief that their data would be safe/unsafe on a 5-point scale and to expand on ``anything else'' that contributes to those feelings in a free-response field. Finally, we added 2 True/False questions about the topics covered in the FAQ.

\Paragraph{Constructing the FAQ.}
Our FAQ is based on the questions participants asked in Survey 1.
Our FAQ answers three questions:
\begin{inparaenum}[(1)]
    \item How do TEEs work?
    \item How do we know the TEE is working correctly?
    \item How are TEEs used in real life?
\end{inparaenum}
The answer to the first question includes additional technical details about how TEEs work, specifically Arm TrustZone~\cite{trustzone} and Intel SGX~\cite{sgx}. To answer the second question, we described attestation and mentioned that researchers are continuing to develop ways to ensure the applications running in the TEE work as expected. Finally, we used authentication in Android~\cite{authentication,gatekeeper} as an example of a real TEE use case in the answer to the third question. The complete FAQ text may be found in Appendix~\ref{app:faq}. 
We also provided links to the resources cited in this paragraph (plus general information about confidential computing~\cite{confidential}) at the end of the survey. We did not keep track of whether participants clicked the links.

\subsection{Recruitment}
\label{sec:method-recruit}
Data collection took place January through May 2024. We used the same recruitment process for both surveys.
We recruited 469 Prolific participants for Survey 1 and 501 for the second using quotas~\cite{prolificquota} to ensure approximately equal numbers of men and women.\footnote{Sample sizes were determined by a rule-of-thumb estimate for the logistic regressions we planned for our analysis~\cite{smeden2019samplesize}.} 
People who participated in Survey 1 were not allowed to participate in Survey 2. Our participants are adults located in the US who are fluent in English. 
We determined compensation rates by piloting our study to estimate survey completion time and our intended hourly rate (\$15 per hour).
We paid participants \$2.50 for Survey 1 (\$15 per hour, median completion time approx. 10 minutes) and \$2.75  for Survey 2 (\$12.35 per hour, median completion time approx. 13 minutes). 

We reviewed results for low-effort or nonsensical free-text responses (none) and removed responses for participants who failed both attention checks (none in Survey 1, 4 in Survey 2). We were left with 469 responses for Survey 1 and 497 for Survey 2. 

\begin{table}
 \centering
 {
\begin{tabular}{lrrrr}
  \toprule
  & \multicolumn{2}{c}{\textbf{Initial Survey}} & \multicolumn{2}{c}{\textbf{Follow-up}} \\
  {} & \textit{n} & \% & \textit{n} & \% \\
  \midrule
  \multicolumn{5}{l}{\textit{Gender}} \\
  Male   & 229 & 48.8\% & 245 & 49.3\% \\
  Female & 228 & 48.6\% & 242 & 48.7\% \\
  Non-binary / third gender & 11 & 2.3\% & 8 & 1.6\% \\
  Prefer not to say & 1 & 0.2\% & 2 & 0.4\% \\
  \midrule
  \multicolumn{5}{l}{\textit{Age}} \\
  18-24 & 74 & 15.8\% & 82 & 16.5\% \\
  25-34 & 151 & 32.2\% & 164 & 33.0\% \\
  35-44 & 119 & 25.4\% & 121 & 24.3\% \\
  45-54 & 63 & 13.4\% & 81 & 16.3\% \\
  55+ & 62 & 13.2\% & 48 & 9.7\% \\
  Prefer not to say & 0 & 0.0\% & 1 & 0.2\% \\
  \midrule
  \multicolumn{5}{l}{\textit{Highest Education Achieved}} \\
  Less than high school & 9 & 1.9\% & 5 & 1.0\% \\
  High school or equivalent & 161 & 34.3\% & 184 & 37.0\% \\
  Bachelor or associate degree & 207 & 44.1\% & 225 & 45.3\% \\
  Graduate degree & 90 & 19.2\% & 81 & 16.3\% \\
  Prefer not to say & 2 & 0.4\% & 2 & 0.4\% \\
  \midrule
  \multicolumn{5}{l}{\textit{Familiar with TEEs?}} \\
  Yes & 35 & 7.5\% & 36 & 7.2\% \\
  No & 434 & 92.5\% & 461 & 92.8\% \\
  \midrule
  \multicolumn{5}{l}{\textit{Experience in Computing?}} \\
  Yes & 95 & 20.3\% & 107 & 21.5\% \\
  No & 374 & 79.7\% & 390 & 78.5\% \\
  \midrule
  \multicolumn{5}{l}{\textit{Experience With Smart Homes?}} \\
  Yes & 382 & 81.4\% & 429 & 86.3\% \\
  No & 87 & 18.6\% & 68 & 13.7\% \\
  \midrule
  \multicolumn{5}{l}{\textit{Experience with Medical Research/Work?}} \\
  Yes & 113 & 24.1\% & 142 & 28.6\% \\
  No & 356 & 75.9\% & 355 & 71.4\% \\
  \midrule
  \textbf{Total} & 469 & 100\% & 497 & 100\% \\
  \bottomrule
\end{tabular}
}
 \vspace{2mm}
 \caption{Participant demographics for both surveys.}
 \label{tab:demographics}
\end{table}

Table~\ref{tab:demographics} shows participant demographics, which are similar for both surveys. 
Participants were balanced across gender, generally young (73.4\% under 45 in Survey 1 and 73.8\% in the second), and college-educated (63.3\% in Survey 1 and 61.6\% in the second). Few participants were familiar with TEEs before taking our survey (around 7\% for both surveys) or have a career or formal education in a computing field (16.4\% in Survey 1 and 21.5\% in the second). Our participants tend to have some experience with smart home devices (81.4\% in Survey 1 and 86.3\% in the second) but not with medical research/work in the medical field (23.2\% and 28.6\%).

\subsection{Qualitative data analysis}
\label{sec:qualanalysis}

Our study has two open-ended questions. 
In the first open-ended question, we ask participants if they have questions about TEEs.
In the second, we ask about aspects of the scenario that contribute to their belief that their data would be safe or unsafe (this question is only in Survey 2). In this section, we describe how we analyzed these questions. Complete codebooks can be found in Appendix~\ref{app:questions}.

\textbf{Questions about TEEs.}
The questions asked by participants in Survey 1 were coded by two authors. Initially, one coder reviewed the participants' answers, constructed the codebook using thematic coding, then trained the other coder on the codebook. Next, the coders independently coded all responses, met to discuss disagreements and update the codebook, then re-coded the answers again. This process was repeated two times until all disagreements were resolved. 
We started with the same codebook for Survey 2 and involved a third author as a coder. 
The same initial coder reviewed responses and trained the other coders on the codebook. Then, all three coders independently coded all answers, meeting to resolve differences and update the codebook. This process was repeated twice until we reached 100\% agreement 
The goal of our coding was to identify themes and our sample was sufficiently small that we could use multiple coders for the entire dataset~\cite{mcdonald2019reliability}.

\textbf{Aspects contributing to feeling data is safe or unsafe. }
The second open-ended question about aspects of safety was analyzed much like the first. One coder started by reading through all the answers, developing a codebook, and training the other two coders. After, each coder independently coded all responses, meeting to resolve differences and update the codebook. This process was repeated twice until all disagreements were resolved.

\subsection{Quantitative data analysis}
\label{sec:quantanalysis}

We performed logistic and ordinal logistic regressions as well as Mann-Whitney U and Wilcoxon signed-rank tests.
The Mann-Whitney U compares those who received TEE information and those who didn’t. The dependent variable was perception of safety. The Wilcoxon signed-rank test was used to compare each participant’s paired scores (mean of scores for Q1-5 vs. mean of scores Q6-10). 
We use logistic regressions instead of a hierarchical model because the treatment (i.e., the structure of the explanations) was not nested.
We keep our predictors consistent across all regressions, except we added the FAQ as a predictor for Survey 2. We choose predictors that allow us to explore the relationship between our explanations and the outcome variables (e.g., whether they answered a True/False question correctly). Our predictors are the \textit{explanation} shown, \textit{computer science experience}, \textit{medical or IoT experience}, and ~\textit{FAQ} condition. To tailor our analyses to the scenarios, we use medical experience as a predictor for medical scenarios and smart home experience for smart home scenarios. 
We selected our models and planned our analyses in advance to limit Type I error rates associated with running multiple tests. Each test covers a different hypothesis,
\new{and we use the Benjamini-Hochberg procedure to correct for multiple hypothesis testing between scenarios.}

The explanations in Survey 1 consisted of all possible variations and combinations of explanations between the first three sentences (a total of 12 different explanations). The baseline for these predictors is \textit{Unsubstantial},  \textit{Technical}, and \textit{No Prevents} for each of the explanation sentences. In Survey 2, we had four possible explanations shown, and the baseline is \textit{None} (no explanation).
The baseline for medical and smart home experience is False (no experience).

~\textbf{Comprehension questions.} We assess user understanding based on a set of True/False questions (10 in Survey 1 and 12 in Survey 2). To analyze these binary outcomes, we performed logistic regressions for each comprehension question in each scenario (e.g., 10 questions $\times$ 2 scenarios $=$ 20 models for Survey 1) and \new{use the Benjamini-Hochberg procedure to correct between scenarios.}
Because we used different models for medical and home IoT scenarios, each participant was in the data set exactly one time, so we do not need to account for repeated measures in our model.
The coefficients represent the odds of the outcome occurring for a one-unit increase in the predictor.
A coefficient greater than zero indicates that the predictor increases the odds of the outcome variable to 1 (a correct answer). Conversely, a coefficient less than zero would indicate a negative impact. Additionally, we assessed the significance of each predictor by looking at the p-value with a significance level of 0.05.

~\textbf{Safety and willingness questions.}
We also asked participants about their perceptions of safety and willingness to engage with our scenario. These questions did not have binary answers. Instead, participants answered using a 3-to-5-point Likert scale. To ensure consistency across models, we binned all Likert scale data used in statistical analysis into 3 levels.  
For the willingness to engage with our scenario, there was no need to re-bin, we had ``Would not'' (baseline), ``Maybe would,'' and, ``Definitely would.'' 
For the safety perceptions, we binned all answers into ``Not at all safe'' (baseline), ``Somewhat safe,'' and ``Safe'' (from binning ``Mostly safe'' with ``Completely safe.'').

To analyze this data, we used an ordinal logistic regression. 
We conducted one regression per question, per scenario, using the same predictors as the comprehension questions  \new{and  Benjamini-Hochberg procedure to correct between scenarios.} Because we used different models for medical and home IoT scenarios, each participant was in the data set exactly one time, so we do not need to account for repeated measures in our models.
We conducted Brant tests to ensure the proportional odds assumption for all predictors. The results indicated that the proportional odds assumption holds for all predictors ($p=0.05$).
Interpreting ordinal logistic regressions is similar to binary logistic regressions. The difference is that a coefficient greater than zero indicates that the predictor increases the odds ratio of the outcome variable reaching or exceeding a higher category when compared to the baseline---for example, a positive coefficient for our willingness questions would indicate the participant is more willing to use the technology, while a negative coefficient indicates they are less willing. We keep the same significance level of 0.05.

We also wanted to understand the aspects of the scenario that affect perceptions of safety with vs. without information about TEEs.
For this, we used the Mann-Whitney U test, a nonparametric test that allows for the comparison of median ranks between two independent groups, even with non-normal data. The groups being compared received different information about TEEs (i.e., the test did not involve repeated measures) and we repeated the test for each scenario.
To compare the average score for questions \Qone{}-\Qnine{} to 
questions \Qfive{}-\Qten{} we use a Wilcoxon signed rank, a nonparametric test to compare the median of the differences between two groups. We examine each scenario separately, \new{with the Benjamini-Hochberg between scenarios}, and a significance level of 0.05.

\subsection{Limitations}
We chose an IoT and Medical scenario to reflect real-life situations where our participants could make choices about the use of a TEE-enhanced technology. However, these scenarios, while designed to be realistic, might not fully capture the complexity of a real-world context and may not be representative of the entire range of contexts where users may need some understanding of TEEs.
We evaluated our comprehension questions for clarity, though it is still possible that participants interpreted the questions differently from us, especially given the inherent constraints of the True/False format.
The study relies on self-reported data, which may be affected by social desirability bias or participants' willingness to disclose their true thoughts and feelings. We tried to mitigate this limitation by ensuring the confidentiality of the participants. We also checked to make sure participants had read and understood the scenarios we were asking about. Moreover, within the online crowdsourcing platforms available, Prolific seems to be one of the most reliable~\cite{douglas2023data,tang2022replication}.
Our sample of participants is skewed young and may not represent the larger population.
While we made efforts to ensure high data quality and received very few responses that raised concerns about LLM-usage, we cannot be sure that participants did not use LLMs.

\subsection{Ethical Considerations}
The surveys began with a consent form that included information about potential risks and benefits, notice of voluntary participation, the right to ask questions, and contact information for the authors' institution(s).
The surveys and consent forms were approved by the IRB at the authors' institution(s). The only personally identifiable data collected were Prolific IDs for recruiting and paying participants and IP addresses for bot detection. Data for subjects who did not consent or withdrew from the study were deleted from Qualtrics. Prolific IDs and IP addresses were removed from data after paying participants prior to further analysis.

\section{Survey 1 Results}
\label{sec:results}
\begin{figure*}[t]
 \centering
 \includegraphics[alt={A bar chart showing hte percent of participants answering each comprehension question correctly. The results for Survey 1 are shown in pink, the results for Survey 2 are shown in blue. The pink and blue bars are similar sizes for each question. The chart also shows the question text and expected answers: Q1 The general public can access your data (False), Q2 A researcher/employee on an unrelated team can access your data (False), Q3 A hacker can use a bug outside the TEE to access your data (False), Q4 A malicious program outside the TEE can access your data (False), Q5 Other researchers/employees working on the same computer can access your data (False), Q6 A person on the research/development team can access your data (True), Q7 If a mistake is made collecting your data, your data could be incorrect (True), Q8 Your data could be used for other projects in the future (True), Q9 The TEE ensures the program running inside produces the correct result (False), Q10 A person on the research/development team can steal your data and sell it (True), Q11 The PIN you use to unlock an Android phone is verified in a TEE (True), Q12 We cannot be sure the TEE is configured correctly (False).}, scale=0.45]{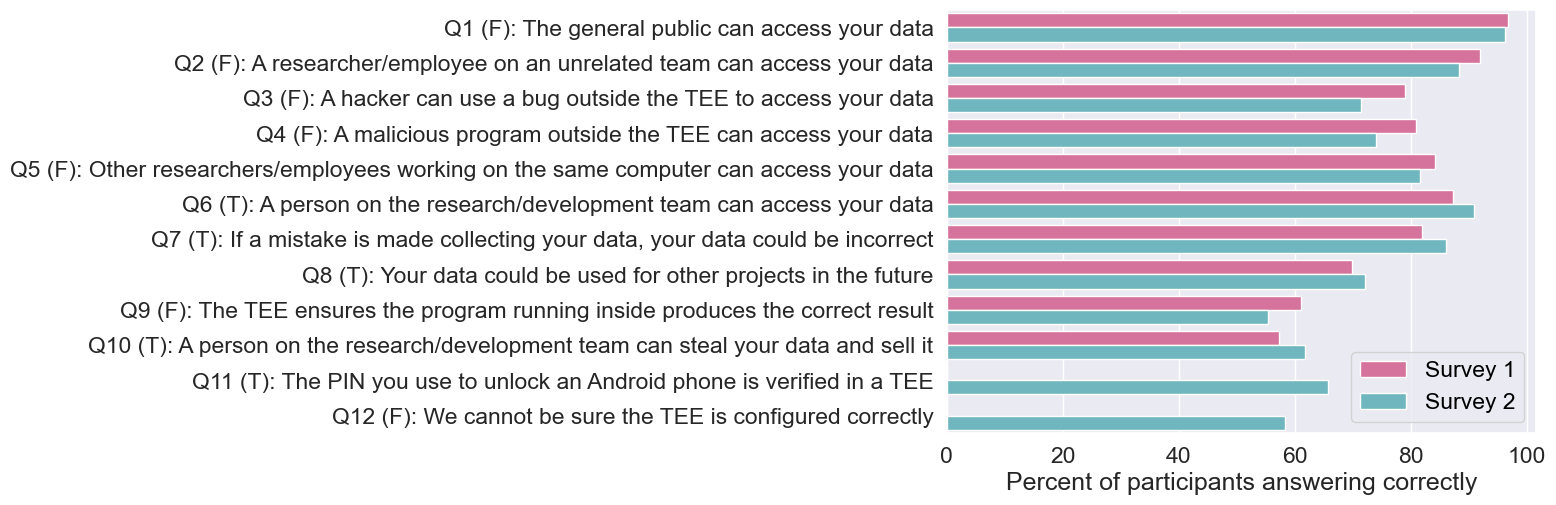}
 \caption{Overall scores for the True/False comprehension questions in both surveys. 
 \Qone{}-\Qnine{} are about features of TEEs,  \Qfive{}-\Qten{} are about limitations of TEEs, and \Qeleven{} and \Qtwelve{} are questions that can be answered based on information in the FAQ and appear in Survey 2 only. The correct answer for each question is shown in parentheses by the question number.}
 \label{fig:comp}
 \end{figure*}
 
In this section, we summarize the results of Survey 1 evaluating candidate TEE explanations.

\subsection{RQ1: Factors Influencing Comprehension}
\label{sec:results-survey}

To measure comprehension of TEE concepts, we asked participants a series of True/False questions. Each question has a correct answer. We evaluated their responses for correctness and summarize the scores 
in Figure~\ref{fig:comp}. The full regression table can be found in Table~\ref{tab:stats-medical} and Table~\ref{tab:stats-iot}, and scores per scenario and TEE explanation can be found in Appendix~\ref{app:followup-stats}. 

\begin{table*}[htp!]
\centering
\centering
\setlength\tabcolsep{1.5pt}
\renewcommand{\arraystretch}{1}
\begin{tabular}{lcccccccccc}
& \multicolumn{10}{c}{\textbf{Medical Scenario Without AI}} \\
\textbf{{ \begin{tabular}{@{}c@{}}  
  Variable \\
  {\scriptsize  McFadden's $R^2$}
  \end{tabular}} }  & 
 \multicolumn{1}{c}{   \begin{tabular}{@{}c@{}}  
  \Qone \\
  {\scriptsize 0.0625	 }
  \end{tabular}} &
 \multicolumn{1}{c}{   \begin{tabular}{@{}c@{}}  
  \Qtwo \\
  {\scriptsize 0.1039	 }
  \end{tabular}} & 
\multicolumn{1}{c}{ \begin{tabular}{@{}c@{}}   \Qthree \\
  {\scriptsize 0.0292		 }
  \end{tabular}} & 
\multicolumn{1}{c}{\begin{tabular}{@{}c@{}}  \Qfour  \\
  {\scriptsize 0.0715 }
  \end{tabular}} & 
\multicolumn{1}{c}{\begin{tabular}{@{}c@{}}  \Qnine  \\
  {\scriptsize 0.02418	}
  \end{tabular}} & 
\multicolumn{1}{c}{\begin{tabular}{@{}c@{}}  \Qfive \\
  {\scriptsize 0.0220 }
  \end{tabular}} &  
\multicolumn{1}{c}{\begin{tabular}{@{}c@{}}  \Qsix \\
  {\scriptsize 0.0204}
  \end{tabular}} & 
\multicolumn{1}{c}{\begin{tabular}{@{}c@{}}  \Qseven \\
  {\scriptsize 0.0124}
  \end{tabular}} & 
\multicolumn{1}{c}{\begin{tabular}{@{}c@{}}  \Qeight  \\
  {\scriptsize 0.0318	 }
  \end{tabular}} & 
\multicolumn{1}{c}{\begin{tabular}{@{}c@{}}  \Qten  \\
  {\scriptsize 0.0300	 }
  \end{tabular}}  \\
\midrule
\multicolumn{11}{l}{\quad\textit{Explanation sentence 1 [Baseline = Unsubstantial]}} \\
  Hardware & 
  \begin{tabular}{@{}c@{}} 
  $1.52$ \\[-2pt]
  {\scriptsize }
  \end{tabular} &
  \begin{tabular}{@{}c@{}} 
  $0.38$ \\[-2pt]
  {\scriptsize }
  \end{tabular} &
  \begin{tabular}{@{}c@{}} 
  $1.32$ \\[-2pt]
  {\scriptsize }
  \end{tabular} &
  \begin{tabular}{@{}c@{}} 
  $0.99$ \\[-2pt]
  {\scriptsize }
  \end{tabular} &
  \begin{tabular}{@{}c@{}}   
    $0.43$ \\[-2pt]
  {\scriptsize }
  \end{tabular}& 
   \begin{tabular}{@{}c@{}}  
    $1.03$ \\[-2pt]
  {\scriptsize }
  \end{tabular}&
    \begin{tabular}{@{}c@{}} 
    $0.47$\\[-2pt]
  {\scriptsize }
  \end{tabular}&
    \begin{tabular}{@{}c@{}} 
    $0.91$\\[-2pt]
  {\scriptsize }
  \end{tabular}&
    \begin{tabular}{@{}c@{}} 
    \cellcolor{negative_medium}$0.41^{**}$ \\[-2pt]
  {\scriptsize [0.210, 0.770]	 }
  \end{tabular}& 
    \begin{tabular}{@{}c@{}} 
    $1.28$\\[-2pt]
  {\scriptsize }
  \end{tabular}
    \\
  Trust & 
  \begin{tabular}{@{}c@{}} 
    $1.46$\\[-2pt]
  {\scriptsize }
  \end{tabular} & 
    \begin{tabular}{@{}c@{}} 
    $0.54$ \\[-2pt]
  {\scriptsize }
  \end{tabular}& 
    \begin{tabular}{@{}c@{}} 
    $1.97$ \\[-2pt]
  {\scriptsize }
  \end{tabular}& 
    \begin{tabular}{@{}c@{}} 
    $0.99$ \\[-2pt]
  {\scriptsize }
  \end{tabular}& 
    \begin{tabular}{@{}c@{}} 
    $0.56$\\[-2pt]
  {\scriptsize }
  \end{tabular}& 
    \begin{tabular}{@{}c@{}} 
    $1.51$ \\[-2pt]
  {\scriptsize }
  \end{tabular}&
    \begin{tabular}{@{}c@{}} 
    $0.70$\\[-2pt]
  {\scriptsize }
  \end{tabular}&
    \begin{tabular}{@{}c@{}} 
    $1.51$\\[-2pt]
  {\scriptsize }
  \end{tabular}&
    \begin{tabular}{@{}c@{}} 
    $0.79$\\[-2pt]
  {\scriptsize }
  \end{tabular}&
    \begin{tabular}{@{}c@{}} 
    $0.93$\\[-2pt]
  {\scriptsize }
  \end{tabular}
    \\
\midrule
\multicolumn{11}{l}{\quad\textit{Explanation sentence 2 [Baseline = Technical]}} \\
  Non-technical &  
  \begin{tabular}{@{}c@{}} 
    $1.17$ \\[-2pt]
  {\scriptsize }
  \end{tabular}&  
  \begin{tabular}{@{}c@{}} 
    \cellcolor{positive_medium}$8.50^{**}$ \\[-2pt]
  {\scriptsize [2.177, 56.657]	 }
  \end{tabular}&  
  \begin{tabular}{@{}c@{}} 
    $0.93$ \\[-2pt]
  {\scriptsize }
  \end{tabular}&  
  \begin{tabular}{@{}c@{}} 
    $1.19$\\[-2pt]
  {\scriptsize }
  \end{tabular} & 
  \begin{tabular}{@{}c@{}} 
    $1.39$\\[-2pt]
  {\scriptsize }
  \end{tabular}& 
  \begin{tabular}{@{}c@{}} 
    $1.63$ \\[-2pt]
  {\scriptsize }
  \end{tabular}& 
  \begin{tabular}{@{}c@{}} 
    $0.89$ \\[-2pt]
  {\scriptsize }
  \end{tabular}& 
  \begin{tabular}{@{}c@{}} 
    $1.15$\\[-2pt]
  {\scriptsize }
  \end{tabular}& 
  \begin{tabular}{@{}c@{}} 
    $1.08$\\[-2pt]
  {\scriptsize }
  \end{tabular}& 
  \begin{tabular}{@{}c@{}} 
    $1.22$\\[-2pt]
  {\scriptsize }
  \end{tabular}
    \\
\midrule
\multicolumn{11}{l}{\quad\textit{Explanation sentence 3 [Baseline = No Prevents]}} \\
  Prevents &  
  \begin{tabular}{@{}c@{}} 
    $0.68$\\[-2pt]
  {\scriptsize }
  \end{tabular} &  
  \begin{tabular}{@{}c@{}} 
    $1.11$\\[-2pt]
  {\scriptsize }
  \end{tabular} &  
  \begin{tabular}{@{}c@{}} 
    $1.89$ \\[-2pt]
  {\scriptsize }
  \end{tabular}&  
  \begin{tabular}{@{}c@{}} 
    \cellcolor{positive_dark}$3.67^{***}$\\[-2pt]
  {\scriptsize [1.778, 8.137]	 }
  \end{tabular} &  
  \begin{tabular}{@{}c@{}} 
    $1.04$\\[-2pt]
  {\scriptsize }
  \end{tabular}& 
  \begin{tabular}{@{}c@{}}  
    $0.90$\\[-2pt]
  {\scriptsize }
  \end{tabular} & 
  \begin{tabular}{@{}c@{}} 
    $0.74$\\[-2pt]
  {\scriptsize }
  \end{tabular}& 
  \begin{tabular}{@{}c@{}} 
    $0.99$\\[-2pt]
  {\scriptsize }
  \end{tabular}& 
  \begin{tabular}{@{}c@{}} 
    $1.14$\\[-2pt]
  {\scriptsize }
  \end{tabular}& 
  \begin{tabular}{@{}c@{}} 
    $0.53$\\[-2pt]
  {\scriptsize }
  \end{tabular}
    \\
\midrule
  Medical experience &  
  \begin{tabular}{@{}c@{}} 
    $0.24$\\[-2pt]
  {\scriptsize }
  \end{tabular} &  
  \begin{tabular}{@{}c@{}} 
    $0.92$\\[-2pt]
  {\scriptsize }
  \end{tabular} &  
  \begin{tabular}{@{}c@{}} 
    $1.09$\\[-2pt]
  {\scriptsize }
  \end{tabular} & 
  \begin{tabular}{@{}c@{}} 
    $2.12$\\[-2pt]
  {\scriptsize }
  \end{tabular} &  
  \begin{tabular}{@{}c@{}} 
    $1.34$\\[-2pt]
  {\scriptsize }
  \end{tabular}&  
  \begin{tabular}{@{}c@{}} 
    $0.84$\\[-2pt]
  {\scriptsize }
  \end{tabular} & 
  \begin{tabular}{@{}c@{}} 
    $1.02$\\[-2pt]
  {\scriptsize }
  \end{tabular}& 
  \begin{tabular}{@{}c@{}} 
    $1.02$\\[-2pt]
  {\scriptsize }
  \end{tabular}& 
  \begin{tabular}{@{}c@{}} 
    $1.31$\\[-2pt]
  {\scriptsize }
  \end{tabular}& 
  \begin{tabular}{@{}c@{}} 
    $1.40$\\[-2pt]
  {\scriptsize }
  \end{tabular}
    \\
  CS experience &  
  \begin{tabular}{@{}c@{}} 
    $0.73$\\[-2pt]
  {\scriptsize }
  \end{tabular} &  
  \begin{tabular}{@{}c@{}} 
    $1.63$\\[-2pt]
  {\scriptsize }
  \end{tabular} &  
  \begin{tabular}{@{}c@{}} 
    $0.97$ \\[-2pt]
  {\scriptsize }
  \end{tabular}&  
  \begin{tabular}{@{}c@{}} 
    $1.21$ \\[-2pt]
  {\scriptsize }
  \end{tabular}& 
  \begin{tabular}{@{}c@{}} 
    $1.42$\\[-2pt]
  {\scriptsize }
  \end{tabular}&  
  \begin{tabular}{@{}c@{}} 
    $1.17$ \\[-2pt]
  {\scriptsize }
  \end{tabular}& 
  \begin{tabular}{@{}c@{}} 
    $1.08$\\[-2pt]
  {\scriptsize }
  \end{tabular}& 
  \begin{tabular}{@{}c@{}} 
    $1.70$\\[-2pt]
  {\scriptsize }
  \end{tabular}& 
  \begin{tabular}{@{}c@{}} 
    $1.45$\\[-2pt]
  {\scriptsize }
  \end{tabular}& 
  \begin{tabular}{@{}c@{}} 
    $0.93$\\[-2pt]
  {\scriptsize }
  \end{tabular}
    \\
\midrule& \multicolumn{10}{c}{\textbf{Medical Scenario With AI}} \\
\textbf{Variable} & 
 \multicolumn{1}{c}{   \begin{tabular}{@{}c@{}}  
  \Qone \\
  {\scriptsize 0.1536	 }
  \end{tabular}} &
 \multicolumn{1}{c}{   \begin{tabular}{@{}c@{}}  
  \Qtwo \\
  {\scriptsize 0.1397	 }
  \end{tabular}} & 
\multicolumn{1}{c}{ \begin{tabular}{@{}c@{}}   \Qthree \\
  {\scriptsize 0.0140		 }
  \end{tabular}} & 
\multicolumn{1}{c}{\begin{tabular}{@{}c@{}}  \Qfour  \\
  {\scriptsize 0.0660 }
  \end{tabular}} & 
\multicolumn{1}{c}{\begin{tabular}{@{}c@{}}  \Qnine  \\
  {\scriptsize 0.0408	}
  \end{tabular}} & 
\multicolumn{1}{c}{\begin{tabular}{@{}c@{}}  \Qfive \\
  {\scriptsize 0.0220 }
  \end{tabular}} &  
\multicolumn{1}{c}{\begin{tabular}{@{}c@{}}  \Qsix \\
  {\scriptsize 0.0273}
  \end{tabular}} & 
\multicolumn{1}{c}{\begin{tabular}{@{}c@{}}  \Qseven \\
  {\scriptsize 0.0420}
  \end{tabular}} & 
\multicolumn{1}{c}{\begin{tabular}{@{}c@{}}  \Qeight  \\
  {\scriptsize 0.0513	 }
  \end{tabular}} & 
\multicolumn{1}{c}{\begin{tabular}{@{}c@{}}  \Qten  \\
  {\scriptsize 0.0613	 }
  \end{tabular}}  \\
\midrule
\multicolumn{11}{l}{\quad\textit{Explanation sentence 1 [Baseline = Unsubstantial]}} \\
  Hardware & 
  \begin{tabular}{@{}c@{}} 
    $1.03$\\[-2pt]
  {\scriptsize }
  \end{tabular} &  
  \begin{tabular}{@{}c@{}} 
    $4.95$\\[-2pt]
  {\scriptsize }
  \end{tabular} &  
  \begin{tabular}{@{}c@{}} 
    $1.07$\\[-2pt]
  {\scriptsize }
  \end{tabular} &  
  \begin{tabular}{@{}c@{}} 
    $1.12$ \\[-2pt]
  {\scriptsize }
  \end{tabular}&  
  \begin{tabular}{@{}c@{}} 
    $1.12$\\[-2pt]
  {\scriptsize }
  \end{tabular}& 
  \begin{tabular}{@{}c@{}}  
    $2.83$\\[-2pt]
  {\scriptsize }
  \end{tabular} & 
  \begin{tabular}{@{}c@{}} 
    $1.86$\\[-2pt]
  {\scriptsize }
  \end{tabular}& 
  \begin{tabular}{@{}c@{}} 
    $1.51$\\[-2pt]
  {\scriptsize }
  \end{tabular}& 
  \begin{tabular}{@{}c@{}} 
    $0.88$\\[-2pt]
  {\scriptsize }
  \end{tabular}& 
  \begin{tabular}{@{}c@{}} 
    $0.73$\\[-2pt]
  {\scriptsize }
  \end{tabular}
    \\
  Trust &  
  \begin{tabular}{@{}c@{}} 
    $0.17$ \\[-2pt]
  {\scriptsize }
  \end{tabular}&  
  \begin{tabular}{@{}c@{}} 
    $0.46$ \\[-2pt]
  {\scriptsize }
  \end{tabular}&  
  \begin{tabular}{@{}c@{}} 
    $0.86$ \\[-2pt]
  {\scriptsize }
  \end{tabular}&  
  \begin{tabular}{@{}c@{}} 
    $0.84$ \\[-2pt]
  {\scriptsize }
  \end{tabular}&  
  \begin{tabular}{@{}c@{}} 
    $1.16$\\[-2pt]
  {\scriptsize }
  \end{tabular}&  
  \begin{tabular}{@{}c@{}} 
    $1.43$\\[-2pt]
  {\scriptsize }
  \end{tabular} & 
  \begin{tabular}{@{}c@{}} 
    $1.63$\\[-2pt]
  {\scriptsize }
  \end{tabular}& 
  \begin{tabular}{@{}c@{}} 
    $1.20$\\[-2pt]
  {\scriptsize }
  \end{tabular}& 
  \begin{tabular}{@{}c@{}} 
    $0.84$\\[-2pt]
  {\scriptsize }
  \end{tabular}& 
  \begin{tabular}{@{}c@{}} 
    $1.04$\\[-2pt]
  {\scriptsize }
  \end{tabular}
    \\
\midrule
\multicolumn{11}{l}{\quad\textit{Explanation sentence 2 [Baseline = Technical]}} \\
  Non-technical &  
  \begin{tabular}{@{}c@{}} 
    $0.79$ \\[-2pt]
  {\scriptsize }
  \end{tabular}&  
  \begin{tabular}{@{}c@{}} 
    $1.60$ \\[-2pt]
  {\scriptsize }
  \end{tabular}&  
  \begin{tabular}{@{}c@{}} 
    $1.11$\\[-2pt]
  {\scriptsize }
  \end{tabular} &  
  \begin{tabular}{@{}c@{}} 
    $0.63$ \\[-2pt]
  {\scriptsize }
  \end{tabular}&  
  \begin{tabular}{@{}c@{}} 
    $1.54$\\[-2pt]
  {\scriptsize }
  \end{tabular}& 
  \begin{tabular}{@{}c@{}}  
    $0.89$ \\[-2pt]
  {\scriptsize }
  \end{tabular}& 
  \begin{tabular}{@{}c@{}} 
    $0.63$\\[-2pt]
  {\scriptsize }
  \end{tabular} & 
  \begin{tabular}{@{}c@{}} 
    $0.61$\\[-2pt]
  {\scriptsize }
  \end{tabular}& 
  \begin{tabular}{@{}c@{}} 
    $0.58$\\[-2pt]
  {\scriptsize }
  \end{tabular}& 
  \begin{tabular}{@{}c@{}} 
    $2.03$\\[-2pt]
  {\scriptsize }
  \end{tabular}
    \\
\midrule
\multicolumn{11}{l}{\quad\textit{Explanation sentence 3 [Baseline = No Prevents]}} \\
  Prevents &  
  \begin{tabular}{@{}c@{}} 
    $0.57$\\[-2pt]
  {\scriptsize }
  \end{tabular} &  
  \begin{tabular}{@{}c@{}} 
    $0.39$\\[-2pt]
  {\scriptsize }
  \end{tabular} &  
  \begin{tabular}{@{}c@{}} 
    $1.70$\\[-2pt]
  {\scriptsize }
  \end{tabular}&  
  \begin{tabular}{@{}c@{}} 
    \cellcolor{positive_medium}$2.83^{**}$ \\[-2pt]
  {\scriptsize [1.367, 6.167]	 }
  \end{tabular}&  
  \begin{tabular}{@{}c@{}} 
    $1.27$\\[-2pt]
  {\scriptsize }
  \end{tabular}&  
  \begin{tabular}{@{}c@{}} 
    $0.96$ \\[-2pt]
  {\scriptsize }
  \end{tabular}& 
  \begin{tabular}{@{}c@{}} 
    $1.75$\\[-2pt]
  {\scriptsize }
  \end{tabular}& 
  \begin{tabular}{@{}c@{}} 
    $1.67$\\[-2pt]
  {\scriptsize }
  \end{tabular}& 
  \begin{tabular}{@{}c@{}} 
    $0.54$\\[-2pt]
  {\scriptsize }
  \end{tabular}& 
  \begin{tabular}{@{}c@{}} 
    $1.01$\\[-2pt]
  {\scriptsize }
  \end{tabular}
    \\
\midrule
  Medical experience &  
  \begin{tabular}{@{}c@{}} 
    $1.38$\\[-2pt]
  {\scriptsize }
  \end{tabular}& 
  \begin{tabular}{@{}c@{}} 
    $2.20$\\[-2pt]
  {\scriptsize }
  \end{tabular} &  
  \begin{tabular}{@{}c@{}} 
    $1.23$\\[-2pt]
  {\scriptsize }
  \end{tabular} & 
  \begin{tabular}{@{}c@{}} 
    $1.01$ \\[-2pt]
  {\scriptsize }
  \end{tabular}&  
  \begin{tabular}{@{}c@{}} 
    $0.79$\\[-2pt]
  {\scriptsize }
  \end{tabular}&  
  \begin{tabular}{@{}c@{}} 
    $1.31$ \\[-2pt]
  {\scriptsize }
  \end{tabular}& 
  \begin{tabular}{@{}c@{}} 
    $0.95$\\[-2pt]
  {\scriptsize }
  \end{tabular}& 
  \begin{tabular}{@{}c@{}} 
    $2.75$\\[-2pt]
  {\scriptsize }
  \end{tabular}& 
  \begin{tabular}{@{}c@{}} 
    $3.10$\\[-2pt]
  {\scriptsize }
  \end{tabular}& 
  \begin{tabular}{@{}c@{}} 
    $\cellcolor{positive_dark}3.60^{***}$\\[-2pt]
  {\scriptsize [1.759, 7.857]}
  \end{tabular}
    \\
  CS experience &  
  \begin{tabular}{@{}c@{}} 
    $0.23$\\[-2pt]
  {\scriptsize }
  \end{tabular} &     
  \begin{tabular}{@{}c@{}} 
    $0.44$ \\[-2pt]
  {\scriptsize } 
  \end{tabular} &  
  \begin{tabular}{@{}c@{}} 
    $0.87$ \\[-2pt]
  {\scriptsize }
  \end{tabular} &  
  \begin{tabular}{@{}c@{}} 
    $0.468$ \\[-2pt]
  {\scriptsize }
  \end{tabular}& 
  \begin{tabular}{@{}c@{}} 
    \cellcolor{negative_light}$0.39^*$\\[-2pt]
  {\scriptsize [0.184, 0.824]}
  \end{tabular}& 
  \begin{tabular}{@{}c@{}} 
    $0.93$\\[-2pt]
  {\scriptsize }
  \end{tabular} & 
  \begin{tabular}{@{}c@{}} 
    $1.27$\\[-2pt]
  {\scriptsize }
  \end{tabular}& 
  \begin{tabular}{@{}c@{}} 
    $1.49$\\[-2pt]
  {\scriptsize }
  \end{tabular}& 
  \begin{tabular}{@{}c@{}} 
    $0.79$\\[-2pt]
  {\scriptsize }
  \end{tabular}& 
  \begin{tabular}{@{}c@{}} 
    $1.30$\\[-2pt]
  {\scriptsize }
  \end{tabular}
    \\
\bottomrule
\multicolumn{11}{l}{$^{***}p < 0.001$; $^{**}p < 0.01$; $^{*}p < 0.05$}
\end{tabular}

\vspace{2mm}
\caption{Regression table for True/False comprehension questions in \textbf{Survey 1} for the Medical Scenarios, where each question in each scenario is a different model (20 models total) \new{ corrected with the Benjamini-Hochberg procedure between scenarios and McFadden's $R^2$ for each model. McFadden's $R^2$ is a pseudo-$R^2$ measure of model fit, where higher values (0.2-0.4) indicate better explanatory power relative to an intercept-only model, though values are typically much smaller than in linear regression.} The numbers are the odds ratios for each predictor, with the baseline used in each model noted in \textit{italics}. Statistical significance is noted with asterisks and shaded cells: blue for positive and orange for negative coefficients. \new{The values shown below each significant odds ratio are the 95\% confidence intervals (CI), indicating the range of plausible effect sizes.}}
\label{tab:stats-medical}
\end{table*}

\begin{table*}[htp!]
\centering
\centering
\setlength\tabcolsep{1.5pt}
\renewcommand{\arraystretch}{1}
\begin{tabular}{lcccccccccc}
& \multicolumn{10}{c}{\textbf{Smart Home Scenario Without AI}} \\
\textbf{{ \begin{tabular}{@{}c@{}}  
  Variable \\
  {\scriptsize  McFadden's $R^2$}
  \end{tabular}} }  & 
 \multicolumn{1}{c}{   \begin{tabular}{@{}c@{}}  
  \Qone \\
  {\scriptsize 0.1434	 }
  \end{tabular}} &
 \multicolumn{1}{c}{   \begin{tabular}{@{}c@{}}  
  \Qtwo \\
  {\scriptsize 0.0542	 }
  \end{tabular}} & 
\multicolumn{1}{c}{ \begin{tabular}{@{}c@{}}   \Qthree \\
  {\scriptsize 0.0370		 }
  \end{tabular}} & 
\multicolumn{1}{c}{\begin{tabular}{@{}c@{}}  \Qfour  \\
  {\scriptsize 0.0483 }
  \end{tabular}} & 
\multicolumn{1}{c}{\begin{tabular}{@{}c@{}}  \Qnine  \\
  {\scriptsize 0.0089	}
  \end{tabular}} & 
\multicolumn{1}{c}{\begin{tabular}{@{}c@{}}  \Qfive \\
  {\scriptsize 0.0403 }
  \end{tabular}} &  
\multicolumn{1}{c}{\begin{tabular}{@{}c@{}}  \Qsix \\
  {\scriptsize 0.0184}
  \end{tabular}} & 
\multicolumn{1}{c}{\begin{tabular}{@{}c@{}}  \Qseven \\
  {\scriptsize 0.0171}
  \end{tabular}} & 
\multicolumn{1}{c}{\begin{tabular}{@{}c@{}}  \Qeight  \\
  {\scriptsize 0.0276	 }
  \end{tabular}} & 
\multicolumn{1}{c}{\begin{tabular}{@{}c@{}}  \Qten  \\
  {\scriptsize 0.0117	 }
  \end{tabular}}  \\
\midrule
\multicolumn{11}{l}{\quad\textit{Explanation sentence 1 [Baseline = Unsubstantial]}} \\
  Hardware &
   
  \begin{tabular}{@{}c@{}} 
    $0.96$ \\[-2pt]
  {\scriptsize }
  \end{tabular}&  
  \begin{tabular}{@{}c@{}} 
    $0.52$ \\[-2pt]
  {\scriptsize }
  \end{tabular}&  
  \begin{tabular}{@{}c@{}} 
    $0.76$ \\[-2pt]
  {\scriptsize }
  \end{tabular}&  
  \begin{tabular}{@{}c@{}} 
    $0.70$ \\[-2pt]
  {\scriptsize }
  \end{tabular}&  
  \begin{tabular}{@{}c@{}} 
    $1.02$\\[-2pt]
  {\scriptsize }
  \end{tabular}& 
  \begin{tabular}{@{}c@{}}  
    $1.21$\\[-2pt]
  {\scriptsize }
  \end{tabular} & 
  \begin{tabular}{@{}c@{}} 
    $0.90$\\[-2pt]
  {\scriptsize }
  \end{tabular}& 
  \begin{tabular}{@{}c@{}} 
    $0.90$\\[-2pt]
  {\scriptsize }
  \end{tabular}& 
  \begin{tabular}{@{}c@{}} 
    $0.92$\\[-2pt]
  {\scriptsize }
  \end{tabular}& 
  \begin{tabular}{@{}c@{}} 
    $1.38$\\[-2pt]
  {\scriptsize }
  \end{tabular}
    \\
  Trust &  
  \begin{tabular}{@{}c@{}} 
    $0.62$ \\[-2pt]
  {\scriptsize }
  \end{tabular}&  
  \begin{tabular}{@{}c@{}} 
    $0.63$ \\[-2pt]
  {\scriptsize }
  \end{tabular}&  
  \begin{tabular}{@{}c@{}} 
    $1.39$ \\[-2pt]
  {\scriptsize }
  \end{tabular}&  
  \begin{tabular}{@{}c@{}} 
    $1.21$ \\[-2pt]
  {\scriptsize }
  \end{tabular}&  
  \begin{tabular}{@{}c@{}} 
    $0.75$\\[-2pt]
  {\scriptsize }
  \end{tabular}& 
  \begin{tabular}{@{}c@{}} 
    $0.93$\\[-2pt]
  {\scriptsize }
  \end{tabular} & 
  \begin{tabular}{@{}c@{}} 
    $0.71$\\[-2pt]
  {\scriptsize }
  \end{tabular}& 
  \begin{tabular}{@{}c@{}} 
    $0.99$\\[-2pt]
  {\scriptsize }
  \end{tabular}& 
  \begin{tabular}{@{}c@{}} 
    $0.88$\\[-2pt]
  {\scriptsize }
  \end{tabular}& 
  \begin{tabular}{@{}c@{}} 
    $1.21$\\[-2pt]
  {\scriptsize }
  \end{tabular}
    \\
\midrule
\multicolumn{11}{l}{\quad\textit{Explanation sentence 2 [Baseline = Technical]}} \\
  Non-technical &  
  \begin{tabular}{@{}c@{}} 
    $0.84$\\[-2pt]
  {\scriptsize }
  \end{tabular} &  
  \begin{tabular}{@{}c@{}} 
    $2.56$\\[-2pt]
  {\scriptsize }
  \end{tabular} &  
  \begin{tabular}{@{}c@{}} 
    $1.26$\\[-2pt]
  {\scriptsize }
  \end{tabular} &  
  \begin{tabular}{@{}c@{}} 
    $0.92$ \\[-2pt]
  {\scriptsize }
  \end{tabular}& 
  \begin{tabular}{@{}c@{}} 
    $1.04$\\[-2pt]
  {\scriptsize }
  \end{tabular}& 
  \begin{tabular}{@{}c@{}}  
    $1.32$\\[-2pt]
  {\scriptsize }
  \end{tabular} & 
  \begin{tabular}{@{}c@{}} 
    $0.87$\\[-2pt]
  {\scriptsize }
  \end{tabular} & 
  \begin{tabular}{@{}c@{}} 
    $1.12$\\[-2pt]
  {\scriptsize }
  \end{tabular}& 
  \begin{tabular}{@{}c@{}} 
    $0.78$\\[-2pt]
  {\scriptsize }
  \end{tabular}& 
  \begin{tabular}{@{}c@{}} 
    $1.27$\\[-2pt]
  {\scriptsize }
  \end{tabular}
    \\
\midrule
\multicolumn{11}{l}{\quad\textit{Explanation sentence 3 [Baseline = No Prevents]}} \\
  Prevents &  
  \begin{tabular}{@{}c@{}} 
    $0.37$\\[-2pt]
  {\scriptsize }
  \end{tabular} &  
  \begin{tabular}{@{}c@{}} 
    $0.99$\\[-2pt]
  {\scriptsize }
  \end{tabular} &  
  \begin{tabular}{@{}c@{}} 
    $1.07$\\[-2pt]
  {\scriptsize }
  \end{tabular} &  
  \begin{tabular}{@{}c@{}} 
    $1.67$\\[-2pt]
  {\scriptsize }
  \end{tabular} &  
  \begin{tabular}{@{}c@{}} 
    $1.17$\\[-2pt]
  {\scriptsize }
  \end{tabular}&  
  \begin{tabular}{@{}c@{}} 
    $1.03$\\[-2pt]
  {\scriptsize }
  \end{tabular} & 
  \begin{tabular}{@{}c@{}} 
    $0.59$\\[-2pt]
  {\scriptsize }
  \end{tabular}& 
  \begin{tabular}{@{}c@{}} 
    $1.03$\\[-2pt]
  {\scriptsize }
  \end{tabular}& 
  \begin{tabular}{@{}c@{}} 
    $1.12$\\[-2pt]
  {\scriptsize }
  \end{tabular}& 
  \begin{tabular}{@{}c@{}} 
    $1.01$\\[-2pt]
  {\scriptsize }
  \end{tabular}
    \\
\midrule
  Smart home experience &  
  \begin{tabular}{@{}c@{}} 
    $0.00$ \\[-2pt]
  {\scriptsize }
  \end{tabular}&  
  \begin{tabular}{@{}c@{}} 
    $0.23$ \\[-2pt]
  {\scriptsize }
  \end{tabular}&  
  \begin{tabular}{@{}c@{}} 
    $0.28$\\[-2pt]
  {\scriptsize }
  \end{tabular} & 
  \begin{tabular}{@{}c@{}} 
    $0.36$ \\[-2pt]
  {\scriptsize }
  \end{tabular}&  
  \begin{tabular}{@{}c@{}} 
    $0.63$\\[-2pt]
  {\scriptsize }
  \end{tabular}&  
  \begin{tabular}{@{}c@{}}
    $0.57$\\[-2pt]
  {\scriptsize }
  \end{tabular} & 
  \begin{tabular}{@{}c@{}} 
    $0.60$\\[-2pt]
  {\scriptsize }
  \end{tabular}& 
  \begin{tabular}{@{}c@{}} 
    $0.73$\\[-2pt]
  {\scriptsize }
  \end{tabular}& 
  \begin{tabular}{@{}c@{}} 
    $0.45$\\[-2pt]
  {\scriptsize }
  \end{tabular}& 
  \begin{tabular}{@{}c@{}} 
    $0.97$\\[-2pt]
  {\scriptsize }
  \end{tabular}
    \\
  CS experience &  
  \begin{tabular}{@{}c@{}} 
    $0.17$\\[-2pt]
  {\scriptsize }
  \end{tabular} &  
  \begin{tabular}{@{}c@{}} 
    $1.51$ \\[-2pt]
  {\scriptsize }
  \end{tabular}&  
  \begin{tabular}{@{}c@{}}   
    $0.70$ \\[-2pt]
  {\scriptsize }
  \end{tabular}&  
  \begin{tabular}{@{}c@{}} 
    $0.59$ \\[-2pt]
  {\scriptsize }
  \end{tabular}& 
  \begin{tabular}{@{}c@{}} 
    $0.85$\\[-2pt]
  {\scriptsize }
  \end{tabular}&  
  \begin{tabular}{@{}c@{}} 
    $0.37$ \\[-2pt]
  {\scriptsize }
  \end{tabular}& 
  \begin{tabular}{@{}c@{}} 
    $0.95$\\[-2pt]
  {\scriptsize }
  \end{tabular}& 
  \begin{tabular}{@{}c@{}} 
    $0.51$\\[-2pt]
  {\scriptsize }
  \end{tabular}& 
  \begin{tabular}{@{}c@{}} 
    $1.93$\\[-2pt]
  {\scriptsize }
  \end{tabular}& 
  \begin{tabular}{@{}c@{}} 
    $0.65$\\[-2pt]
  {\scriptsize }
  \end{tabular}
    \\
\midrule
& \multicolumn{10}{c}{\textbf{Smart Home Scenario With AI}} \\
\textbf{Variable} & 
 \multicolumn{1}{c}{   \begin{tabular}{@{}c@{}}  
  \Qone \\
  {\scriptsize 0.0757	 }
  \end{tabular}} &
 \multicolumn{1}{c}{   \begin{tabular}{@{}c@{}}  
  \Qtwo \\
  {\scriptsize 0.0750	 }
  \end{tabular}} & 
\multicolumn{1}{c}{ \begin{tabular}{@{}c@{}}   \Qthree \\
  {\scriptsize 0.0166	 }
  \end{tabular}} & 
\multicolumn{1}{c}{\begin{tabular}{@{}c@{}}  \Qfour  \\
  {\scriptsize 0.0492 }
  \end{tabular}} & 
\multicolumn{1}{c}{\begin{tabular}{@{}c@{}}  \Qnine  \\
  {\scriptsize 0.0312	}
  \end{tabular}} & 
\multicolumn{1}{c}{\begin{tabular}{@{}c@{}}  \Qfive \\
  {\scriptsize 0.0124 }
  \end{tabular}} &  
\multicolumn{1}{c}{\begin{tabular}{@{}c@{}}  \Qsix \\
  {\scriptsize 0.0306}
  \end{tabular}} & 
\multicolumn{1}{c}{\begin{tabular}{@{}c@{}}  \Qseven \\
  {\scriptsize 0.0154}
  \end{tabular}} & 
\multicolumn{1}{c}{\begin{tabular}{@{}c@{}}  \Qeight  \\
  {\scriptsize 0.0223	 }
  \end{tabular}} & 
\multicolumn{1}{c}{\begin{tabular}{@{}c@{}}  \Qten  \\
  {\scriptsize 0.0240	 }
  \end{tabular}}  \\
\midrule
\multicolumn{11}{l}{\quad\textit{Explanation sentence 1 [Baseline = Unsubstantial]}} \\
  Hardware &  
  \begin{tabular}{@{}c@{}} 
    $0.44$\\[-2pt]
  {\scriptsize }
  \end{tabular} &  
  \begin{tabular}{@{}c@{}} 
    $0.90$\\[-2pt]
  {\scriptsize }
  \end{tabular} &  
  \begin{tabular}{@{}c@{}} 
    $1.36$\\[-2pt]
  {\scriptsize }
  \end{tabular} &  
  \begin{tabular}{@{}c@{}} 
    $1.48$\\[-2pt]
  {\scriptsize }
  \end{tabular} &  
  \begin{tabular}{@{}c@{}} 
    $1.34$\\[-2pt]
  {\scriptsize }
  \end{tabular}& 
  \begin{tabular}{@{}c@{}} 
    $0.70$ \\[-2pt]
  {\scriptsize }
  \end{tabular}& 
  \begin{tabular}{@{}c@{}} 
    $0.66$\\[-2pt]
  {\scriptsize }
  \end{tabular}& 
  \begin{tabular}{@{}c@{}} 
    $0.90$\\[-2pt]
  {\scriptsize }
  \end{tabular}& 
  \begin{tabular}{@{}c@{}} 
    $0.61$\\[-2pt]
  {\scriptsize }
  \end{tabular}& 
  \begin{tabular}{@{}c@{}} 
    $0.66$\\[-2pt]
  {\scriptsize }
  \end{tabular}
    \\
  Trust & 
\begin{tabular}{@{}c@{}} 
  
    $0.18$\\[-2pt]
  {\scriptsize }
  \end{tabular} &  
  \begin{tabular}{@{}c@{}} 
    $0.35$ \\[-2pt]
  {\scriptsize }
  \end{tabular}&  
  \begin{tabular}{@{}c@{}} 
    $1.42$\\[-2pt]
  {\scriptsize }
  \end{tabular} &  
  \begin{tabular}{@{}c@{}} 
    $0.81$ \\[-2pt]
  {\scriptsize }
  \end{tabular}&  
  \begin{tabular}{@{}c@{}} 
    $0.80$\\[-2pt]
  {\scriptsize }
  \end{tabular}& 
  \begin{tabular}{@{}c@{}}  
    $0.80$ \\[-2pt]
  {\scriptsize }
  \end{tabular}& 
  \begin{tabular}{@{}c@{}} 
    $1.17$\\[-2pt]
  {\scriptsize }
  \end{tabular}& 
  \begin{tabular}{@{}c@{}} 
    $1.49$\\[-2pt]
  {\scriptsize }
  \end{tabular}& 
  \begin{tabular}{@{}c@{}} 
    $0.50$\\[-2pt]
  {\scriptsize }
  \end{tabular}& 
  \begin{tabular}{@{}c@{}} 
    $1.02$\\[-2pt]
  {\scriptsize }
  \end{tabular}
    \\
\midrule
\multicolumn{11}{l}{\quad\textit{Explanation sentence 2 [Baseline = Technical]}} \\
  Non-technical & 
  \begin{tabular}{@{}c@{}} 
    $1.14$ \\[-2pt]
  {\scriptsize }
  \end{tabular}&  
  \begin{tabular}{@{}c@{}} 
    $2.03$\\[-2pt]
  {\scriptsize }
  \end{tabular} &  
  \begin{tabular}{@{}c@{}} 
    $1.15$ \\[-2pt]
  {\scriptsize }
  \end{tabular}&  
  \begin{tabular}{@{}c@{}} 
    $0.70$ \\[-2pt]
  {\scriptsize }
  \end{tabular}&  
  \begin{tabular}{@{}c@{}} 
    $0.95$\\[-2pt]
  {\scriptsize }
  \end{tabular}& 
  \begin{tabular}{@{}c@{}} 
    $1.02$\\[-2pt]
  {\scriptsize }
  \end{tabular} & 
  \begin{tabular}{@{}c@{}} 
    $0.54$\\[-2pt]
  {\scriptsize }
  \end{tabular} & 
  \begin{tabular}{@{}c@{}} 
    $1.23$\\[-2pt]
  {\scriptsize }
  \end{tabular}& 
  \begin{tabular}{@{}c@{}} 
    $1.12$\\[-2pt]
  {\scriptsize }
  \end{tabular}& 
  \begin{tabular}{@{}c@{}} 
    $1.42$\\[-2pt]
  {\scriptsize }
  \end{tabular}
    \\
\midrule
\multicolumn{11}{l}{\quad\textit{Explanation sentence 3 [Baseline = No Prevents]}} \\
  Prevents &  
  \begin{tabular}{@{}c@{}} 
    $0.55$\\[-2pt]
  {\scriptsize }
  \end{tabular} &  
  \begin{tabular}{@{}c@{}} 
    $2.08$\\[-2pt]
  {\scriptsize }
  \end{tabular} &  
  \begin{tabular}{@{}c@{}} 
    $0.68$ \\[-2pt]
  {\scriptsize }
  \end{tabular}&  
  \begin{tabular}{@{}c@{}} 
    \cellcolor{positive_medium}$2.69^{**}$ \\[-2pt]
  {\scriptsize[1.347, 5.571]	 }
  \end{tabular}&  
  \begin{tabular}{@{}c@{}} 
    $2.23$
    \\[-2pt]
  {\scriptsize }
  \end{tabular} &  

   \begin{tabular}{@{}c@{}} 
    $1.16$ \\[-2pt]
  {\scriptsize }
  \end{tabular}& 
  \begin{tabular}{@{}c@{}} 
    $1.30$\\[-2pt]
  {\scriptsize }
  \end{tabular}& 
  \begin{tabular}{@{}c@{}} 
    $1.12$\\[-2pt]
  {\scriptsize }
  \end{tabular}& 
  \begin{tabular}{@{}c@{}} 
    $0.78$\\[-2pt]
  {\scriptsize }
  \end{tabular}& 
  \begin{tabular}{@{}c@{}} 
    $0.67$\\[-2pt]
  {\scriptsize }
  \end{tabular}
    \\
\midrule
  Smart home experience &  
  \begin{tabular}{@{}c@{}} 
    $0.55$\\[-2pt]
  {\scriptsize }
  \end{tabular} &  
  \begin{tabular}{@{}c@{}} 
    $2.08$\\[-2pt]
  {\scriptsize }
  \end{tabular} &  
  \begin{tabular}{@{}c@{}} 
    $0.68$\\[-2pt]
  {\scriptsize }
  \end{tabular} &  
  \begin{tabular}{@{}c@{}} 
    $1.26$\\[-2pt]
  {\scriptsize }
  \end{tabular} &  
  \begin{tabular}{@{}c@{}} 
    $1.20$\\[-2pt]
  {\scriptsize }
  \end{tabular}& 
  \begin{tabular}{@{}c@{}} 
    $0.52$\\[-2pt]
  {\scriptsize } 
  \end{tabular} &
  \begin{tabular}{@{}c@{}}
    $1.06$\\[-2pt]
  {\scriptsize }
  \end{tabular}& 
  \begin{tabular}{@{}c@{}} 
    $0.70$\\[-2pt]
  {\scriptsize }
  \end{tabular}& 
  \begin{tabular}{@{}c@{}} 
    $1.01$\\[-2pt]
  {\scriptsize }
  \end{tabular}& 
  \begin{tabular}{@{}c@{}} 
    $0.61$\\[-2pt]
  {\scriptsize }
  \end{tabular}
    \\
  CS experience & 
  \begin{tabular}{@{}c@{}} 
    $0.45$\\[-2pt]
  {\scriptsize }
  \end{tabular} & 
  \begin{tabular}{@{}c@{}} 
    $0.63$\\[-2pt]
  {\scriptsize }
  \end{tabular} &  
  \begin{tabular}{@{}c@{}} 
    $1.17$\\[-2pt]
  {\scriptsize }
  \end{tabular} &  
  \begin{tabular}{@{}c@{}} 
    $1.26$\\[-2pt]
  {\scriptsize }
  \end{tabular} & 
  \begin{tabular}{@{}c@{}} 
    $0.68$\\[-2pt]
  {\scriptsize }
  \end{tabular}& 
  \begin{tabular}{@{}c@{}}  
    $1.07$\\[-2pt]
  {\scriptsize }
  \end{tabular} & 
  \begin{tabular}{@{}c@{}} 
    $1.39$\\[-2pt]
  {\scriptsize }
  \end{tabular}& 
  \begin{tabular}{@{}c@{}} 
    $1.27$\\[-2pt]
  {\scriptsize }
  \end{tabular}& 
  \begin{tabular}{@{}c@{}} 
    $1.52$\\[-2pt]
  {\scriptsize }
  \end{tabular}& 
  \begin{tabular}{@{}c@{}} 
    $1.34$\\[-2pt]
  {\scriptsize }
  \end{tabular}
    \\
\bottomrule
\multicolumn{11}{l}{$^{***}p < 0.001$; $^{**}p < 0.01$; $^{*}p < 0.05$}
\end{tabular}

\vspace{2mm}
\caption{Regression table for True/False comprehension questions in \textbf{Survey 1} for the IoT Scenarios, where each question in each scenario is a different model (20 models total) \new{ corrected with the Benjamini-Hochberg procedure between scenarios and McFadden's $R^2$ for each model. McFadden's $R^2$ is a pseudo-$R^2$ measure of model fit, where higher values (0.2-0.4) indicate better explanatory power relative to an intercept-only model, though values are typically much smaller than in linear regression.} The numbers are the odds ratios for each predictor, with the baseline used in each model noted in \textit{italics}. Statistical significance is noted with asterisks and shaded cells: blue for positive and orange for negative coefficients. \new{The values shown below each significant odds ratio are the 95\% confidence intervals (CI), indicating the range of plausible effect sizes.}}
\label{tab:stats-iot}
\end{table*}

\Paragraph{Participants are less likely to recognize TEE limitations.}
Looking at the overall trend in comprehension scores across all conditions, shown in Figure~\ref{fig:comp}, participants are more likely to correctly answer \Qone{} through \Qnine{} (86.5\% correct overall) than \Qfive{} through \Qten{} (71.5\% overall). 
We also tested this hypothesis with a Wilcoxon signed-rank test comparing the average score for questions about limitations, to the average
score for questions about features in Survey 1. We found a significant difference between these two groups for the two smart home scenarios (without AI \(p\ < 0.001\), with AI \(p < 0.01\)) and for the medical scenario without AI (\(p < 0.001\)). Complete results are shown in Table~\ref{tab:features_limitations_stats} (similar results hold for Survey 2, shown in Table~\ref{tab:features_limitations_stats_1st} in Appendix~\ref{app:followup-stats}).
We hypothesize that this difference is because \Qone{} through \Qnine{} ask about TEE capabilities, while \Qfive{} through \Qten{} ask about TEE limitations. Another possible explanation is that \Qone{} through \Qnine{} can be answered by recalling facts from the explanations, while \Qfive{} through \Qten{} require more inference.

\Paragraph{TEE  \textit{Prevents} and \textit{Non-technical} explanations can improve comprehension.}
As described in Section~\ref{sec:quantanalysis}, we ran 40 regression models to predict the relationship between each explanation factor and experience on each of the four scenarios and 10 comprehension questions.
We found that some explanations are better than others. 
The \textit{Non-technical} explanation is especially good at explaining who is (or is not) allowed to access the data in the TEE (\Qtwo{}) while the \textit{Prevents} explanation is best at explaining that the TEE protects against malicious software on the rest of the computer (\Qfour{}).
The effects for the \textit{Non-technical} explanation hold for all scenarios and are significant for the medical scenario without AI (\(p < 0.01\)). Similarly, the effects for the \textit{Prevents} explanation hold across all scenarios for \Qfour{} and are significant for both medical scenarios (\(p < 0.001\) without AI, \(p < 0.01\) with AI) and the smart home scenario with AI (\(p < 0.01\)).

\subsection{RQ2: Factors Influencing Willingness and Feeling of Safety}
\label{subsec:willingness}

While the \textit{type of technology} we described in the scenario seems to have an effect on participant willingness to use TEE-enhanced technology and belief that TEE-enhanced technology will keep their data safe, this does not seem to be the case for the \textit{TEE explanation}. 
Most explanation predictors are not significant in our regression models, except for the \textit{Non-Technical} explanation, which seems to make participants feel significantly safer in the smart home device scenario without AI (see Table~\ref{tab:trust-stats}).

\Paragraph{TEE explanations seem to have little effect on willingness and feeling of safety.}
Regardless of TEE explanation, our results  
suggest that participants were nearly equally willing to engage with the technology in our scenarios. 20\%-22.4\% said they were ``definitely willing'' and 50.4\%-55.7\% were ``maybe willing'' across all scenarios. Similarly, participants seem to believe their data would be nearly equally safe regardless of how the TEE was explained. 24.0\%-28.3\% said it would be ``completely safe'' and 62.3-66.9\% said it would be ``somewhat'' safe across all scenarios. 
The only explanation that had a significant effect was the 
\textit{Non-technical}  explanation on the perception of safety, and only for the smart home scenario without AI (\(p < 0.01\)).

\subsection{Questions from our participants}
\label{sec:questions}
The survey had two opportunities for participants to ask us questions about TEEs. We received 310 questions from 252 participants. In this section, we describe the most common types of questions asked. Note that participants may have written multiple questions in a single response, with each question potentially having one or more theme. Our codebook describing themes and the frequency of each theme can be found in Appendix{app:questions}.

Our participants had many questions that our TEE explanations did not answer. The most common questions were about TEEs, but there were many other questions about the scenario where the TEE is used, potential risks they might encounter, and what guarantees there are that the TEE will function as described. We also received some comments indicating that participants lack trust in TEEs and data privacy, in general. Participants asked more questions after the medical scenarios, and these questions were more likely to focus on TEEs and potential risks than in the smart home scenarios.

\Paragraph{Attributing quotes to participants.}
When attributing quotes, we report the participant ID, which survey the quote is from, and the treatment they were assigned. For Survey 1, each participant receives three letters, corresponding to the three sentences in the TEE explanation they received: (H)ardware, (T)rust, or (U)nsubstantial; (T)echnical or (N)on-technical; and (P)revents, or (N)o Prevents. For example, (P30S1-HTN) means participant \#30 in Survey 1, who received the \textit{Hardware}, \textit{Technical}, \textit{No Prevents} explanation.

\Paragraph{Questions about TEEs.}
The most common questions we received were about TEEs (143 responses). 49 of these questions asked for more technical details generally: {“How exactly does a TEE work?”} (P55S1-TTP). 
21 participants wanted more information about how the TEE creates an isolated environment: {``How exactly is a TEE isolated from the rest of the computer?''} (P146S1-HNP). 
15 participants wanted more implementation details: {``Is there a second set of RAM with an independent CPU or something?''} (P379S1-HTP). 
In addition, 15 participants wanted more information about what else the machine is capable of:  {``\ldots I presume that means that the [researchers] cannot use any other programs at the same time?''} (P48S1-UTN). 
It is noteworthy that only 13 participants asked questions that should have already been answered by the scenario text or the TEE explanation they received. Since this represents relatively few of the responses we received (less than 5\%), it suggests that most participants were paying attention to our survey.

\begin{table*}[htp!]
\centering
\begin{tabular}{l|llllllll} 
\toprule
& \multicolumn{4}{c}{Medical} & \multicolumn{4}{c}{Smart Home}  \\ 
& \multicolumn{2}{c}{With AI} & \multicolumn{2}{c}{Without AI} & \multicolumn{2}{c}{With AI} & \multicolumn{2}{c}{Without AI}  \\ 
\midrule          
& W & p-value 
& W & p-value 
& W & p-value 
& W & p-value 
              \\ 
\midrule
\textbf{Average Score}        
& 7454  &  $3.97e-03^{***}$                  
& 11862  &  $2.66e-10^{***}$  
& 10020  &  $1.33e-09^{***}$ 
& 15110  &  $4.77e-12^{***}$                      \\
\bottomrule
\multicolumn{9}{l}{$^{***}p < 0.001$; $^{**}p < 0.01$; $^{*}p < 0.05$}
\end{tabular}
\vspace{2mm}
\caption{{Table with results of  Wilcoxon signed-rank tests comparing the average score for questions Q1-Q5, to the average score for questions Q6-Q10 in Study 1. 
Each scenario was treated separately, and corrected with the Benjamini-Hochberg procedure.
Statistical significance is noted with asterisks.}}\label{tab:features_limitations_stats}
\end{table*}

\Paragraph{Questions about the scenarios.}
We received 91 questions about our scenarios. 41 asked about the data involved, including what data is collected, data retention policies, and how/whether the data is anonymized: {``What happens to my data when I no longer wish for it to be stored''} (P149S1-TNN). 
25 questions  were about the people on the research/development team: {``How many people have access to the TEE, what are their qualifications\ldots.''} (P12S1-HTN).

\Paragraph{Questions about risks.}
74 participants asked about potential risks. The most common risk, mentioned by 27 participants, was hackers: {``I get that the program is safe from other possibly malicious programs, but what about hackers''} (P47S1-TTP). 
23 participants were concerned about people behaving maliciously, including the people in the scenario with legitimate access to their data: {``What kind of process ensures that the researchers will not share my data?''} (P181S1-HTN).

\Paragraph{Questions about guarantees or real-world uses.}
51 participants wanted to know how they could be sure the TEE would work: {``\ldots how [is] it guaranteed that it can't be accessed?''} (P87S1-TNN). 
21 had questions about real TEEs, including 11 who asked whether they had ever been involved in a breach: {``I would like to know if there have been cases in the past where TEEs have been hacked''} (P204S1-HTP). 
4 wondered whether TEEs are actually real: {``Is it a real thing? Or a hypothetical idea just for the study?''} (P127S1-TTP).

\begin{table*}[htp!]
\centering
\centering
\setlength\tabcolsep{2.2pt}\renewcommand{\arraystretch}{1.2}
\begin{tabular}{lll|ll|ll|ll}
& \multicolumn{2}{c|}{\textbf{Survey 1}} 
& \multicolumn{2}{c|}{\textbf{Survey 1}} 
& \multicolumn{2}{c|}{\textbf{Survey 1}} 
& \multicolumn{2}{c}{\textbf{Survey 1}} \\
& \multicolumn{2}{c|}{\textbf{Medical Scenario}} 
& \multicolumn{2}{c|}{\textbf{Smart Home Scenario}} 
& \multicolumn{2}{c|}{\textbf{Medical Scenario}} 
& \multicolumn{2}{c}{\textbf{Smart Home Scenario}} \\
& \multicolumn{2}{c|}{\textbf{Without AI}} 
& \multicolumn{2}{c|}{\textbf{Without AI}} 
& \multicolumn{2}{c|}{\textbf{With AI}} 
& \multicolumn{2}{c}{\textbf{With AI}} \\
\midrule

   \begin{tabular}{@{}c@{}} 
  \textbf{Variable} \\[-2pt]
  {\scriptsize R$^2$ Nagelkerke}
  \end{tabular}&

   \begin{tabular}{@{}c@{}} 
\textbf{Willingness} \\[-2pt]
  {\scriptsize 0.0263}
  \end{tabular}&
\begin{tabular}{@{}c@{}} 
\textbf{Safety} \\[-2pt]
  {\scriptsize 0.0083}
  \end{tabular}&
\begin{tabular}{@{}c@{}} 
\textbf{Willingness} \\[-2pt]
  {\scriptsize 0.0747}
  \end{tabular}&
\begin{tabular}{@{}c@{}} 
\textbf{Safety} \\[-2pt]
  {\scriptsize 0.0732}
  \end{tabular}&
\begin{tabular}{@{}c@{}} 
\textbf{Willingness} \\[-2pt]
  {\scriptsize 0.0196}
  \end{tabular}&
\begin{tabular}{@{}c@{}} 
\textbf{Safety} \\[-2pt]
  {\scriptsize 0.0151}
  \end{tabular}&
\begin{tabular}{@{}c@{}} 
\textbf{Willingness}\\[-2pt]
  {\scriptsize 0.0466}
  \end{tabular} &
\begin{tabular}{@{}c@{}} 
\textbf{Safety} \\[-2pt]
  {\scriptsize 0.0351}
  \end{tabular}\\
\multicolumn{3}{l|}{\quad\textit{Expln sentence 1 [Baseline = Unsubstantial]}} &&&&\\
Hardware &
  $\ph1.39$ &
  $\ph0.97$&
  $\ph1.38$&
  $\ph0.95$ &
  $\ph0.99$&
  $\ph1.02$&
  $\ph1.12$&
  $\ph1 .07$
\\
Trust &
  $\ph1.16$ &
  $\ph1.15$&
  $\ph1.43$&
  $\ph1.08$&
  $\ph0.72$&
  $\ph0.74$&
  $\ph0.89$&
  $\ph0.84$
\\
\multicolumn{3}{l|}{\quad\textit{Expln sentence 2 [Baseline = Technical]}} &&&&\\
Non-Technical &
  $\ph0.78$ &
  $\ph0.91$&
  $\ph1.4$&
   \begin{tabular}{@{}c@{}} 
  \cellcolor{positive_medium}$\ph2.39^{**}$ \\[-2pt]
  {\scriptsize [1.354, 4.300]}
  \end{tabular}&
  $\ph1.20$&
  $\ph1.14$&
  $\ph0.93$&
  $\ph0.79$
\\
\multicolumn{3}{l|}{\quad\textit{Expln sentence 3 [Baseline = No Prevents]}} &&&&\\
Prevents &
  $\ph1.31$ &
  $\ph1.26$&
  $\ph1.35$&
  $\ph1.03$&
  $\ph0.84$&
  $\ph0.84$&
  $\ph0.74$&
  $\ph1.51$
\\
\midrule
Medical/Smart home exp &
  $\ph1.54$&
  $\ph0.86$&
   \begin{tabular}{@{}c@{}}  
  \cellcolor{positive_dark}$\ph3.26^{***}$ \\[-2pt]
  {\scriptsize [1.645, 6.586] }
  \end{tabular}
  &
  $\ph0.84$ &
  $\ph1.15$&
  $\ph0.79$&
  \begin{tabular}{@{}c@{}} 
  \cellcolor{positive_medium}$\ph2.34^{**}$\\[-2pt]
  {\scriptsize [1.256, 4.402]}
  \end{tabular}&
  $\ph1.80$
\\
CS Experience &
  $\ph1.01$&
  $\ph1.02$&
  $\ph1.12$&
  $\ph2.12$&
  $\ph1.36$&
  $\ph1.28$&
  $\ph1.14$&
  $\ph1.13$
\\
\midrule
\midrule
& \multicolumn{2}{c|}{\textbf{Survey 2}} 
& \multicolumn{2}{c|}{\textbf{Survey 2}} 
& \multicolumn{2}{c|}{\textbf{Survey 2}} 
& \multicolumn{2}{c}{\textbf{Survey 2}} \\
& \multicolumn{2}{c|}{\textbf{Medical Scenario}} 
& \multicolumn{2}{c|}{\textbf{Smart Home Scenario}} 
& \multicolumn{2}{c|}{\textbf{Medical Scenario}} 
& \multicolumn{2}{c}{\textbf{Smart Home Scenario}} \\
& \multicolumn{2}{c|}{\textbf{Without AI}} 
& \multicolumn{2}{c|}{\textbf{Without AI}} 
& \multicolumn{2}{c|}{\textbf{With AI}} 
& \multicolumn{2}{c}{\textbf{With AI}} \\
\midrule
\textbf{Variable} & 
 \begin{tabular}{@{}c@{}} 
\textbf{Willingness} \\[-2pt]
  {\scriptsize 0.0084}
  \end{tabular}&
\begin{tabular}{@{}c@{}} 
\textbf{Safety} \\[-2pt]
  {\scriptsize 0.0392}
  \end{tabular}&
\begin{tabular}{@{}c@{}} 
\textbf{Willingness} \\[-2pt]
  {\scriptsize 0.1197 }
  \end{tabular}&
\begin{tabular}{@{}c@{}} 
\textbf{Safety} \\[-2pt]
  {\scriptsize 0.1093}
  \end{tabular}&
\begin{tabular}{@{}c@{}} 
\textbf{Willingness} \\[-2pt]
  {\scriptsize 0.0362}
  \end{tabular}&
\begin{tabular}{@{}c@{}} 
\textbf{Safety} \\[-2pt]
  {\scriptsize 0.0302}
  \end{tabular}&
\begin{tabular}{@{}c@{}} 
\textbf{Willingness}\\[-2pt]
  {\scriptsize 0.0771}
  \end{tabular} &
\begin{tabular}{@{}c@{}} 
\textbf{Safety} \\[-2pt]
  {\scriptsize 0.0595}
  \end{tabular}\\
  
\multicolumn{3}{l|}{\quad\textit{TEE Explanation [Baseline = None]}} &&&&\\
Unsubstantial &
  $\ph0.73$ &
  $\ph1.88$ &
  $\ph0.81$ &
  $\ph0.70$ &
  $\ph0.99$ &
  $\ph1.08$ &
  $\ph1.57$ &
  $\ph2.59$ 
\\
Hardware &
  $\ph1.12$ &
  $\ph1.36$ &
  $\ph1.58$ &
  $\ph1.92$ &
  $\ph1.21$ &
  $\ph1.97$ &
  $\ph0.99$ &
  $\ph1.36$ 
\\
Trust &
  $\ph1.02$ &
  $\ph2.05$ &
  $\ph0.87$ &
  $\ph1.08$ &
  $\ph0.69$ &
  $\ph1.55$ &
  $\ph1.04$ &
  $\ph1.57$ 
\\
\midrule
\multicolumn{3}{l|}{\quad\textit{FAQ [Baseline = None]}} &&&&\\
Hidden &
  $\ph0.91$ &
  $\ph1.32$ &
  $\ph1.70$ &
  $\ph1.68$ &
  $\ph0.76$ &
  $\ph1.20$ &
  $\ph0.76$ &
  $\ph1.57$ 
\\
Shown &
  $\ph0.99$ &
  $\ph1.75$ &
  $\ph1.38$ &
  $\ph2.39$ &
  $\ph1.52$ &
  $\ph1.52$ &
  $\ph1.27$ &
  $\ph1.72$ 
\\
\midrule
Medical/Smart home exp &
  $\ph0.95$ &
  $\ph0.80$ &

  \begin{tabular}{@{}c@{}} 
  \cellcolor{positive_dark}$\ph6.23^{***}$\\[-2pt]
  {\scriptsize [1.256, 4.402]}
  \end{tabular}&

  \begin{tabular}{@{}c@{}} 
  \cellcolor{positive_medium}$\ph3.19^{*}$\\[-2pt]
  {\scriptsize [1.256, 4.402]}
  \end{tabular}&

  $\ph0.99$ &
  $\ph0.99$ &
  \begin{tabular}{@{}c@{}} 
  \cellcolor{positive_dark}$\ph3.46^{***}$ \\[-2pt]
  {\scriptsize [1.256, 4.402]}
  \end{tabular}
  &
  $\ph1.68$ 
\\
CS Experience &
  $\ph1.08$ &
  $\ph1.17$ &
  $\ph1.14$ &
  $\ph1.55$ &
  $\ph1.23$ &
  $\ph1.15$ &
  $\ph0.85$ &
  $\ph0.70$ 
\\
\bottomrule
\end{tabular}

\vspace{2mm}
\caption{Regression table for questions about willingness to use technology and belief that the TEE will keep data safe in \textbf{Survey 1 and 2} respectively. The first column is how willing the participant would be to use the TEE-enhanced technology, the second column is the belief that their data will be safe. There is one ordinal logistic regression model for each question in each scenario (24 models total) \new{ corrected with the Benjamini-Hochberg procedure between scenarios and R$^2$ Nagelkerke for each regression}. The numbers in this table are the odds ratios for each predictor, with the baseline explanations used in each model noted in \textit{italics}. Statistical significance is noted with asterisks and shaded cells: blue for positive coefficients and orange for negative. \new{The values shown below each significant odds ratio are the 95\% confidence intervals (CI), indicating the range of plausible effect sizes.}}
\label{tab:trust-stats}
\end{table*} 

\Paragraph{Other concerns.}
43 participants did not ask questions,  instead using the space to share opinions. 10 commented on the scenario {``\ldots I was biased about this to begin with. I don't trust these devices"} (P62S1-TNP). 
16 wrote about technology: {“You do understand that people don't trust technology?”} (P389S1-UTN). 10 people mentioned that they don't trust TEEs: ``I don't trust my information will be secure, especially with the words `trusted environment'~'' (P243S1-HNN).

\section{Survey 2 Results}
\label{sec:follow-up}
Our first survey shows that TEE explanations might be effective at communicating TEE concepts, especially when they are \textit{Non-technical}, mention specific threats a TEE can \textit{Prevent}, and do not expect people to infer new things. However, our explanations had little effect on willingness to use 
TEE-enhanced technology or the belief that the TEE will keep data safe. 
Moreover, the questions we collected suggest that a potential reason that TEE explanations have little effect 
may be that our participants' primary data privacy concerns are beyond the capabilities of TEEs.
In this section, we describe our follow-up survey, Survey 2, where we introduce an FAQ based on the questions asked by participants in our first survey and additional questions related to their belief that the TEE-enhanced technology will keep their data safe. 

\subsection{RQ3: Effect of FAQ}
\label{sec:followup-comp-trust}
We summarize the scores for each question in Figure~\ref{fig:comp}. The full regression table can be found in Table~\ref{tab:stats-followup} and~\ref{tab:stats-followup-iot}.
Having an FAQ seems to have helped participants answer questions about TEE features correctly, but it also seems to have made them less likely to answer questions about TEE limitations correctly. The FAQ had little effect on participants' willingness to adopt TEE-enhanced technology. Still, it did tend to make people feel more confident that their data would be safe when protected by a TEE. 

\begin{table*}[htp!]
\centering
\centering
\setlength\tabcolsep{2.8pt}
\renewcommand{\arraystretch}{1}
\begin{tabular}{lcccccccccccc}
& \multicolumn{10}{c}{\textbf{Medical Scenario Without AI}} \\
\textbf{{ \begin{tabular}{@{}c@{}}  
  Variable \\
  {\scriptsize McFadden's $R^2$}
  \end{tabular}} } & 
\multicolumn{1}{c}{ 
    \begin{tabular}{@{}c@{}}  
  \Qone \\
  {\scriptsize0.1389 }
  \end{tabular}} & 

\multicolumn{1}{c}{
    \begin{tabular}{@{}c@{}}  
  \Qtwo \\
  {\scriptsize0.1389 }
  \end{tabular}} & 
\multicolumn{1}{c}{ \begin{tabular}{@{}c@{}}   \Qthree \\
  {\scriptsize0.1727 }
  \end{tabular}} & 
\multicolumn{1}{c}{\begin{tabular}{@{}c@{}}  \Qfour  \\
  {\scriptsize0.0794 }
  \end{tabular}} & 
\multicolumn{1}{c}{\begin{tabular}{@{}c@{}}  \Qnine  \\
  {\scriptsize0.0639}
  \end{tabular}} & 
\multicolumn{1}{c}{\begin{tabular}{@{}c@{}}  \Qfive \\
  {\scriptsize0.0451 }
  \end{tabular}} &  
\multicolumn{1}{c}{\begin{tabular}{@{}c@{}}  \Qsix \\
  {\scriptsize0.0653 }
  \end{tabular}} & 
\multicolumn{1}{c}{\begin{tabular}{@{}c@{}}  \Qseven \\
  {\scriptsize0.0293 }
  \end{tabular}} & 
\multicolumn{1}{c}{\begin{tabular}{@{}c@{}}  \Qeight  \\
  {\scriptsize0.0145	 }
  \end{tabular}} & 
\multicolumn{1}{c}{\begin{tabular}{@{}c@{}}  \Qten  \\
  {\scriptsize0.0065	 }
  \end{tabular}} & 
\multicolumn{1}{c}{\begin{tabular}{@{}c@{}}  \Qeleven \\
  {\scriptsize0.0372	 }
  \end{tabular}} & 
\multicolumn{1}{c}{\begin{tabular}{@{}c@{}}  \Qtwelve  \\
  {\scriptsize0.3102	 }
  \end{tabular}} \\
\midrule
\multicolumn{13}{l}{\quad\textit{Explanation [Baseline = No Explanation]}} \\
  Unsubstantial &
   \begin{tabular}{@{}c@{}}  
  $0.66$ \\[-2pt]
  {\scriptsize}
  \end{tabular} &
  \begin{tabular}{@{}c@{}}  
  $0.31$  {\scriptsize}
  \end{tabular} &
  \begin{tabular}{@{}c@{}}  
  $2.14$  {\scriptsize}
  \end{tabular} &
  \begin{tabular}{@{}c@{}}  
  $2.01$  {\scriptsize}
  \end{tabular} &
  \begin{tabular}{@{}c@{}}  
  $2.59$  {\scriptsize}
  \end{tabular} &
  \begin{tabular}{@{}c@{}}  
  $0.26$  {\scriptsize}
  \end{tabular} &
  \begin{tabular}{@{}c@{}}  
  $1.42$  {\scriptsize}
  \end{tabular} &
  \begin{tabular}{@{}c@{}}  
  $0.98$  {\scriptsize}
  \end{tabular} &
  \begin{tabular}{@{}c@{}}  
  $1.26$  {\scriptsize}
  \end{tabular} &
  \begin{tabular}{@{}c@{}}  
  $0.67$  {\scriptsize}
  \end{tabular} &
  \begin{tabular}{@{}c@{}}  
  $0.61$  {\scriptsize}
  \end{tabular} & 
  \begin{tabular}{@{}c@{}}  
  $1.23$  {\scriptsize}
  \end{tabular} \\
Hardware &
  \begin{tabular}{@{}c@{}}  
  $1.93$  {\scriptsize}
  \end{tabular} &
  \begin{tabular}{@{}c@{}}  
  $0.76$  {\scriptsize}
  \end{tabular} &
  \begin{tabular}{@{}c@{}}  
  \cellcolor{positive_light}$2.53^{*}$  \\{\scriptsize\shortstack{[1.108,\\6.148]}}
  \end{tabular} &
  \begin{tabular}{@{}c@{}}  
  \cellcolor{positive_light}$4.22^{*}$  \\{\scriptsize\shortstack{[1.627,\\12.425]}}
  \end{tabular} &
  \begin{tabular}{@{}c@{}}  
  $1.92$  {\scriptsize}
  \end{tabular} &
  \begin{tabular}{@{}c@{}}  
  $0.49$  {\scriptsize}
  \end{tabular} &
  \begin{tabular}{@{}c@{}}  
  $1.34$  {\scriptsize}
  \end{tabular} &
  \begin{tabular}{@{}c@{}}  
  $0.88$  {\scriptsize}
  \end{tabular} &
  \begin{tabular}{@{}c@{}}  
  $1.55$  {\scriptsize}
  \end{tabular} &
  \begin{tabular}{@{}c@{}}  
  $0.72$  {\scriptsize}
  \end{tabular} &
  \begin{tabular}{@{}c@{}}  
  $0.58$  {\scriptsize}
  \end{tabular} &
  \begin{tabular}{@{}c@{}}  
  $1.40$  {\scriptsize}
  \end{tabular} \\
Trust &
\begin{tabular}{@{}c@{}}  
   $0.71$ {\scriptsize}
  \end{tabular}&
  \begin{tabular}{@{}c@{}}  
  $0.19$ {\scriptsize}
  \end{tabular}&
  \begin{tabular}{@{}c@{}}  
  \cellcolor{positive_medium}$3.90^{**}$ \\
{\scriptsize\shortstack{[1.662,\\9.756]}}
  \end{tabular}&
  \begin{tabular}{@{}c@{}}  
  \cellcolor{positive_light}$2.69^*$ \\ {\scriptsize\shortstack{[1.168,\\6.573]}}
  \end{tabular}&
  \begin{tabular}{@{}c@{}}  
  $1.65$ {\scriptsize}
  \end{tabular}& 
  \begin{tabular}{@{}c@{}}  
  $0.61$ {\scriptsize}
  \end{tabular}&
  \begin{tabular}{@{}c@{}}  
  $0.90$ {\scriptsize}
  \end{tabular}&
  \begin{tabular}{@{}c@{}}  
  $0.82$ {\scriptsize}
  \end{tabular}&
  \begin{tabular}{@{}c@{}}  
  $1.15$ {\scriptsize}
  \end{tabular}&
  \begin{tabular}{@{}c@{}}  
  $0.70$ {\scriptsize}
  \end{tabular}&
  \begin{tabular}{@{}c@{}}  
  $0.67$ {\scriptsize}
  \end{tabular}&
  \begin{tabular}{@{}c@{}}  
  $1.17$ {\scriptsize}
  \end{tabular} \\
\midrule
\multicolumn{13}{l}{\quad\textit{FAQ [Baseline = No FAQ]}} \\
Hidden &
   \begin{tabular}{@{}c@{}}  
  $7 \times 10^7${\scriptsize}
  \end{tabular}&
  \begin{tabular}{@{}c@{}}  
  $1.34${\scriptsize}
  \end{tabular}&
  \begin{tabular}{@{}c@{}}  
  \cellcolor{positive_medium}$3.63^{**}$\\
  {\scriptsize\shortstack{[1.651,\\8.538]}}
  \end{tabular}&
  \begin{tabular}{@{}c@{}}  
  $1.39${\scriptsize}
  \end{tabular}&
  \begin{tabular}{@{}c@{}}  
  $2.05${\scriptsize}
  \end{tabular}& 
  \begin{tabular}{@{}c@{}}  
  $1.27${\scriptsize}
  \end{tabular}&
  \begin{tabular}{@{}c@{}}  
  $ 0.47${\scriptsize}
  \end{tabular}&
  \begin{tabular}{@{}c@{}}  
  $1.13${\scriptsize}
  \end{tabular}&
  \begin{tabular}{@{}c@{}}  
  $0.93${\scriptsize}
  \end{tabular}&
  \begin{tabular}{@{}c@{}}  
  $0.44${\scriptsize}
  \end{tabular}&
  \begin{tabular}{@{}c@{}}  
  \cellcolor{positive_dark}$19.30^{***}$ \\{\scriptsize\shortstack{[8.913,\\45.145]}}
  \end{tabular}&
  \begin{tabular}{@{}c@{}}  
  \cellcolor{positive_dark}$6.23^{***}$ \\
  {\scriptsize\shortstack{[3.154,\\12.785]}}
  \end{tabular}
   \\
Shown &
\begin{tabular}{@{}c@{}}  
  $1.70$
  {\scriptsize}
  \end{tabular} &
  \begin{tabular}{@{}c@{}}  
  $2.32$
  {\scriptsize}
  \end{tabular} &
  \begin{tabular}{@{}c@{}}  
  $1.45$
  {\scriptsize}
  \end{tabular} &
  \begin{tabular}{@{}c@{}}  
  $1.97$
  {\scriptsize}
  \end{tabular} &
  \begin{tabular}{@{}c@{}}  
  $2.03$
  {\scriptsize}
  \end{tabular} & 
  \begin{tabular}{@{}c@{}}  
  $0.75$
  {\scriptsize}
  \end{tabular} &
  \begin{tabular}{@{}c@{}}  
  $0.37$
  {\scriptsize}
  \end{tabular} &
  \begin{tabular}{@{}c@{}}  
  $0.74$
  {\scriptsize}
  \end{tabular} &
  \begin{tabular}{@{}c@{}}  
  $1.20$
  {\scriptsize}
  \end{tabular} &
  \begin{tabular}{@{}c@{}}  
  \cellcolor{negative_medium}$0.36^{*}$\\
  {\scriptsize\shortstack{[0.183,\\0.708]}}
  \end{tabular} &
  \begin{tabular}{@{}c@{}}  
  \cellcolor{positive_dark}$23.81^{***}$\\
  {\scriptsize\shortstack{[10.978,\\56.001]	}}
  \end{tabular} &
  \begin{tabular}{@{}c@{}}  
  \cellcolor{positive_dark}$6.42^{***}$\\
  {\scriptsize\shortstack{[3.331,\\12.939]}}
  \end{tabular} 
  \\
\midrule
Medical Exp &
 \begin{tabular}{@{}c@{}} 
  $0.38$
  {\scriptsize}
  \end{tabular} &
  \begin{tabular}{@{}c@{}}  
  $2.64$
  {\scriptsize}
  \end{tabular} &
  \begin{tabular}{@{}c@{}}  
  $0.89$
  {\scriptsize}
  \end{tabular} &
  \begin{tabular}{@{}c@{}}  
  $0.58$
  {\scriptsize}
  \end{tabular} &
  \begin{tabular}{@{}c@{}}  
  $0.63$
  {\scriptsize}
  \end{tabular} & 
  \begin{tabular}{@{}c@{}}  
  $0.47$
  {\scriptsize}
  \end{tabular} &
  \begin{tabular}{@{}c@{}}  
  $1.04$
  {\scriptsize}
  \end{tabular} &
  \begin{tabular}{@{}c@{}}  
  $1.52$
  {\scriptsize}
  \end{tabular} &
  \begin{tabular}{@{}c@{}}  
  $0.90$
  {\scriptsize}
  \end{tabular} &
  \begin{tabular}{@{}c@{}}  
  $1.13$
  {\scriptsize}
  \end{tabular} &
  \begin{tabular}{@{}c@{}}  
  $1.36$
  {\scriptsize}
  \end{tabular} &
  \begin{tabular}{@{}c@{}}  
  $0.90$
  {\scriptsize}
  \end{tabular}  \\
CS Exp &
\begin{tabular}{@{}c@{}}  
  $0.43$
  {\scriptsize}
  \end{tabular} &
  \begin{tabular}{@{}c@{}}  
  $0.91$
  {\scriptsize}
  \end{tabular} &
  \begin{tabular}{@{}c@{}}  
  $1.11$
  {\scriptsize}
  \end{tabular} &
  \begin{tabular}{@{}c@{}}  
  $1.40$
  {\scriptsize}
  \end{tabular} &
  \begin{tabular}{@{}c@{}}  
  $1.09$
  {\scriptsize}
  \end{tabular} & 
  \begin{tabular}{@{}c@{}}  
  $0.68$
  {\scriptsize}
  \end{tabular} &
  \begin{tabular}{@{}c@{}}  
  $0.90$
  {\scriptsize}
  \end{tabular} &
  \begin{tabular}{@{}c@{}}  
  $1.22$
  {\scriptsize}
  \end{tabular} &
  \begin{tabular}{@{}c@{}}  
  $1.14$
  {\scriptsize}
  \end{tabular} &
  \begin{tabular}{@{}c@{}}  
  $0.98$
  {\scriptsize}
  \end{tabular} &
  \begin{tabular}{@{}c@{}}  
  $1.04$
  {\scriptsize}
  \end{tabular} &
  \begin{tabular}{@{}c@{}}  
  $0.71$
  {\scriptsize}
  \end{tabular} \\
\midrule & \multicolumn{10}{c}{\textbf{Medical Scenario With AI}} \\
\textbf{Variable} & 

 \multicolumn{1}{c}{   \begin{tabular}{@{}c@{}}  
  \Qone \\
  {\scriptsize0.0636	 }
  \end{tabular}} &
 \multicolumn{1}{c}{   \begin{tabular}{@{}c@{}}  
  \Qtwo \\
  {\scriptsize0.0634	 }
  \end{tabular}} & 
\multicolumn{1}{c}{ \begin{tabular}{@{}c@{}}   \Qthree \\
  {\scriptsize0.0811		 }
  \end{tabular}} & 
\multicolumn{1}{c}{\begin{tabular}{@{}c@{}}  \Qfour  \\
  {\scriptsize0.0770 }
  \end{tabular}} & 
\multicolumn{1}{c}{\begin{tabular}{@{}c@{}}  \Qnine  \\
  {\scriptsize0.0443	}
  \end{tabular}} & 
\multicolumn{1}{c}{\begin{tabular}{@{}c@{}}  \Qfive \\
  {\scriptsize0.1080 }
  \end{tabular}} &  
\multicolumn{1}{c}{\begin{tabular}{@{}c@{}}  \Qsix \\
  {\scriptsize0.0348}
  \end{tabular}} & 
\multicolumn{1}{c}{\begin{tabular}{@{}c@{}}  \Qseven \\
  {\scriptsize0.0554 }
  \end{tabular}} & 
\multicolumn{1}{c}{\begin{tabular}{@{}c@{}}  \Qeight  \\
  {\scriptsize0.0342	 }
  \end{tabular}} & 
\multicolumn{1}{c}{\begin{tabular}{@{}c@{}}  \Qten  \\
  {\scriptsize0.0234	 }
  \end{tabular}} & 
\multicolumn{1}{c}{\begin{tabular}{@{}c@{}}  \Qeleven \\
  {\scriptsize0.2677	 }
  \end{tabular}} & 
\multicolumn{1}{c}{\begin{tabular}{@{}c@{}}  \Qtwelve  \\
  {\scriptsize0.0854	 }
  \end{tabular}} \\

\midrule
\multicolumn{13}{l}{\quad\textit{Explanation [Baseline = No Explanation]}} \\
  Unsubstantial &
  \begin{tabular}{@{}c@{}}  
  $3.56$
  {\scriptsize}
  \end{tabular} &
  \begin{tabular}{@{}c@{}}  
  $1.48$
  {\scriptsize}
  \end{tabular} &
  \begin{tabular}{@{}c@{}}  
  $1.75$
  {\scriptsize}
  \end{tabular} &
  \begin{tabular}{@{}c@{}}  
  \cellcolor{positive_light}$3.35^{*}$
  \\ {\scriptsize\shortstack{[1.434,\\8.192]}}
  \end{tabular} &
  \begin{tabular}{@{}c@{}}  
  \cellcolor{positive_light}$3.60^*$ \\
  {\scriptsize\shortstack{[1.252,\\11.981]}}
  \end{tabular} & 
  \begin{tabular}{@{}c@{}}  
  $2.51$
  {\scriptsize}
  \end{tabular} &
  \begin{tabular}{@{}c@{}}  
  $2.86$
  {\scriptsize}
  \end{tabular} &
  \begin{tabular}{@{}c@{}}  
  $1.31$
  {\scriptsize}
  \end{tabular} &
  \begin{tabular}{@{}c@{}}  
  $3.03$
  {\scriptsize}
  \end{tabular} &
  \begin{tabular}{@{}c@{}}  
  $1.57$
  {\scriptsize}
  \end{tabular} &
  \begin{tabular}{@{}c@{}}  
  $0.55$
  {\scriptsize}
  \end{tabular} &
  \begin{tabular}{@{}c@{}}  
  $0.58$
  {\scriptsize}
  \end{tabular} \\
Hardware &
  \begin{tabular}{@{}c@{}}  
  $5.00$
  {\scriptsize}
  \end{tabular} &
  \begin{tabular}{@{}c@{}}  
  $2.41$ 
  {\scriptsize}
  \end{tabular} &
  \begin{tabular}{@{}c@{}}  
  $2.20$
  {\scriptsize}
  \end{tabular} &
  \begin{tabular}{@{}c@{}}  
  \cellcolor{positive_light}$2.41^*$\\
  {\scriptsize\shortstack{[1.106,\\5.405]}}
  \end{tabular} &
  \begin{tabular}{@{}c@{}}  
  $2.23$
  {\scriptsize}
  \end{tabular} &
  \begin{tabular}{@{}c@{}}  
  $2.53$
  {\scriptsize}
  \end{tabular} &
  \begin{tabular}{@{}c@{}}  
  $1.73$
  {\scriptsize}
  \end{tabular} &
  \begin{tabular}{@{}c@{}}  
  $1.55$
  {\scriptsize}
  \end{tabular} &
  \begin{tabular}{@{}c@{}}  
  $1.43$
  {\scriptsize}
  \end{tabular} &
  \begin{tabular}{@{}c@{}}  
  $0.99$
  {\scriptsize}
  \end{tabular} &
  \begin{tabular}{@{}c@{}}  
  $0.92$
  {\scriptsize}
  \end{tabular} &
  \begin{tabular}{@{}c@{}}  
  $0.79$
  {\scriptsize}
  \end{tabular} \\
Trust &
\begin{tabular}{@{}c@{}}  
  $1.27$
  {\scriptsize}
  \end{tabular}&
  \begin{tabular}{@{}c@{}}  
  $1.21$
  {\scriptsize}
  \end{tabular} &
  \begin{tabular}{@{}c@{}}  
  $1.68$
  {\scriptsize}
  \end{tabular} &
  \begin{tabular}{@{}c@{}}  
  \cellcolor{positive_light}$2.69^*$
  \\ {\scriptsize\shortstack{[1.168,\\6.573]	}}
  \end{tabular} &
  \begin{tabular}{@{}c@{}}  
  $0.99$
  {\scriptsize}
  \end{tabular} & 
  \begin{tabular}{@{}c@{}}  
  $5.26$
  {\scriptsize}
  \end{tabular} &
  \begin{tabular}{@{}c@{}}  
  $1.88$
  {\scriptsize}
  \end{tabular} &
  \begin{tabular}{@{}c@{}}  
  $1.39$
  {\scriptsize}
  \end{tabular} &
  \begin{tabular}{@{}c@{}}  
  $2.03$
  {\scriptsize}
  \end{tabular} &
  \begin{tabular}{@{}c@{}}  
  $1.42$
  {\scriptsize}
  \end{tabular} &
  \begin{tabular}{@{}c@{}}  
  $1.14$
  {\scriptsize}
  \end{tabular} &
  \begin{tabular}{@{}c@{}}  
  $0.53$
  {\scriptsize}
  \end{tabular} \\
\midrule
\multicolumn{13}{l}{\quad\textit{FAQ [Baseline = No FAQ]}} \\
Hidden &
\begin{tabular}{@{}c@{}}  
  $0.99$
  {\scriptsize}
  \end{tabular} &
  \begin{tabular}{@{}c@{}}  
  $2.16$
  {\scriptsize}
  \end{tabular} &
  \begin{tabular}{@{}c@{}}  
  $0.87$
  {\scriptsize}
  \end{tabular} &
  \begin{tabular}{@{}c@{}}  
  $1.22$
  {\scriptsize}
  \end{tabular} &
  \begin{tabular}{@{}c@{}}  
  $1.28$
  {\scriptsize}
  \end{tabular} & 
  \begin{tabular}{@{}c@{}}  
  $0.24$
  {\scriptsize}
  \end{tabular} &
  \begin{tabular}{@{}c@{}}  
  $1.54$
  {\scriptsize}
  \end{tabular} &
  \begin{tabular}{@{}c@{}}  
  \cellcolor{negative_medium}$0.26^{*}$\\
  {\scriptsize\shortstack{[0.088,\\0.651]}}
  \end{tabular} &
  \begin{tabular}{@{}c@{}}  
  $0.79$
  {\scriptsize}
  \end{tabular} &
  \begin{tabular}{@{}c@{}}  
  $0.62$
  {\scriptsize}
  \end{tabular} &
  \begin{tabular}{@{}c@{}}  
  \cellcolor{positive_dark}$ 9.97^{***}$\\
  {\scriptsize\shortstack{[4.897,\\21.427]}	}
  \end{tabular} &
  \begin{tabular}{@{}c@{}}  
  \cellcolor{positive_dark}$3.32^{***}$ \\
  {\scriptsize\shortstack{[1.747,\\6.451]	}}
  \end{tabular} \\
Shown &
\begin{tabular}{@{}c@{}}  
  $1.60$
  {\scriptsize}
  \end{tabular} &
  \begin{tabular}{@{}c@{}}  
  \cellcolor{positive_light}$3.71^*$ \\
  {\scriptsize\shortstack{[1.435,\\10.840]}}
  \end{tabular} &
  \begin{tabular}{@{}c@{}}  
  $1.14$
  {\scriptsize}
  \end{tabular} &
  \begin{tabular}{@{}c@{}}  
  $1.28$
  {\scriptsize}
  \end{tabular} &
  \begin{tabular}{@{}c@{}}  
  $1.06$
  {\scriptsize}
  \end{tabular} & 
  \begin{tabular}{@{}c@{}}  
  $0.58$
  {\scriptsize}
  \end{tabular} &
  \begin{tabular}{@{}c@{}}  
  $2.14$
  {\scriptsize}
  \end{tabular} &
  \begin{tabular}{@{}c@{}}  
  $0.44$
  {\scriptsize}
  \end{tabular} &
  \begin{tabular}{@{}c@{}}  
  $0.92$
  {\scriptsize}
  \end{tabular} &
  \begin{tabular}{@{}c@{}}  
  $0.86$
  {\scriptsize}
  \end{tabular} &
  \begin{tabular}{@{}c@{}}  
  \cellcolor{positive_dark}$26.31^{***}$\\
  {\scriptsize\shortstack{[10.998,\\71.892]	}}
  \end{tabular} &
  \begin{tabular}{@{}c@{}}  
  \cellcolor{positive_dark}$4.76^{***}$\\
  {\scriptsize\shortstack{[2.441,\\9.589]}	}
  \end{tabular} \\
\midrule
Medical Exp &
  \begin{tabular}{@{}c@{}}  
  $0.50$
  {\scriptsize}
  \end{tabular} &
  \begin{tabular}{@{}c@{}}  
  $0.61$
  {\scriptsize}
  \end{tabular} &
  \begin{tabular}{@{}c@{}}  
  $0.70$
  {\scriptsize}
  \end{tabular} &
  \begin{tabular}{@{}c@{}}  
  \cellcolor{negative_light}$0.46^*$\\
  {\scriptsize\shortstack{[0.240,\\0.870]	 }}
  \end{tabular} &
  \begin{tabular}{@{}c@{}}  
  $0.78$
  {\scriptsize}
  \end{tabular} & 
  \begin{tabular}{@{}c@{}}  
  $0.53$
  {\scriptsize}
  \end{tabular} &
  \begin{tabular}{@{}c@{}}  
  $1.28$
  {\scriptsize}
  \end{tabular} &
  \begin{tabular}{@{}c@{}}  
  $1.75$
  {\scriptsize}
  \end{tabular} &
  \begin{tabular}{@{}c@{}}  
  $0.55$
  {\scriptsize}
  \end{tabular} &
  \begin{tabular}{@{}c@{}}  
  $1.57$
  {\scriptsize}
  \end{tabular} &
  \begin{tabular}{@{}c@{}}  
  $0.96$
  {\scriptsize}
  \end{tabular} &
  \begin{tabular}{@{}c@{}}  
  $0.66$
  {\scriptsize}
  \end{tabular} \\
CS Exp &
  \begin{tabular}{@{}c@{}}  
  $2.32$
  {\scriptsize}
  \end{tabular} &
  \begin{tabular}{@{}c@{}}  
  $0.88$
  {\scriptsize}
  \end{tabular} &
  \begin{tabular}{@{}c@{}}  
  $0.64$
  {\scriptsize}
  \end{tabular} &
  \begin{tabular}{@{}c@{}}  
  \cellcolor{negative_light}$0.39^{*}$ \\
  {\scriptsize\shortstack{[0.208,\\0.757]	}}
  \end{tabular} &
  \begin{tabular}{@{}c@{}}  
  $1.13$
  {\scriptsize}
  \end{tabular} & 
  \begin{tabular}{@{}c@{}}  
  $0.41$
  {\scriptsize}
  \end{tabular} &
  \begin{tabular}{@{}c@{}}  
  $1.03$
  {\scriptsize}
  \end{tabular} &
  \begin{tabular}{@{}c@{}}  
  $1.88$
  {\scriptsize}
  \end{tabular} &
  \begin{tabular}{@{}c@{}}  
  $1.12$
  {\scriptsize}
  \end{tabular} &
  \begin{tabular}{@{}c@{}}  
  $1.03$
  {\scriptsize}
  \end{tabular} &
  \begin{tabular}{@{}c@{}}  
  $1.72$
  {\scriptsize}
  \end{tabular} &
  \begin{tabular}{@{}c@{}}  
  $0.84$
  {\scriptsize}
  \end{tabular} \\

\bottomrule
\multicolumn{11}{l}{$^{***}p < 0.001$; $^{**}p < 0.01$; $^{*}p < 0.05$}
\end{tabular}

\vspace{2mm}
\caption{Regression table for True/False comprehensions in \textbf{Survey 2} for the Medical Scenarios. There is one logistic regression model for each question in each scenario (24 models total) \new{ corrected with the Benjamini-Hochberg procedure between scenarios and McFadden's $R^2$ for each model. McFadden's $R^2$ is a pseudo-$R^2$ measure of model fit, where higher values (0.2-0.4) indicate better explanatory power relative to an intercept-only model, though values are typically much smaller than in linear regression.} The numbers in this table are the odds ratios for each predictor, with the baseline explanations used in each model noted in \textit{italics}. Statistical significance is noted with asterisks and shaded cells: blue for positive coefficients and orange for negative coefficients. \new{The values shown below each significant odds ratio are the 95\% confidence intervals (CI), indicating the range of plausible effect sizes.}}
\vspace{+30pt}
\label{tab:stats-followup}
\end{table*}

\begin{table*}[htp!]
\centering
\centering
\setlength\tabcolsep{2.8pt}
\renewcommand{\arraystretch}{0.9}
\begin{tabular}{lcccccccccccc} & \multicolumn{10}{c}{\textbf{Smart Home Scenario Without AI}} \\
\textbf{{ \begin{tabular}{@{}c@{}}  
  Variable \\
  {\scriptsize  McFadden's $R^2$}
  \end{tabular}} }  & 
 \multicolumn{1}{c}{   \begin{tabular}{@{}c@{}}  
  \Qone \\
  {\scriptsize0.1039 }
  \end{tabular}} &
 \multicolumn{1}{c}{   \begin{tabular}{@{}c@{}}  
  \Qtwo \\
  {\scriptsize0.0299	 }
  \end{tabular}} & 
\multicolumn{1}{c}{ \begin{tabular}{@{}c@{}}   \Qthree \\
  {\scriptsize0.0414		 }
  \end{tabular}} & 
\multicolumn{1}{c}{\begin{tabular}{@{}c@{}}  \Qfour  \\
  {\scriptsize0.0541 }
  \end{tabular}} & 
\multicolumn{1}{c}{\begin{tabular}{@{}c@{}}  \Qnine  \\
  {\scriptsize0.0506	}
  \end{tabular}} & 
\multicolumn{1}{c}{\begin{tabular}{@{}c@{}}  \Qfive \\
  {\scriptsize0.0310 }
  \end{tabular}} &  
\multicolumn{1}{c}{\begin{tabular}{@{}c@{}}  \Qsix \\
  {\scriptsize0.0238}
  \end{tabular}} & 
\multicolumn{1}{c}{\begin{tabular}{@{}c@{}}  \Qseven \\
  {\scriptsize0.0362 }
  \end{tabular}} & 
\multicolumn{1}{c}{\begin{tabular}{@{}c@{}}  \Qeight  \\
  {\scriptsize0.0340	 }
  \end{tabular}} & 
\multicolumn{1}{c}{\begin{tabular}{@{}c@{}}  \Qten  \\
  {\scriptsize0.0180	 }
  \end{tabular}} & 
\multicolumn{1}{c}{\begin{tabular}{@{}c@{}}  \Qeleven \\
  {\scriptsize0.3042	 }
  \end{tabular}} & 
\multicolumn{1}{c}{\begin{tabular}{@{}c@{}}  \Qtwelve  \\
  {\scriptsize0.0798	 }
  \end{tabular}}  \\
\midrule
\multicolumn{13}{l}{\quad\textit{Explanation [Baseline = No Explanation]}} \\
Unsubstantial &
  \begin{tabular}{@{}c@{}}  
  $0.23$
  {\scriptsize}
  \end{tabular} &
  \begin{tabular}{@{}c@{}}  
  $0.99$
  {\scriptsize}
  \end{tabular} &
  \begin{tabular}{@{}c@{}}  
  $2.18$
  {\scriptsize}
  \end{tabular} &
  \begin{tabular}{@{}c@{}}  
  \cellcolor{positive_light}$2.92^*$ \\
  {\scriptsize\shortstack{[1.296,\\ 6.864]}}
  \end{tabular} &
  \begin{tabular}{@{}c@{}}  
  $0.59$
  {\scriptsize}
  \end{tabular} & 
  \begin{tabular}{@{}c@{}}  
  $1.07$
  {\scriptsize}
  \end{tabular} &
  \begin{tabular}{@{}c@{}}  
  $1.21$
  {\scriptsize}
  \end{tabular} &
  \begin{tabular}{@{}c@{}}  
  $1.17$
  {\scriptsize}
  \end{tabular} &
  \begin{tabular}{@{}c@{}}  
  $0.95$
  {\scriptsize}
  \end{tabular} &
  \begin{tabular}{@{}c@{}}  
  $1.42$
  {\scriptsize}
  \end{tabular} &
  \begin{tabular}{@{}c@{}}  
  $0.53$
  {\scriptsize}
  \end{tabular} &
  \begin{tabular}{@{}c@{}}  
  $0.59$
  {\scriptsize}
  \end{tabular} \\
Hardware &
  \begin{tabular}{@{}c@{}}  
  $0.30$
  {\scriptsize}
  \end{tabular} &
  \begin{tabular}{@{}c@{}}  
  $1.68$
  {\scriptsize}
  \end{tabular} &
  \begin{tabular}{@{}c@{}}  
  $2.69$
  {\scriptsize}
  \end{tabular} &
  \begin{tabular}{@{}c@{}}  
  \cellcolor{positive_light}$2.86^*$\\
  {\scriptsize\shortstack{[1.263,\\ 6.683]	}}
  \end{tabular} &
  \begin{tabular}{@{}c@{}}  
  $1.22$
  {\scriptsize}
  \end{tabular} & 
  \begin{tabular}{@{}c@{}}  
  $0.68$
  {\scriptsize}
  \end{tabular} &
  \begin{tabular}{@{}c@{}}  
  $0.98$
  {\scriptsize}
  \end{tabular} &
  \begin{tabular}{@{}c@{}}  
  $0.66$
  {\scriptsize}
  \end{tabular} &
  \begin{tabular}{@{}c@{}}  
  $0.90$
  {\scriptsize}
  \end{tabular} &
  \begin{tabular}{@{}c@{}}  
  $1.17$
  {\scriptsize}
  \end{tabular} &
  \begin{tabular}{@{}c@{}}  
  $0.58$
  {\scriptsize}
  \end{tabular} &
  \begin{tabular}{@{}c@{}}  
  $0.81$
  {\scriptsize}
  \end{tabular} \\
Trust &
  \begin{tabular}{@{}c@{}}  
  $0.54$
  {\scriptsize}
  \end{tabular} &
  \begin{tabular}{@{}c@{}}  
  $1.43$
  {\scriptsize}
  \end{tabular} &
  \begin{tabular}{@{}c@{}}  
  $1.84$
  {\scriptsize}
  \end{tabular} &
  \begin{tabular}{@{}c@{}}  
  $2.01$
  {\scriptsize}
  \end{tabular} &
  \begin{tabular}{@{}c@{}}  
  $0.90$
  {\scriptsize}
  \end{tabular} & 
  \begin{tabular}{@{}c@{}}  
  $2.18$
  {\scriptsize}
  \end{tabular} &
  \begin{tabular}{@{}c@{}}  
  $1.63$
  {\scriptsize}
  \end{tabular} &
  \begin{tabular}{@{}c@{}}  
  $1.39$
  {\scriptsize}
  \end{tabular} &
  \begin{tabular}{@{}c@{}}  
  $0.87$
  {\scriptsize}
  \end{tabular} &
  \begin{tabular}{@{}c@{}}  
  $1.31$
  {\scriptsize}
  \end{tabular} &
  \begin{tabular}{@{}c@{}}  
  $0.92$
  {\scriptsize}
  \end{tabular} &
  \begin{tabular}{@{}c@{}}  
  $0.74$
  {\scriptsize}
  \end{tabular} \\
\midrule
\multicolumn{13}{l}{\quad\textit{FAQ [Baseline = No FAQ]}} \\
Hidden &
  \begin{tabular}{@{}c@{}}  
  $2.16$
  {\scriptsize}
  \end{tabular} &
  \begin{tabular}{@{}c@{}}  
  $1.15$
  {\scriptsize}
  \end{tabular} &
  \begin{tabular}{@{}c@{}}  
  $1.11$
  {\scriptsize}
  \end{tabular} &
  \begin{tabular}{@{}c@{}}  
  $1.36$
  {\scriptsize}
  \end{tabular} &
  \begin{tabular}{@{}c@{}}  
  $1.62$
  {\scriptsize}
  \end{tabular} & 
  \begin{tabular}{@{}c@{}}  
  $1.14$
  {\scriptsize}
  \end{tabular} &
  \begin{tabular}{@{}c@{}}  
  $0.95$
  {\scriptsize}
  \end{tabular} &
  \begin{tabular}{@{}c@{}}  
  $1.08$
  {\scriptsize}
  \end{tabular} &
  \begin{tabular}{@{}c@{}}  
  $0.59$
  {\scriptsize}
  \end{tabular} &
  \begin{tabular}{@{}c@{}}  
  $0.54$
  {\scriptsize}
  \end{tabular} &
  \begin{tabular}{@{}c@{}}  
  \cellcolor{positive_dark}$19.89^{***}$\\  {\scriptsize\shortstack{[9.156,\\ 46.606]	 }}
  \end{tabular} &
  \begin{tabular}{@{}c@{}}  
  \cellcolor{positive_dark}$3.13^{***}$\\
{\scriptsize\shortstack{[1.688,\\ 5.889]	} }
  \end{tabular} \\
Shown &
  \begin{tabular}{@{}c@{}}  $5.00$
  {\scriptsize}
  \end{tabular} &
  \begin{tabular}{@{}c@{}}  
  $3.10$
  {\scriptsize}
  \end{tabular} &
  \begin{tabular}{@{}c@{}}  
  $1.14$
  {\scriptsize}
  \end{tabular} &
  \begin{tabular}{@{}c@{}}  
  $1.21$
  {\scriptsize}
  \end{tabular} &
  \begin{tabular}{@{}c@{}}  
  $2.53$
  {\scriptsize}
  \end{tabular} & 
  \begin{tabular}{@{}c@{}}  
  $1.99$
  {\scriptsize}
  \end{tabular} &
  \begin{tabular}{@{}c@{}}  
  $0.46$
  {\scriptsize}
  \end{tabular} &
  \begin{tabular}{@{}c@{}}  
  $0.68$
  {\scriptsize}
  \end{tabular} &
  \begin{tabular}{@{}c@{}}  
  $0.81$
  {\scriptsize}
  \end{tabular} &
  \begin{tabular}{@{}c@{}}  
  $0.80$
  {\scriptsize}
  \end{tabular} &
  \begin{tabular}{@{}c@{}}  
  \cellcolor{positive_dark}$18.73^{***}$\\  {\scriptsize\shortstack{[8.354,\\ 46.535]	}}
  \end{tabular} &
  \begin{tabular}{@{}c@{}}  
  \cellcolor{positive_dark}$4.18^{***}$\\
{\scriptsize\shortstack{[2.145,\\8.367]	}}
  \end{tabular} \\
\midrule
Smart Home Experience &
  \begin{tabular}{@{}c@{}}  
  $1.54$
  {\scriptsize}
  \end{tabular} &
  \begin{tabular}{@{}c@{}}  
  $1.02$
  {\scriptsize}
  \end{tabular} &
  \begin{tabular}{@{}c@{}}  
  $1.97$
  {\scriptsize}
  \end{tabular} &
  \begin{tabular}{@{}c@{}}  
  $2.12$
  {\scriptsize}
  \end{tabular} &
  \begin{tabular}{@{}c@{}}  
  $2.08$
  {\scriptsize}
  \end{tabular} & 
  \begin{tabular}{@{}c@{}}  
  $1.05$
  {\scriptsize}
  \end{tabular} &
  \begin{tabular}{@{}c@{}}  
  $1.02$
  {\scriptsize}
  \end{tabular} &
  \begin{tabular}{@{}c@{}}  
  $1.05$
  {\scriptsize}
  \end{tabular} &
  \begin{tabular}{@{}c@{}}  
  $0.74$
  {\scriptsize}
  \end{tabular} &
  \begin{tabular}{@{}c@{}}  
  $0.73$
  {\scriptsize}
  \end{tabular} &
  \begin{tabular}{@{}c@{}}  
  $1.42$
  {\scriptsize}
  \end{tabular} &
  \begin{tabular}{@{}c@{}}  
  $1.17$
  {\scriptsize}
  \end{tabular} \\
CS Exp &
  \begin{tabular}{@{}c@{}}  
  $0.35$
  {\scriptsize}
  \end{tabular} &
  \begin{tabular}{@{}c@{}}  
  $1.23$
  {\scriptsize}
  \end{tabular} &
  \begin{tabular}{@{}c@{}}  
  $0.58$
  {\scriptsize}
  \end{tabular} &
  \begin{tabular}{@{}c@{}}  
  $0.54$
  {\scriptsize}
  \end{tabular} &
  \begin{tabular}{@{}c@{}}  
  $0.68$
  {\scriptsize}
  \end{tabular} &
  \begin{tabular}{@{}c@{}}  
  $0.76$
  {\scriptsize}
  \end{tabular} &
  \begin{tabular}{@{}c@{}}  
  $0.84$
  {\scriptsize}
  \end{tabular} &
  \begin{tabular}{@{}c@{}}  
  $2.10$
  {\scriptsize}
  \end{tabular} &
  \begin{tabular}{@{}c@{}}  
  \cellcolor{positive_light}$2.36^{*}$ \\
  {\scriptsize\shortstack{[1.273,\\ 4.469]	}}
  \end{tabular} &  
  \begin{tabular}{@{}c@{}}  
  $1.15$
  {\scriptsize}
  \end{tabular} &
  \begin{tabular}{@{}c@{}}  
  $1.67$
  {\scriptsize}
  \end{tabular} &
  \begin{tabular}{@{}c@{}}  
  $0.66$
  {\scriptsize}
  \end{tabular} \\
\midrule
& \multicolumn{10}{c}{\textbf{Smart Home Scenario With AI}} \\
\textbf{Variable} & 

 \multicolumn{1}{c}{   \begin{tabular}{@{}c@{}}  
  \Qone \\
  {\scriptsize0.0833	 }
  \end{tabular}} &
 \multicolumn{1}{c}{   \begin{tabular}{@{}c@{}}  
  \Qtwo \\
  {\scriptsize0.0722	 }
  \end{tabular}} & 
\multicolumn{1}{c}{ \begin{tabular}{@{}c@{}}   \Qthree \\
  {\scriptsize0.0698		 }
  \end{tabular}} & 
\multicolumn{1}{c}{\begin{tabular}{@{}c@{}}  \Qfour  \\
  {\scriptsize0.0580 }
  \end{tabular}} & 
\multicolumn{1}{c}{\begin{tabular}{@{}c@{}}  \Qnine  \\
  {\scriptsize0.0834	}
  \end{tabular}} & 
\multicolumn{1}{c}{\begin{tabular}{@{}c@{}}  \Qfive \\
  {\scriptsize0.0379 }
  \end{tabular}} &  
\multicolumn{1}{c}{\begin{tabular}{@{}c@{}}  \Qsix \\
  {\scriptsize0.0249}
  \end{tabular}} & 
\multicolumn{1}{c}{\begin{tabular}{@{}c@{}}  \Qseven \\
  {\scriptsize0.0282 }
  \end{tabular}} & 
\multicolumn{1}{c}{\begin{tabular}{@{}c@{}}  \Qeight  \\
  {\scriptsize0.03178	 }
  \end{tabular}} & 
\multicolumn{1}{c}{\begin{tabular}{@{}c@{}}  \Qten  \\
  {\scriptsize0.0834	 }
  \end{tabular}} & 
\multicolumn{1}{c}{\begin{tabular}{@{}c@{}}  \Qeleven \\
  {\scriptsize0.3149	 }
  \end{tabular}} & 
\multicolumn{1}{c}{\begin{tabular}{@{}c@{}}  \Qtwelve  \\
  {\scriptsize0.0672	 }
  \end{tabular}}  \\
\midrule
\multicolumn{13}{l}{\quad\textit{Explanation [Baseline = No Explanation]}} \\
Unsubstantial &
  \begin{tabular}{@{}c@{}}  
  $2.44$
  {\scriptsize}
  \end{tabular} &
  \begin{tabular}{@{}c@{}}  
  $2.41$
  {\scriptsize}
  \end{tabular} &
  \begin{tabular}{@{}c@{}}  
  \cellcolor{positive_medium}$3.46^{**}$ \\
  {\scriptsize\shortstack{[1.593,\\ 7.734]	 }}
  \end{tabular} &
  \begin{tabular}{@{}c@{}}  
  \cellcolor{positive_light}$2.77^*$\\
  {\scriptsize\shortstack{[1.280,\\ 6.235]	}}
  \end{tabular} &
  \begin{tabular}{@{}c@{}}  
  \cellcolor{positive_medium}$5.16^{**}$\\
  {\scriptsize\shortstack{[2.069,\\ 14.358]	}}
  \end{tabular} & 
  \begin{tabular}{@{}c@{}}  
  $0.39$
  {\scriptsize}
  \end{tabular} &
  \begin{tabular}{@{}c@{}}  
  $0.99$
  {\scriptsize}
  \end{tabular} &
  \begin{tabular}{@{}c@{}}  
  $0.86$
  {\scriptsize}
  \end{tabular} &
  \begin{tabular}{@{}c@{}}  
  $0.65$
  {\scriptsize}
  \end{tabular} &
  \begin{tabular}{@{}c@{}}  
  $0.99$
  {\scriptsize}
  \end{tabular} &
  \begin{tabular}{@{}c@{}}  
  $0.63$
  {\scriptsize}
  \end{tabular} &
  \begin{tabular}{@{}c@{}}  
  $0.94$
  {\scriptsize}
  \end{tabular} \\
Hardware &
  \begin{tabular}{@{}c@{}}  
  $1.90$
  {\scriptsize}
  \end{tabular} &
  \begin{tabular}{@{}c@{}}  
  $2.05$
  {\scriptsize}
  \end{tabular} &
  \begin{tabular}{@{}c@{}}  
  \cellcolor{positive_light}$3.00^{*}$ \\
{\scriptsize\shortstack{[1.364, \\6.837]	 }}
  \end{tabular} &
  \begin{tabular}{@{}c@{}}  
  \cellcolor{positive_medium}$3.06^{**}$\\
{\scriptsize\shortstack{[1.372,\\ 7.218]}	}
  \end{tabular} &
  \begin{tabular}{@{}c@{}}  
  \cellcolor{positive_light}$3.22^{*}$\\
{\scriptsize\shortstack{[1.375,\\ 8.011]}	}
  \end{tabular} &
  \begin{tabular}{@{}c@{}}  
  $0.93$
  {\scriptsize}
  \end{tabular} &
  \begin{tabular}{@{}c@{}}  
  $1.14$
  {\scriptsize}
  \end{tabular} &
  \begin{tabular}{@{}c@{}}  
  $0.89$
  {\scriptsize}
  \end{tabular} &
  \begin{tabular}{@{}c@{}}  
  $1.21$
  {\scriptsize}
  \end{tabular} &
  \begin{tabular}{@{}c@{}}  
  $0.50$
  {\scriptsize}
  \end{tabular} &
  \begin{tabular}{@{}c@{}}  
  $0.71$
  {\scriptsize}
  \end{tabular} &
  \begin{tabular}{@{}c@{}}  
  $1.06$
  {\scriptsize}
  \end{tabular} \\
Trust &
  \begin{tabular}{@{}c@{}}  
  $0.51$
  {\scriptsize}
  \end{tabular} &
  \begin{tabular}{@{}c@{}}  
  $0.95$
  {\scriptsize}
  \end{tabular} &
  \begin{tabular}{@{}c@{}}  
  \cellcolor{positive_medium}$3.13^{*}$ \\
  {\scriptsize\shortstack{ [1.426,\\ 7.080]	} }
  \end{tabular} &
  \begin{tabular}{@{}c@{}}  
  \cellcolor{positive_light}$2.29^*$\\
{\scriptsize\shortstack{ [1.065, \\5.099]}	}
  \end{tabular} &
  \begin{tabular}{@{}c@{}}  
  $2.16$
  {\scriptsize}
  \end{tabular} &
  \begin{tabular}{@{}c@{}}  
  $0.53$
  {\scriptsize}
  \end{tabular} &
  \begin{tabular}{@{}c@{}}  
  $1.63$
  {\scriptsize}
  \end{tabular} &
  \begin{tabular}{@{}c@{}}  
  $0.78$
  {\scriptsize}
  \end{tabular} &
  \begin{tabular}{@{}c@{}}  
  $1.19$
  {\scriptsize}
  \end{tabular} &
  \begin{tabular}{@{}c@{}}  
  $0.71$
  {\scriptsize}
  \end{tabular} &
  \begin{tabular}{@{}c@{}}  
  $0.68$
  {\scriptsize}
  \end{tabular} &
  \begin{tabular}{@{}c@{}}  
  $0.86$
  {\scriptsize}
  \end{tabular} \\
\midrule
\multicolumn{13}{l}{\quad\textit{FAQ [Baseline = No FAQ]}} \\
Hidden &
  \begin{tabular}{@{}c@{}}  
  $1.05$
  {\scriptsize}
  \end{tabular} &
  \begin{tabular}{@{}c@{}}  
  $0.83$
  {\scriptsize}
  \end{tabular} &
  \begin{tabular}{@{}c@{}}  
  $1.07$
  {\scriptsize}
  \end{tabular} &
  \begin{tabular}{@{}c@{}}  
  $1.17$
  {\scriptsize}
  \end{tabular} &
  \begin{tabular}{@{}c@{}}  
  $0.75$
  {\scriptsize}
  \end{tabular} & 
  \begin{tabular}{@{}c@{}}  
  $0.72$
  {\scriptsize}
  \end{tabular} &
  \begin{tabular}{@{}c@{}}  
  $0.58$
  {\scriptsize}
  \end{tabular} &
  \begin{tabular}{@{}c@{}}  
  $0.77$
  {\scriptsize}
  \end{tabular} &
  \begin{tabular}{@{}c@{}}  
  $0.98$
  {\scriptsize}
  \end{tabular} &
  \begin{tabular}{@{}c@{}}  
  $0.53$
  {\scriptsize}
  \end{tabular} &
  \begin{tabular}{@{}c@{}}  
  \cellcolor{positive_dark}$12.55^{***}$\\
  {\scriptsize\shortstack{ [5.774,\\ 29.187]	 }}
  \end{tabular} &
  \begin{tabular}{@{}c@{}}  
  \cellcolor{positive_dark}$3.49^{***}$\\
  {\scriptsize\shortstack{ [1.786,\\ 6.971] }	 }
  \end{tabular} \\
Shown &
  \begin{tabular}{@{}c@{}}  
  $0.84$
  {\scriptsize}
  \end{tabular} &
  \begin{tabular}{@{}c@{}}  
  $0.49$
  {\scriptsize}
  \end{tabular} &
  \begin{tabular}{@{}c@{}}  
  $2.36$
  {\scriptsize}
  \end{tabular} &
  \begin{tabular}{@{}c@{}}  
  $1.26$
  {\scriptsize}
  \end{tabular} &
  \begin{tabular}{@{}c@{}}  
  $1.15$
  {\scriptsize}
  \end{tabular} &
  \begin{tabular}{@{}c@{}}  
  $0.54$
  {\scriptsize}
  \end{tabular} &
  \begin{tabular}{@{}c@{}}  
  $0.67$
  {\scriptsize}
  \end{tabular} &
  \begin{tabular}{@{}c@{}}  
  $0.44$
  {\scriptsize}
  \end{tabular} &
  \begin{tabular}{@{}c@{}}  
  $0.69$
  {\scriptsize}
  \end{tabular} &
  \begin{tabular}{@{}c@{}}  
  $0.60$
  {\scriptsize}
  \end{tabular} &
  \begin{tabular}{@{}c@{}}  
  \cellcolor{positive_dark}$43.38^{***}$\\
  {\scriptsize\shortstack{[16.814, \\131.972]	}}
  \end{tabular} &
  \begin{tabular}{@{}c@{}}  
  \cellcolor{positive_dark}$3.67^{***}$\\
  {\scriptsize\shortstack{[1.921,\\ 7.109]	}}
  \end{tabular} \\
\midrule
Smart Home Experience &
  \begin{tabular}{@{}c@{}}  
  $3.25$
  {\scriptsize}
  \end{tabular} &
  \begin{tabular}{@{}c@{}}  
  $0.27$
  {\scriptsize}
  \end{tabular} &
  \begin{tabular}{@{}c@{}}  
  $1.27$
  {\scriptsize}
  \end{tabular} &
  \begin{tabular}{@{}c@{}}  
  $0.59$
  {\scriptsize}
  \end{tabular} &
  \begin{tabular}{@{}c@{}}  
  $0.28$
  {\scriptsize}
  \end{tabular} &
  \begin{tabular}{@{}c@{}}  
  $0.38$
  {\scriptsize}
  \end{tabular} &
  \begin{tabular}{@{}c@{}}  
  $0.37$
  {\scriptsize}
  \end{tabular} &
  \begin{tabular}{@{}c@{}}  
  $0.50$
  {\scriptsize}
  \end{tabular} &
  \begin{tabular}{@{}c@{}}  
  $0.47$
  {\scriptsize}
  \end{tabular} &  
  \begin{tabular}{@{}c@{}}  
  $0.66$
  {\scriptsize}
  \end{tabular} &
  \begin{tabular}{@{}c@{}}  
  $2.18$
  {\scriptsize}
  \end{tabular} &
  \begin{tabular}{@{}c@{}}  
  $0.85$
  {\scriptsize}
  \end{tabular} \\
CS Exp &
  \begin{tabular}{@{}c@{}}  
  $0.24$
  {\scriptsize}
  \end{tabular} &
  \begin{tabular}{@{}c@{}}  
  $0.53$
  {\scriptsize}
  \end{tabular} &
  \begin{tabular}{@{}c@{}}  
  $0.71$
  {\scriptsize}
  \end{tabular} &
  \begin{tabular}{@{}c@{}}  
  $0.55$
  {\scriptsize}
  \end{tabular} &
  \begin{tabular}{@{}c@{}}  
  $1.04$
  {\scriptsize}
  \end{tabular} &
  \begin{tabular}{@{}c@{}}  
  $1.27$
  {\scriptsize}
  \end{tabular} &
  \begin{tabular}{@{}c@{}}  
  $0.79$
  {\scriptsize}
  \end{tabular} &
  \begin{tabular}{@{}c@{}}  
  $0.83$
  {\scriptsize}
  \end{tabular} &
  \begin{tabular}{@{}c@{}}  
  $1.68$
  {\scriptsize}
  \end{tabular} &
  \begin{tabular}{@{}c@{}}  
  $1.34$
  {\scriptsize}
  \end{tabular} &
  \begin{tabular}{@{}c@{}}  
  $1.54$
  {\scriptsize}
  \end{tabular} &
  \begin{tabular}{@{}c@{}}  
  $0.66$    
  {\scriptsize}
  \end{tabular} \\
\bottomrule
\multicolumn{11}{l}{$^{***}p < 0.001$; $^{**}p < 0.01$; $^{*}p < 0.05$}
\end{tabular}

\vspace{2mm}
\caption{ Regression table for True/False comprehensions questions in \textbf{Survey 2} for the IoT Scenarios. There is one logistic regression model for each question in each scenario (24 models total) \new{ corrected with the Benjamini-Hochberg procedure between scenarios and McFadden's $R^2$ for each model. McFadden's $R^2$ is a pseudo-$R^2$ measure of model fit, where higher values (0.2-0.4) indicate better explanatory power relative to an intercept-only model, though values are typically much smaller than in linear regression.} The numbers in this table are the odds ratios for each predictor, with the baseline explanations used in each model noted in \textit{italics}. Statistical significance is noted with asterisks and shaded cells: blue for positive coefficients and orange for negative coefficients. \new{The values shown below each significant odds ratio are the 95\% confidence intervals (CI), indicating the range of plausible effect sizes.}}
\vspace{+30pt}
\label{tab:stats-followup-iot}
\end{table*}

\Paragraph{Participants did interact with the FAQ.}
To determine whether people were reading the FAQ, we added two comprehension questions. \Qeleven{} asked about real-world use of TEEs (``How are TEEs used in real life?'' in the FAQ) and \Qtwelve{} asks whether we can know that a TEE is configured correctly (``How do we know the TEE is working correctly?''). In both cases, participants were significantly more likely to answer the question correctly if they had an FAQ than if they didn't (Table~\ref{tab:stats-followup} and~\ref{tab:stats-followup-iot}). 77\% of participants who received a \textit{Hidden} FAQ expanded the questions at least once.

\Paragraph{The FAQ has a mixed effect on comprehension.}
Participants with an FAQ were more likely to correctly answer some of the questions about TEE features (\Qone{}-\Qnine{}) or the FAQ-specific questions (\Qeleven{}-\Qtwelve{}). The difference between the \textit{Shown} FAQ and \textit{None} FAQ condition was statistically significant for \Qtwo{} for the medical scenario with AI (\(p<0.05\)) while the \textit{Hidden} FAQ condition was significantly better than the \textit{None} FAQ condition for \Qthree{} for the medical scenario without AI (\(p<0.01\)). Meanwhile, both \textit{Shown} and \textit{Hidden} FAQ conditions were better than the \textit{None} FAQ for \Qeleven{} and \Qtwelve{} (statistically significant for all scenarios, \(p < 0.001\)). 
When the FAQ helped participants answer the question correctly, we found similar results for both types of FAQ presentations, 
except for the FAQ-specific questions, where the \textit{Shown} FAQ was better than the \textit{Hidden} one.

Interestingly, having an FAQ made it more likely that participants would answer some of the questions about TEE limitations (\Qfive{}-\Qten{}) \textit{incorrectly} than if they didn't have an FAQ at all. The difference between the \textit{Shown} FAQ condition and \textit{None} FAQ condition was statistically significant for \Qten{} for the medical scenario without AI (\(p<0.05\)).
The \textit{Hidden} FAQ condition was significantly worse than the \textit{None} FAQ condition for \Qseven{} for the medical scenario with AI (\(p<0.05\)).

\Paragraph{Having an FAQ or explanation seems to have little effect on willingness to use technology or confidence that data will be safe.}
Similar to the findings in Survey 1, where we saw almost no impact from different explanations, in Survey 2, we see that explanations and FAQ do not seem to significantly affect participants' willingness to use TEE-enabled technology or their belief that their data will be safe (see Table~\ref{tab:trust-stats}).

\subsection{RQ4: Aspects Contributing to Safety}
\label{sec:aspects}
For each scenario, we asked participants which aspects of the scenario contribute to their belief that their data would be safe (or unsafe), including: the use of a TEE, that a hospital (or company, depending on the scenario) is collecting the data, the people on the team, what data is collected, and the purpose of the data collection.
We also asked if any other aspects of the scenario not already mentioned contributed to their belief that their data would be safe or unsafe. 
Overall, we found that providing information about TEEs (by giving them an explanation \textit{or} an FAQ) seems to make people more confident that the TEE would keep their data safe. There were many other aspects of the scenario that people were concerned about that TEE explanations and FAQs did not address. 
The results in this section are supported by a Wilcoxon signed-rank test, see Table~\ref{tab:ascects_statistical_test} for details.

\begin{table*}[htp!]
\centering
\begin{tabular}{l|llllllll} 
\toprule
& \multicolumn{4}{c}{Medical} & \multicolumn{4}{c}{Smart Home}  \\ 
& \multicolumn{2}{c}{With AI} & \multicolumn{2}{c}{Without AI} & \multicolumn{2}{c}{With AI} & \multicolumn{2}{c}{Without AI}  \\ 
\midrule
Aspects                     
& W & p-value                 
& W & p-value                 
& W & p-value                    
& W & p-value                     \\ 
\midrule
\textbf{Use of TEE}        
& 1421  &  $0.0129^{**}$                  
& 1638.5  & $0.001^{***}$  
& 1483.5  &  $0.003^{***}$ 
& 1547.5  &  $0.003^{***}$        \\
\textbf{Purpose}          
& 2168  &  0.795                   
&  2689 &   0.871                   
& 2231  &   0.818                    
& 2258.5  &  0.489                           \\
\textbf{Data Collect}      
& 2258  & 0.864                     
& 2815.5   &  0.871       
&  2363 & 0.818                       
& 2384  & 0.541                            \\
\textbf{Hospital/Company}  
& 2090  &    0.772                    
&  3285 &    0.161     
& 2637.5  &   0.818                     
&  2263 &   0.487                          \\
\textbf{People on the team}
& 1965  & 0.612                      
& 3412.5  &  0.103        
& 2580  &  0.818                       
& 2309.5  &   0.489                          \\
\bottomrule
\multicolumn{9}{l}{$^{***}p < 0.001$; $^{**}p < 0.01$; $^{*}p < 0.05$}
\end{tabular}
\vspace{2mm}
\caption{{Table with results of  Mann-Whitney U tests for the aspects contributing to the belief that the data would be safe/unsafe, with information about the TEE (TEE explanation or FAQ) vs no TEE information provided. Each scenario was treated separately, \new{and corrected with the Benjamini-Hochberg procedure}.
Statistical significance is noted with asterisks.}}\label{tab:ascects_statistical_test}
\end{table*}

\Paragraph{Attributing quotes to participants.}
We attribute quotes using a similar strategy as Survey 1, except that, here, the treatment is represented using two letters. The first letter is the TEE explanation: (H)ardware, (T)rust, (U)nsubstantial, or (X) for no explanation. The second letter is the FAQ condition: (H)idden, (S)hown, or (X) for no FAQ. 
For example, (P68S2-HX) is participant \#68 in Survey 2, who received the \textit{Hardware} explanation and the \textit{None} FAQ condition.

\Paragraph{Explaining TEEs seems to make people more confident the TEE will keep their data safe.}
In the group that had access to information about TEEs, 80.3\% said the use of a TEE made them feel their data was definitely or somewhat safe, while only 52.3\% of those who had no access to information about TEEs said the use of a TEE made them feel their data was definitely or somewhat safe (\(p < 0.01\) for the medical scenario with AI and \(p < 0.001\) for all other scenarios).

\Paragraph{TEE information seems to have little effect on other aspects.}
For aspects other than the use of a TEE, providing a TEE explanation or FAQ seems to make little difference to participants' feelings of safety. 
For example, 61.6\% of participants reported feeling definitely or somewhat safe about the purpose of data collection when they had information about the TEE vs. 57.8\% without information. 
For the other aspects, providing information about the TEE made people somewhat \textit{less} sure their data would be safe. The place where information made the biggest difference was when we asked about the people involved in the scenario. Here, 50.8\% of participants with information about the TEE reported the people made them feel their data would be definitely or somewhat safe, while 54.4\% without information about the TEE said the same. 
None of these differences were statistically significant.

\Paragraph{Other aspects of the scenario mentioned by participants.}
Here, we describe some of the most common aspects, not already discussed above (many participants used this opportunity to expand on their previous answers). 
We received 660 responses total from 382 participants, where each response might mention one or more aspects of the scenario. 392 responses mentioned at least one aspect contributing to the feeling their data would be unsafe and 249 responses mentioned aspects contributing to the feeling their data would be safe.

Some aspects of the scenario participants reported contributing to the feeling that data would be unsafe include: prior experiences with/knowledge of breaches (42 responses), the future use clause in the scenario text (28 responses), the belief that some of the data was being collected unnecessarily (24 responses), the use of AI in the scenario (19 responses), the risk that there could be a bug in the TEE code (14 responses), and the risk that their data would be sold (12 responses). 5 people were concerned that we mentioned future research: {``The fact that it says researchers are looking for new ways to verify the program is working correctly. That makes me a little hesitant. Sounds like there are still bugs\ldots''} (P144S2-TS). 3 people seemed suspicious about being told to ``trust'' the technology: ``Comes across a bit like: `Yeah trust me bro your medical records are totally safe bro, trust me, bro there's an acronym. You like acronyms right man?'~'' (P237S2-TX). 

Most aspects participants reported contributing to the feeling that their data would be safe were repeated from previous questions. The most common new aspect contributing to the feeling of safety is the perception that the data being collected is not interesting enough to an attacker anyway (25 responses).

\subsection{More questions from our participants}
\label{sec:questions-f}
Similar to the first survey, we gave participants two opportunities to ask us questions they have about TEEs and received 267 responses. 
In this section, we summarize the most common questions we received, following the same structure as Section~\ref{sec:questions}, and how questions differed between participants who did and did not have access to an FAQ. We began with the same codebook as in the first survey, with only a few additional codes emerging during the analysis. 
The codebook describing all of the themes and how frequently they occurred may be found in Appendix~\ref{app:questions}.

Giving participants an FAQ made them less likely to ask questions about TEEs or the scenario, which were the most common kinds of questions we received overall. Some other questions were more common from participants who received an FAQ, like asking for more examples or about guarantees.

\Paragraph{An FAQ seems to reduce questions about TEEs.}
As in the first survey, we received the most questions (117 responses) about TEEs themselves. Although only 34\% of people did not receive an FAQ, 45\% of the questions about TEEs came from people who did not receive an FAQ. The most common questions were, again, asking for more information about how the TEE (31 responses) or its isolation mechanism (7 responses) work. Unlike in the first survey, we also saw 17 people asking what a TEE is, more generally: {``What is a TEE??''} (P351S2-XX). 
Other common questions requested more implementation details (14 responses) or compared TEEs to other technologies (8 responses). 
We also had 13 requests for a less technical explanation: {``Need more details about how they work in general without the use of complicated verbiage.''} (P52S2-XH). Most of these (69\% of the requests) came from people who were forced to wait on the FAQ page.

\Paragraph{An FAQ seems to reduce questions about people in the scenario.}
We also received 74 questions about the scenario. Again, 45\% of the questions about the scenario came from people who did not receive an FAQ. The most common questions were about the people involved in the scenario (28 responses) or the data (25 responses). A disproportionate 57\% of questions about people come from the participants who did not receive an FAQ, while the questions about data are more evenly distributed between FAQ conditions.

\Paragraph{A hidden FAQ seems to reduce questions about hackers.}
60 participants had questions about the risks they might encounter. Similar to above, 43\% of these questions came from people who did not receive an FAQ. Unlike other questions, though, participants were least likely to ask questions about hackers (28 responses, total) if they got the expandable FAQ: 18\% of questions about hackers came from the hidden FAQ condition, while 43\% came from the shown FAQ and 39\% from the no FAQ condition. Questions about people behaving maliciously (12 responses) were nearly evenly distributed between FAQ conditions. The remaining questions about risks disproportionately came from the people who did not receive an FAQ (54\% of the remaining questions asked about risks).

\Paragraph{Questions about guarantees or real-world uses seem to be more common with an FAQ.}
We received 20 questions about guarantees and 29 questions about real-world uses for TEEs. Both questions were more common with an FAQ than without. 45\% of questions about guarantees came from participants in the hidden FAQ condition and 45\% of questions about real uses of TEEs came from participants who were shown the FAQ. It is possible that some questions came from participants who wanted to write something but couldn't think of anything else to ask: {``I can't think of any more questions. Maybe, would be nice to see more real world examples''} (P22S2-XS).

\Paragraph{A hidden FAQ also seems to reduce other concerns.}
Similar to our initial survey, 41 participants did not ask a question but used the space to share other thoughts. The most common thoughts were general distrust (23 responses), followed by opinions about the scenario (9 responses). The participants receiving the hidden FAQ condition seemed to be the least likely to use this space to express distrust (these account for 17\% of the 23 responses), while the remaining questions were nearly evenly distributed between the shown FAQ and no FAQ conditions  (43\% and 39\%, respectively).

\section{Discussion}
\label{sec:discussion}
In this section, we make recommendations for explaining technical concepts to non-experts, navigate the (seemingly) contradictory results between our study and prior work~\cite{pittTEE}, and highlight opportunities for future research. 

\subsection{Explaining technical concepts to non-experts}

\Paragraph{Avoid technical jargon.}
Our results in Section~\ref{sec:results-survey} echo prior work~\cite{zhang2022usable,wu2020risk,distler2020making} on the importance of avoiding jargon when explaining technical concepts to non-experts.
This was also mentioned by participants reading the supplementary technical details we introduced with the FAQ: 
{``Need more details about how they work in general without the use of complicated verbiage.''} (P52S2-XH).

\Paragraph{Be direct and tell users what you want them to know.} 
In Section~\ref{sec:results-survey}, we found that people were more likely to answer questions correctly when the answer was in the explanation or scenario text directly than questions where people had to generalize what they learned and \textit{infer} the answers. 
For example, participants in the \textit{Prevents} TEE explanation condition were told that the TEE can protect against malicious software on the computer. This group was significantly more likely to answer the question about malicious programs (\Qfour{}) correctly in three of the four scenarios because the explanation they received gave them the answer to the question.
On the other hand, while 87.2\% of Survey 1 participants knew that the research/development team could access the data (\Qfive{}), only 57.2\% used that knowledge in \Qten{} to infer that the same group of people could \textit{steal} their data and use it for personal gain. It is possible that our participants had trouble inferring that even people authorized to access their data might use it for malicious purposes.

\Paragraph{Don't tell people what technology they should trust.} 
In Section~\ref{sec:aspects}, we showed that explaining TEEs does little to address some of the concerns our participants have about technology. In fact, in some cases, information about TEEs made people slightly \textit{more} skeptical. 
One reason for this might be that TEEs do not address all security and privacy threats, so explaining them, even if they are explained well, does not address all of the concerns people have. 
These concerns could also explain why some participants were wary of the word ``trust'' in ``Trusted Execution Environment'' as we noted in Sections~\ref{sec:questions} and~\ref{sec:aspects}.
Because security technologies are often orthogonal to the concerns people shared with us, it could be counterproductive for users already feeling skeptical if these solutions are marketed to them as trusted: ``I don't trust my information will be secure, especially with the words `trusted environment'~'' (P243S1-HNN).

\subsection{Comparing our findings to prior work}
One of our motivations was the finding from prior work~\cite{pittTEE} that cloud-based TEEs can make people more comfortable sharing data with home IoT, especially if they understand what a TEE is. This study focused on the factors impacting comfort, while we focused on effective strategies for explaining TEEs. On the surface, our finding that the TEE explanation has little effect on participants' willingness to use TEE-enhanced technology or their perception of safety (Section~\ref{subsec:willingness} and Table~\ref{tab:trust-stats}) seem to contradict these results. One explanation for this difference could simply be the methodological differences between the previous study and ours.

\subsection{Future research}
\Paragraph{Explaining limitations, not just features.}
Our explanations appear to be better at describing the protections provided by TEEs than their limitations. In fact, despite ensuring our explanations faithfully represent TEE security features, participants believed TEEs offer protections that they do not, such as guaranteeing the results of a computation are correct or preventing people with legitimate access from selling their data (\Qeight{}{} and \Qten{} and in Sections~\ref{sec:results-survey} and~\ref{sec:followup-comp-trust}). A similar phenomenon was observed in prior work~\cite{akgul2021evaluating}. 
More research is necessary to understand how to highlight the limitations of security technology. This research needs to measure comprehension, willingness to use, and beliefs about safety. Another possible explanation for the results in Section~\ref{sec:results-survey} is that the questions about TEE limitations required participants to infer more from the explanations, while questions about TEE capabilities could mostly be answered by the TEE explanations directly. More research is also needed to understand which played a bigger factor in the differences in comprehension scores.

\Paragraph{Investigating showing vs. hiding the FAQ.}
In Section~\ref{sec:questions}, we explained that many participants in Survey 1 asked for more technical details about TEEs. In Survey 2, we provided participants with an FAQ, and in Section~\ref{sec:questions-f}, we explained that it did lead to fewer questions. 
However, we also saw that the technical details could be overwhelming to some and that the \textit{Shown} FAQ was more effective for some, \textit{but not all}, of the True/False comprehension questions (Section~\ref{sec:followup-comp-trust}). More research is needed to understand why different FAQ models perform differently for some comprehension questions. One hypothesis is that hiding the FAQ allows people to focus their attention on the relevant information, but also makes it more likely that they won't read it at all. Future work could also shed light on how we might balance the trade-offs between providing additional technical details to those who want them and hiding them from those who find them unnecessary.

\Paragraph{Revisit prior work with our enhanced explanations.}
As explained above, the methodological differences between our study and the one from prior work~\cite{pittTEE} likely explain the seemingly contradictory results about user comfort. Nevertheless, it would be useful to repeat their study using our most effective explanations to see if their results can be reproduced. Repeating the study using their methodology would enable direct comparisons and allow us to better understand why even our best explanations seem to have little effect on user comfort.

\subsection{How much do users need to know about TEEs?}
Our results provide some 
insights into how we can communicate about technical security concepts more effectively, but suggest that understanding TEEs does not impact decisions about whether or not to use TEE-enhanced technology. 
While we started off with the hypothesis that understanding TEEs would improve users' trust in technology, our participants' responses drive home the point that, as some of our participants correctly realized, knowledge that a system uses a TEE is insufficient to draw conclusions that a user's data will be adequately protected. We might imagine that the TEE is just one of several components that are being used to protect user data in our scenarios and we could potentially provide a much more detailed explanation of all the protective components to assure users that their data is safe, or to highlight exactly what risks they might face. But this begs the question of whether we should really expect users to understand the inner workings of a security system, or if it should simply be offered to improve transparency around data privacy. 

Ultimately, decisions about TEEs are still best left to experts, not end users. Experts' choices about whether and how to use TEEs should revolve around the technology they are developing and the data they require, not whether the TEE would make users more willing to use the technology.

\section{Conclusion}
\label{sec:conclusion}
In this study, we evaluated strategies for explaining TEEs. Some were more effective at enhancing understanding than others. Our findings highlight the importance of avoiding technical jargon and directly communicating what people should learn.
On the other hand, we found that our explanations have limited effects on willingness to use technology or the feeling of safety, likely because TEEs do not address many of the privacy concerns our participants have. Our results provide insights into how we can communicate more effectively about technical security concepts, but also suggest that explaining security technology might not resolve the concerns users have about data privacy.

\section*{Acknowledgments}
The authors would like to thank the people who took the time to participate in our study and make this research possible. 
This work supported by the National Science Foundation [NSF 2207216]; and the 
Fundação para a Ciência e a Tecnologia, I.P. (Portuguese Foundation for Science and
Technology) [PRT/BD/153739/2021].

\bibliographystyle{plain}
\bibliography{main}

\begin{appendices}
\section{Existing TEE explanations}
\label{app:wild-explanations}
This section includes the final list of the sources of explanations gathered in the wild, the explanations themselves, and the respective URLs. All URLs are truncated for formatting purposes, but the links in the URLs lead to the pages with the explanations.

\Paragraph{1. Google cloud}
From: \href{https://cloud.google.com/confidential-computing/confidential-vm/docs/about-cvm}{cloud.google.com}

``TEEs are secure and isolated environments that prevent unauthorized access or modification of applications and data while they are in use.''

\Paragraph{2. Intel SGX}
From: \href{https://www.intel.com/content/www/us/en/developer/tools/software-guard-extensions/linux-overview.html#:~:text=The%20SDK%20is%20a%20collection,SGX%20in%20C%2FC%2B%2B.}{intel.com/}

``A trusted execution environment is a secure area of a main processor. It helps protect the code and data loaded inside it with respect to confidentiality and integrity. Data integrity prevents unauthorized entities from outside the TEE from altering data, while code integrity prevents code in the TEE from being replaced or modified by unauthorized entities, which may also be the computer owner itself as in certain DRM schemes described in SGX. This is done by implementing unique, immutable, and confidential architectural security such as Intel Software Guard Extensions (Intel SGX) which offers hardware-based memory encryption that isolates specific application code and data in memory. Intel SGX allows user-level code to allocate private regions of memory, called enclaves, which are designed to be protected from processes running at higher privilege levels. A TEE as an isolated execution environment provides security features such as isolated execution, integrity of applications executing with the TEE, along with confidentiality of their assets. In general terms, the TEE offers an execution space that provides a higher level of security for trusted applications running on the device than a rich operating system and more functionality than a 'secure element'.''

\Paragraph{3. NVIDIA}
From: \href{https://blogs.nvidia.com/blog/2023/03/01/what-is-confidential-computing/}{blogs.nvidia.com/}

``Using cryptographic keys linked to the processors, confidential computing creates a trusted execution environment or secure enclave. That safe digital space supports a cryptographically signed proof, called attestation, that the hardware and firmware is correctly configured to prevent the viewing or alteration of their data or application code.'' 

\Paragraph{4. Forbes}
From: \href{https://www.forbes.com/sites/forbestechcouncil/2020/01/07/trusted-execution-environments-a-primary-or-secondary-security-mechanism/?sh=529c738ecb3e}{forbes.com}

``A trusted execution environment is a protected area on the hardware where code can be run securely and in isolation. Code running inside the environment should not be able to be viewed or modified, even if an attacker is able to run malicious code with full permissions on the same processor. As such, trusted execution environments have the potential to significantly boost the security of our systems.'' 

\Paragraph{5. AWS}
From: \href{https://docs.aws.amazon.com/panorama/latest/dev/security-features.html}{docs.aws.amazon.com}

``Trusted execution environment – The appliance uses a trusted execution environment (TEE) based on ARM TrustZone, with isolated storage, memory, and processing resources. Keys and other sensitive data stored in the trust zone can only be accessed by a trusted application, which runs in a separate operating system within the TEE. The AWS Panorama Appliance software runs in the untrusted Linux environment alongside application code. It can only access cryptographic operations by making a request to the secure application.'' 

\Paragraph{6. Wikipedia}
From: \href{https://en.m.wikipedia.org/wiki/Trusted_execution_environment}{en.m.wikipedia.org}

``A trusted execution environment (TEE) is a secure area of a main processor. It helps code and data loaded inside it to be protected with respect to confidentiality and integrity. Data integrity prevents unauthorized entities from outside the TEE from altering data, while code integrity prevents code in the TEE from being replaced or modified by unauthorized entities, which may also be the computer owner itself as in certain DRM schemes described in SGX. This is done by implementing unique, immutable, and confidential architectural security such as Intel Software Guard Extensions (Intel SGX) which offers hardware-based memory encryption that isolates specific application code and data in memory. Intel SGX allows user-level code to allocate private regions of memory, called enclaves, which are designed to be protected from processes running at higher privilege levels. A TEE as an isolated execution environment provides security features such as isolated execution, integrity of applications executing with the TEE, along with confidentiality of their assets. In general terms, the TEE offers an execution space that provides a higher level of security for trusted applications running on the device than a rich operating system (OS) and more functionality than a secure element (SE).'' 

\Paragraph{7. TrustSonic}
From: \href{https://www.trustonic.com/technical-articles/what-is-a-trusted-execution-environment-tee/}{trustonic.com} through Google search

``A Trusted Execution Environment (TEE) is an environment for executing code, in which those executing the code can have high levels of trust in that surrounding environment, because it can ignore threats from the rest of the device.''

\Paragraph{8. DualityTech}
From: \href{https://dualitytech.com/blog/what-is-a-trusted-execution-environment-tee/}{dualitytech.com}  through Google search.

``A Trusted Execution Environment is a secure area inside the main processor where code is executed and data is processed in an isolated private enclave such that it is invisible or inaccessible to external parties. The technology protects data by ensuring no other application can access it, and both insider and outsider threats can’t compromise it even if the operating system is compromised. This level of security is equivalent to what existing classic cryptography methods such as symmetric-key encryption, hashing and digital signature, provide.'' 

\Paragraph{9. Android}
From: \href{https://source.android.com/docs/security/features/trusty}{source.android.com} through Google search.

``Trusty is a secure Operating System (OS) that provides a Trusted Execution Environment (TEE) for Android. The Trusty OS runs on the same processor as the Android OS, but Trusty is isolated from the rest of the system by both hardware and software. Trusty and Android run parallel to each other. Trusty has access to the full power of a device’s main processor and memory but is completely isolated. Trusty's isolation protects it from malicious apps installed by the user and potential vulnerabilities that may be discovered in Android.'' 

\Paragraph{10. Evervault}
From: \href{https://evervault.com/blog/what-is-a-trusted-execution-environment-tee}{evervault.com} through Google search.

``A TEE is an environment for executing code in which those running the code can have high levels of trust in that surrounding environment because it is insulated from the rest of the device. A Trusted Execution Environment (TEE), also known as a Secure Enclave, is a highly constrained compute environment that allows for cryptographic verification (attestation) of the code being executed. TEEs are designed with no persistent storage, no shell access, and no network connectivity by default. As a result, they provide a completely isolated environment with heavily restricted external access, making it possible to run sensitive workloads securely.'' 

\Paragraph{11. Piwik}
From: \href{https://piwik.pro/blog/what-is-a-trusted-execution-environment/}{piwik.pro} through Google search.

``A trusted execution environment (TEE) is a secure area of a main processor that guarantees optimal protection for highly sensitive data in all its states, with respect to confidentiality and integrity. TEE can be used on-premises, in the cloud or within embedded hardware platforms. For example, marketing analytics software processes sensitive data about clients and visitors and can keep such data safe during processing by deploying this application in a TEE.''

\Paragraph{12. PheonixNap}
From: \href{https://phoenixnap.com/blog/trusted-execution-environment}{phoenixnap.com} through Google search.

``Trusted Execution Environments (TEEs) are CPU-encrypted isolated private enclaves inside the memory, used for protecting data in use at the hardware level. While the sensitive data is inside an enclave, unauthorized entities cannot  remove it, modify it, or add more data to it. The contents of an enclave remain invisible and inaccessible to external parties, protected  against outsider and insider threats. As a result, a TEE ensures the following:
Data integrity, Code integrity, Data confidentiality.
Depending on the vendor and the underlying technology, TEEs can enable additional features''

\Paragraph{13. Reddit Post}
From: \href{https://www.reddit.com/r/degoogle/comments/rfpcy7/i_once_worked_for_a_trusted_execution_environment/}{reddit.com} through Google search.

``The TEE is essentially another, totally separate, totally isolated OS. It runs in parallel with, for example, Android. The TEE is specifically named, because it is way more secure than your REE (your Android, your iOS).''

\Paragraph{14. Personal blog}
From: \href{https://sergioprado.blog/introduction-to-trusted-execution-environment-tee-arm-trustzone/}{sergioprado.blog} through Google search.

``A Trusted Execution Environment (TEE) is an environment where the code executed and the data accessed is isolated and protected in terms of confidentiality (no one have access to the data) and integrity (no one can change the code and its behavior).''

\Paragraph{15. Medium}
From: \href{https://medium.com/geekculture/trusted-execution-environment-tee-implementations-drawbacks-8a434301d679}{medium.com} through Google search.

``A Trusted Execution Environment (TEE) is an environment in which the executed code and the data that is accessed are physically isolated and confidentially protected so that no one without integrity can access the data or change the code or its behavior. We are not aware of many devices in the US that use trusted execution environments, including smartphones, set-top boxes, video game consoles, and Smart TVs. A TEE is a secure and integrity-protected processing environment that consists of processing, and storage capabilities.''

\Paragraph{16. Red Hat}
From: \href{https://next.redhat.com/2019/12/02/current-trusted-execution-environment-landscape/}{next.redhat.com} through Google search.

``Trusted Execution Environments (TEEs) are a fairly new technological approach to addressing some of these problems. They allow you to run applications within a set of memory pages that are encrypted by the host CPU in such a way even the owner of the host system is supposed to be unable to peer into or modify the running processes in the TEE instance. All TEEs provide confidentiality guarantees for code and data running within them, meaning that the running workload can’t be seen from outside the TEE. Some TEEs offer memory integrity protection, which prevents the data loaded into the TEE from being modified from the outside (we will come back to this below). As expected, none provide guaranteed availability, since lower stack levels must still be able to control scheduling and TEE launch, and can block system calls.''

\Paragraph{17. Scientific Publication}
From ~\cite{Sommerhalder2023} through Google search.

``Trusted execution environments are secure areas of central processors or devices that execute code with higher security than the rest of the device. Security is provided by encrypted memory regions called enclaves. Because the environment is isolated from the rest of the device, it is not affected by infection or compromise of the device. Trusted execution environments have applications for different usages, such as mobile phones, cloud data processing, or cryptocurrencies. Furthermore, since Trusted execution environments are part of a standard chipset, this inexpensive technology can be leveraged across many devices, resulting in increased security, especially in the mobile sector and IoT products.''

\Paragraph{18. Scientific Publication}
From ~\cite{pereira2021towards} through Google search.

``Trusted Execution Environments (TEEs) are used to protect sensitive data and run secure execution for security-critical applications, by providing an environment isolated from the rest of the system. However, over the last few years, TEEs have been proven weak, as either TEEs built upon security-oriented hardware extensions (e.g., Arm TrustZone) or resorting to dedicated secure elements were exploited multiple times.''

\Paragraph{19. CSRC}
From: \href{https://csrc.nist.gov/glossary/term/trusted_execution_environment}{csrc.nist.gov} through Google search.

An area or enclave protected by a system processor.

\Paragraph{20. Scientific Publication}
From ~\cite{suzaki2021ts} through Google search.

``A trusted execution environment (TEE) is a new hardware security feature that is isolated from a normal OS (i.e., rich execution environment (REE)). The TEE enables us to run a critical process, but the behavior is invisible from the normal OS, which makes it difficult to debug and tune the performance. In addition, the hardware/software architectures of TEE are different on CPUs. For example, Intel SGX allows user-mode only, although Arm TrustZone and RISC-V Keystone run a trusted OS. In addition, each TEE has each SDK for programming. Each SDK offers own APIs and makes difficult to write a common program.'' 

\Paragraph{21. Scientific Publication}
From ~\cite{hoang2020quick} through Google search.

``The Trusted Execution Environment (TEE) offers a software platform for secure applications. The TEE offers a memory isolation scheme and software authentication from a high privilege mode. The procedure uses different algorithms such as hashes and signatures, to authenticate the application to secure. Although the TEE hardware has been defined for memory isolation, the security algorithms often are executed using software implementations.''

\Paragraph{22. Scientific Publication}
From ~\cite{chakrabarti2020trusted} through Google search.

``TEEs ensure that code outside of the TEE, including the operating system and hypervisor, cannot compromise the execution integrity and confidentiality of programs run inside the TEE. Based on hardware-rooted trust, TEEs additionally allow to prove the integrity of such execution even to remote third parties (remote attestation). By using TEEs that protect not only against software attackers but also hardware attackers, even the cloud provider is moved out of the trust domain. By leveraging such hardware-based TEEs, there is an alternate approach on building secure multiparty computation toolkits''

\Paragraph{23. Linkedin Post}
From: \href{https://www.linkedin.com/pulse/tee-trusted-execution-environment-amit-nadiger}{linkedin.com} through Google search.

``A Trusted Execution Environment (TEE) is a secure and isolated area within a computer or mobile device's hardware that can run code and processes with higher levels of security and privacy than the device's main operating system. TEEs are designed to protect sensitive data and processes, such as encryption keys, biometric authentication, and digital payments, from malware, hackers, and other threats. TEEs are typically implemented in microprocessors with hardware-level security features, such as ARM TrustZone or Intel SGX, that provide a separate, isolated environment within the device's main operating system. This isolated environment is designed to prevent unauthorized access to sensitive data or processes and to ensure that code and data executed in the TEE cannot be tampered with or observed by the main operating system or any other software running on the device.''

\Paragraph{24. NVIDIA Jetson}
From: \href{https://docs.nvidia.com/jetson/archives/l4t-archived/l4t-325/index.html#page/Tegra%20Linux%20Driver%20Package%20Development%20Guide/trusty.html}{docs.nvidia.com} through Google search.

``TEE provides an execution environment that includes security features to ensure code and data on a device is protected.''

\Paragraph{25. ACSM}
From: \href{https://australiancybersecuritymagazine.com.au/confidential-computing-enforces-the-trusted-execution-environment-tee/}{australiancybersecuritymagazine.com.au} through Google search.

``A Trusted Execution Environment (TEE) is an environment that offers a level of assurance of data integrity, data confidentiality, and code integrity. A hardware-based TEE uses the techniques to provide increased security guarantees for code execution and data protection within that environment.''

\Paragraph{26. Scientific Publication}
From ~\cite{busch2020unearthing} through Google search.

``TEEs are an integral part of the security architecture of mobile devices. They provide an execution context where security-critical services, such as user authentication, mobile payment, and digital rights management, can run isolated from the Rich Operating System (Rich OS). ''

\Paragraph{27. Scientific Publication}
From ~\cite{han2023mytee} through Google search.

``The trusted execution environment (TEE) is one of the
reasonable security measures for protecting security-critical
services and data on embedded devices. Particularly, considering ARM’s high market share in mobile and embedded devices (90\% for mobile, IoT, and in-vehicle systems), TrustZone,
the security extension of ARM processor, is the most potent
measure to enable TEEs on embedded devices. It provides
various security extensions, such as separating the security
states of the CPU, hardware-based memory access control, and
secure IO.''

\Paragraph{28. Global Platform}
From: \href{https://globalplatform.org/resource-publication/what-is-a-trusted-execution-environment-tee-past-present-future/}{globalplatform.org} through Google search.

``The Trusted Execution Environment (TEE) is a secure area in a device that ensures sensitive data is stored, processed, and protected in an isolated and trusted environment. As such, it offers protection against attacks generated in the rest of the device and even other actors inside the TEE.''

\Paragraph{29. IBM}
From: \href{https://www.ibm.com/docs/sk/rsct/3.2?topic=concepts-rsct-in-aix-trusted-execution-environment}{ibm.com} through Google search.

``The AIX security feature of Trusted Execution (TE) environment provides protection to the installed components and software applications. This security feature is achieved by maintaining and validating integrity of each component of the system and installed supported applications. The kernel trusts and invokes only those objects that the kernel can validate the integrity successfully. Any untrusted object is denied permission to execute. The TE feature of AIX protects the system against many malware that might gain access to the system and infect legitimate system or application components, causing unauthorized execution of the malware code along with the legitimate application.'' 

\Paragraph{30. ArgusSec}
From: \href{https://argus-sec.com/blog/standards-and-compliance/trusted-execution-environment-tee/}{argus-sec.com} through Google search.

``The Trusted Execution Environment (TEE) is a secure area that resides alongside the Rich Execution Environment (REE) of the main processor in connected devices, most notably smartphones. The purpose of the TEE is to provide a trusted and isolated environment in which sensitive data and assets can be stored, and trusted code executed, protecting these sensitive assets and Trusted Applications (TAs) from any software attacks generated within the REE.'' 

\Paragraph{31. Applus}
From: \href{https://www.appluslaboratories.com/global/en/what-we-do/service-sheet/trusted-execution-environment-(tee)}{appluslaboratories.com} through Google search.

``The TEE is a secure execution environment that runs in parallel with the operating system of the device (e.g. Android) and where only authorised and reliable applications are run (trusted apps). The TEE uses software and hardware security resources to protect the applications which are being executed in the TEE. This increases the security in the storage and processing of the sensitive data managed by the trusted applications. Furthermore, the TEE provides secure applications with a standardized set of routines and functions (APIs) that facilitate their development. This solution is applicable not only to smart phones but also to other devices such as tablets, Smart TV, set-boxes and other products that manage sensitive data and are connected to the Internet (Internet of things).'' 

\Paragraph{32. Azeria Labs}
From: \href{https://azeria-labs.com/trusted-execution-environments-tee-and-trustzone/}{azeria-labs.com} through Google search.

``Modern operating system kernels are also large, and have similar problems avoiding memory-corruption vulnerabilities. Isolating these kernels is a lot more complicated than for normal programs. To do this, device developers can make use of a Trusted Execution Environment (TEE). These TEEs isolate critical code and data away from the main operating system, so that even if the main operating system is compromised, the data and code residing inside the TEE remains isolated. Use-cases for TEEs include verifying the integrity of the operating system itself, managing user credentials such as via a fingerprint sensor, and the storage and management of device encryption keys. High-value assets are not limited to kernel-mode components. Video Digital Rights Management (DRM) applications, banking applications and secure messengers may also want to protect their code and data from devices that may have malware installed.'' 

\Paragraph{33. Bitfount}
From: \href{https://www.bitfount.com/pets-explained/trusted-execution-environments}{bitfount.com} through Google search.

``Trusted Execution Environments (TEEs) are one mechanism for enabling multiple parties to collaboratively do computation. As the name suggests, the security depends on the computation running in an environment that all the parties trust. Imagine a clean room or bunker where everyone knows data can come in but only information they are comfortable with goes out.'' 

\Paragraph{34. Cryptologie}
From: \href{https://www.cryptologie.net/article/501/hardware-solutions-to-highly-adversarial-environments-part-3-trusted-execution-environment-tee-sgx-trustzone-and-hardware-security-tokens/}{cryptologie.net} through Google search.

``Trusted Execution Environment (TEE) is a concept that extends the instruction set of a processor to allow for programs to run in a separate secure environment. The separation between this secure environment and the ones we are used to deal with already (often called rich execution environment) is done via hardware. So what ends up happening is that modern CPUs run both a normal OS as well as a secure OS simultaneously. Both have their own set of registers but share most of the rest of the CPU architecture (and of course system). By using clever CPU-enforced logic, data from the secure world cannot be accessed from the normal world. Due to TEE being implemented directly on the main processor, not only does it mean a TEE is a faster and cheaper product than a TPM or secure element, it also comes for free in a lot of modern CPUs.'' 

\Paragraph{35. Stack Exchange}
From: \href{https://security.stackexchange.com/questions/233279/what-trusted-execution-environment-tee-solutions-exist-for-mobile-devices}{security.stackexchange.com} through Google search.

``A trusted execution environment (TEE) provides a way for one to deploy tamper-proof programs on a device. The most prominent example of TEEs seem to be Intel SGX for PCs.'' 

\Paragraph{36. NXP}
From: \href{https://www.nxp.com/design/training/trusted-execution-environment-getting-started-with-op-tee-on-i-mx-processors:TIP-TRUSTED-EXECUTION-ENVIRONMENT-GETTING-STARTED}{nxp.com}, through Google search.

``Building more secure embedded system products starts with utilizing advanced hardware security features such as a trusted execution environment (TEE). A TEE isolates sensitive data and processes from non-secure processes, creating secure or trusted zones and non-secure, non-trusted zones in your embedded system product.''

\subsection{Example of a misleading explanation}

From: \Paragraph{NY Times}
\href{https://www.nytimes.com/2019/11/19/technology/artificial-intelligence-dawn-song.html}{nytimes.com}

``A trusted execution environment that protects software from most kinds of attack. Within the secure enclave, bits of computer code, called smart contracts, allow data owners to control who has access to their data and how it is used.'' 

\section{Candidate Trusted Execution Environment Explanations}
\label{app:tee}
Candidate TEE explanations are composed of 2-3 sentences, where
each sentence has a different theme. We evaluate every combination 
of the 2-3 sentences in our surveys. Each theme is shown below in
\emph{italics}, followed by the corresponding sentence from our 
evaluation.

\Paragraph{Sentence 1: Introducing TEEs}
~\\\emph{Hardware}: A Trusted Execution Environment (TEE) is a 
technique for running programs and interacting with data securely
using a protected area of the physical computer.
\\\emph{Trust}: A Trusted Execution Environment (TEE) is a 
technique for running programs and interacting with data securely,
even if the rest of the computer is not trustworthy.
\\\emph{Unsubstantial}: A Trusted Execution Environment (TEE) is a 
technique for running programs and interacting with data securely.

\Paragraph{Sentence 2: Isolation, confidentiality, and integrity}
~\\\emph{Technical}:
A program running in a TEE is isolated from the rest of the computer 
to protect the confidentiality and integrity of the program and data.
\\\emph{Non-Technical}:
A program running in a TEE is isolated from the rest of the computer 
to allow only authorized people to view or change the program and 
data.

\Paragraph{Sentence 3: (Optional) threat prevented by TEE}
~\\\emph{Prevents}: The TEE protects the program and data even when 
other software on the computer is behaving maliciously.
\\\emph{No Prevents}: (No third sentence)

\section{Complete Scenario Text}
\label{app:scenarios}
This section includes the complete scenario text from our surveys. We use [brackets] to show small differences between scenarios with and without AI (in that order). We note larger differences in (parentheses).

\Paragraph{Medical scenarios}
Suppose you have been invited to participate in a [\textbf{medical research study}/\textbf{medical research study}]. 

(Scenario with AI only)
The goal of the study is to \textbf{develop an AI tool} which can lead to more accurate diagnosis. If the research is successful, the doctor at your local clinic might use the tool during your visits. 
Researchers are asking you to share clinical notes, diagnostic tests, and other related health data from the last 5 years, all identified with your name and address. 

(Scenario without AI only)
The goal of the study is to \textbf{determine where to build a new hospital} by investigating transportation-related barriers that might make it more difficult for some patients to receive routine preventative healthcare. If the research is successful, your city may build a new hospital closer to your home. 
Researchers are asking you to share your name and address, as well as your relevant medical history. 

You are also told:
\begin{itemize}
\item After being collected, your data will be securely transmitted and stored so that only a small team of researchers will have access to your data
\item Everyone else will only have access to the AI tool trained on everyone's combined data
\item In the future, your data may be used for other research studies
\end{itemize}

The researchers also explain that both your data and the [entire AI tool training process/researcher’s calculations] will be \textbf{protected by a TEE}.

\Paragraph{Smart home scenarios}
Suppose you are shopping for a [\textbf{new voice assistant}, like an Amazon Alexa or a Google Home/\textbf{smart light bulb}].

(Scenario with AI only)
You read about one model online that uses AI to process your voice commands. You can also control the device using an app on your phone. 
The data collected by the device includes anything you enter into the app as well as the history of your interactions with the assistant (specifically, your name and address, when you make requests, what requests you make, and your voice), which might be used by the company to continue to \textbf{improve the AI} and could help it recognize your voice better. 

(Scenario without AI only)
You read about one model online that you can control using an app on your phone. 
The data collected by the device includes anything you enter into the app as well as the history of your interactions with the light (specifically, your name and address, when you turn on the light, and what setting you use), which might be used by the company to \textbf{develop new light bulbs} that you might be interested in purchasing. 

You read that:
\begin{itemize}
\item After being collected, your data will be securely transmitted and stored so that only a small team of developers will have access to your data
\item Everyone else will only have access to the AI model trained on everyone's combined data
\item In the future, your data may be used to develop other devices at the same company
\end{itemize}

You also read that both your data and the [entire AI tool training process/developer’s work] will be \textbf{protected by a TEE}.

\section{Survey Instrument}
\label{app:survey}
This section has the detailed survey questions for both surveys. We use [brackets] to indicate places where questions differ between scenarios. Details about survey flow and differences between Survey 1 and 2 are shown in \textbf{bold text} while placeholders for scenario text/TEE explanations/FAQ, and other commentary not in the survey are shown in \textit{italics}. We use $\circ\,$ to list answer choices. All questions are single choice closed-ended questions unless otherwise noted.
Both surveys begin with a consent form explaining the study procedures, participation requirements, possible risks and benefits of participating, compensation rates, information about how data is protected, participants' rights related to voluntary participation, and contact information for the authors' institution(s) if they have questions or concerns. After consenting, we collect the participant's Prolific ID.

\subsection{Scenario Intro}
\label{app:intro}

\textit{Scenario text from Appendix~\ref{app:scenarios}}

\begin{enumerate}
\item How confident are you that you understand the scenario we described above?
  \opt Completely confident 
  \opt Somewhat confident 
  \opt Not at all confident

\item \textbf{Question shown if it is the first scenario}
  How confident are you that you understand what protections a TEE offers?
  \opt Completely confident 
  \opt Somewhat confident 
  \opt Not at all confident
\end{enumerate}

\textit{The following two questions serve as an attention check for the medical scenario with AI}
\begin{enumerate}
\setcounter{enumi}{2}
\item In the scenario above, what is the goal of the medical study?
  If you are not sure, please re-read the scenario above.
  \opt Develop an AI tool for diagnosing patients
  \opt Investigate potential causes of long COVID
  \opt Identify factors contributing to mood disorders in teenagers
  \opt Understand the impact of reliable social support for new mothers

\item \textbf{Question shown if they answer the previous question incorrectly}
  In the scenario above, what is the goal of the medical study?  
  Please re-read the scenario above and try again. This is your second chance to get the question correct.
  \opt Develop an AI tool for diagnosing patients
  \opt Investigate potential causes of long COVID
  \opt Identify factors contributing to mood disorders in teenagers
  \opt Understand the impact of reliable social support for new mothers
\end{enumerate}

\textit{The following two questions serve as an attention check for the medical scenario without AI}
\begin{enumerate}
\setcounter{enumi}{2}
\item  In the scenario above, what is the goal of the medical study?
  If you are not sure, please re-read the scenario above.
  \opt Determine where to build a new hospital  
  \opt Investigate potential causes of long COVID  
  \opt Identify factors contributing to mood disorders in teenagers  
  \opt Understand the impact of reliable social support for new mothers

\item  \textbf{Question shown if they answer the previous question incorrectly}
  In the scenario above, what is the goal of the medical study?  
  Please re-read the scenario above and try again. This is your second chance to get the question correct.
  \opt Determine where to build a new hospital  
  \opt Investigate potential causes of long COVID  
  \opt Identify factors contributing to mood disorders in teenagers  
  \opt Understand the impact of reliable social support for new mothers
\end{enumerate}

\textit{The following two questions serve as an attention check for the smart home scenario with AI}
\begin{enumerate}
\setcounter{enumi}{2}
\item  In the scenario above, what kind of device are you shopping for?
  If you are not sure, please re-read the scenario above. 
  \opt Smart voice assistant  
  \opt Smart refrigerator  
  \opt Smart television
  \opt Smart security camera

\item  \textbf{Question shown if they answer the previous question incorrectly}
  In the scenario above, what kind of device are you shopping for?  
  Please re-read the scenario above and try again. This is your second chance to get the question correct.
  \opt Smart voice assistant  
  \opt Smart refrigerator  
  \opt Smart television
  \opt Smart security camera
\end{enumerate}

\textit{The following two questions serve as an attention check for the smart home scenario without AI}
\begin{enumerate}
\setcounter{enumi}{2}
\item  In the scenario above, what kind of device are you shopping for?
  If you are not sure, please re-read the scenario above. 
  \opt Smart light bulb
  \opt Smart refrigerator
  \opt Smart television
  \opt Smart security camera

\item  \textbf{Question shown if they answer the previous question incorrectly}
  In the scenario above, what kind of device are you shopping for?  
  Please re-read the scenario above and try again. This is your second chance to get the question correct.
  \opt Smart light bulb
  \opt Smart refrigerator
  \opt Smart television
  \opt Smart security camera
\end{enumerate}

\subsection{FAQ Intro}
\textbf{This section is shown in the ``Show'' FAQ condition of Survey 2 only}

On this page, we answer some frequently asked questions about TEEs.

\textit{FAQ text from Appendix~\ref{app:faq}}

\textit{Participants are not allowed to proceed to the next page of the survey until at least 60 seconds have elapsed.}

\subsection{Willingness and Feeling of Safety}
\label{app:trust}
The following questions are about Scenario \textit{Number} and the same TEE description you were shown before.
You may re-read the scenario if you click the ``show/hide scenario'' button.
\textit{Show/Hide button for displaying scenario text}

\textbf{The following is included in the ``Show'' or ``Hide'' FAQ conditions of Survey 2 only:} We've also answered some frequently asked questions about TEEs. You can read the answers by clicking the "Expand" buttons.
\textit{Buttons for displaying the FAQ in the follow-up survey}

\textit{TEE explanation from Appendix~\ref{app:tee}}

\textbf{The following question is for medical scenarios only}
\begin{enumerate}
\setcounter{enumi}{4}
\item Imagine that you have enough free time to participate in the medical study described above. 
  In this scenario, how likely would you be willing to participate in the medical study?
  \opt Definitely would
  \opt Maybe would
  \opt Would not
  \opt Not sure (Why?) \textit{Free-response}
\end{enumerate}

\textbf{The following question is for smart home scenarios only}
\begin{enumerate}
\setcounter{enumi}{4}
\item Imagine that the [voice assistant/smart light bulb] described above is within your budget and you know how to set it up. 
  How likely would you be to purchase the [voice assistant/smart light bulb]?
  \opt Definitely would
  \opt Maybe would
  \opt Would not
  \opt Not sure (Why?) \textit{Free-response}
\end{enumerate}

\begin{enumerate}
\setcounter{enumi}{5}
\item How safe do you think data would be when it is protected by a TEE?
  \opt Completely safe
  \opt Mostly safe (\textbf{Survey 2 only})
  \opt Somewhat safe
  \opt Not at all safe
\end{enumerate}

\textbf{The following two questions are for Survey 2 only}
\begin{enumerate}
\setcounter{enumi}{6}
\item We want to understand which aspects of the scenario make you feel that your data would be safe or unsafe.
  For each of the following aspects of the scenario, tell us how safe it makes you feel your data would be. 
  \opt Definitely safe 
  \opt Somewhat safe
  \opt Neither safe nor unsafe
  \opt Somewhat unsafe
  \opt Definitely unsafe
  \begin{enumerate}
      \item The use of a TEE 
      \item That a [hospital/company] is collecting the data 
      \item The people on the [research/development] team
      \item Which data is collected
      \item The purpose of the data collection 
  \end{enumerate}

\item If there is anything else about the scenario that makes you feel that your data would be safe or unsafe, what is it?
  Does it make you feel that your data would be safe? Or unsafe? \textit{Free-response}
\end{enumerate}

\begin{enumerate}
\setcounter{enumi}{8}
\item What questions do you still have about TEEs, if any? \textit{Free-response}
\end{enumerate}

\textbf{The following question is for Survey 1 only}
\begin{enumerate}
\setcounter{enumi}{9}
\item How much would knowing the answers to these questions change your willingness to [participate in the medical study/purchase the voice assistant/purchase the light bulb]?
  \opt Definitely would change
  \opt Probably would change
  \opt Probably would not change
  \opt Definitely would not change
  \opt I have no questions

\end{enumerate}

\subsection{Comprehension}
\label{app:comp}

\begin{table*}[p]
\centering
\begin{tabular}{@{}lll@{}}
\toprule

\textbf{Q\#} & \textbf{T/F}  & \textbf{Question Text} \\ 
\midrule
    \Qone  & F & 
        A member of the general public can access your data                             \\
    \Qtwo  & F &  
        \_\_\_  can access your data                                                    \\
        && \qquad\textit{Medical}: A hospital employee unrelated to the research team   \\ 
        && \qquad\textit{Smart home}: Someone working at the company on an unrelated team \\
    \Qthree  & F &
        If there were a bug in other software on the computer, outside of the TEE 
        storing your data, then a  \\
        && hacker could use the bug to access your data          \\
    \Qfour  & F & 
        If a disgruntled [\_\_\_]   installed a malicious program on the computer storing 
        your data, then they \\
        && could access your data                                     \\
            && \qquad\textit{Medical}: hospital employee unrelated to the research team  \\
            && \qquad\textit{Smart home}: employee on an unrelated team                  \\
    \Qnine & F &
        Other [\_\_\_] working on different projects on the same computer can access 
        your data                                                                       \\
            && \qquad\textit{Medical}: researchers                                      \\
            && \qquad\textit{Smart home}: developers                                    \\
    \Qfive  & T & 
        A member of the [\_\_\_] can access your data                                   \\
            && \qquad\textit{Medical}: research team                                    \\
            && \qquad\textit{Smart home}: development team                              \\
    \Qsix  & T & 
        If [\_\_\_] makes a mistake collecting your data, then your data could be 
            incorrect                                                                   \\
            && \qquad\textit{Medical}: a member of the research team                    \\
            && \qquad\textit{Smart home}: the light bulb / the voice assistant          \\
    \Qseven  & T & 
        A [\_\_\_] could later use your data to [\_\_\_]                                \\
            && \qquad\textit{Medical without AI}: A member of the research team 
                / choose the location for a new fire station                            \\
            && \qquad\textit{Medical with AI}: A member of the research team 
                / train another AI diagnosis tool for a different \\
            && \qquad\quad medical condition     \\
            && \qquad\textit{Smart home}: Someone on the development team 
                / develop a smart vacuum                                                \\
    \Qeight  & F & 
        The TEE ensures [\_\_\_]                                                        \\
            && \qquad\textit{Medical without AI}: the hospital being constructed will be 
                closer to the patients who most need it                                 \\
            && \qquad\textit{Medical with AI}: the diagnosis made by the AI tool will 
                always be correct                                                       \\
            && \qquad\textit{Smart home without AI}: the new light bulbs will have features 
                relevant to you                                                         \\
            && \qquad\textit{Smart home with AI}: your voice will always be recognized 
                by the improved AI                                                      \\
    \Qten & T & 
        [\_\_\_] could steal your data and sell it on the dark web                      \\
            && \qquad\textit{Medical}: A member of the research team                    \\
            && \qquad\textit{Smart home}: Someone on the development team               \\
    \Qeleven & T &
        When you unlock your Android phone with a PIN, the PIN is verified in a TEE     \\
    \Qtwelve & F &
        We cannot be sure that a TEE is configured correctly                            \\
\\ \bottomrule
\end{tabular}

\caption{Questions for evaluating TEE concept comprehension. The expected answer (\textbf{T}rue or \textbf{F}alse) is shown in the second column. \Qeleven{} and \Qtwelve{} only appear in the follow-up survey. We note the places where the questions differ between scenarios.}
\label{tab:question-text}
\end{table*}

The following questions are about Scenario \textit{Number} and the same TEE description you were shown before.
You may re-read the scenario if you click the ``show/hide scenario'' button.
\textit{Show/Hide button for displaying scenario text}

\textbf{The following is included in the ``Show'' or ``Hide'' FAQ conditions of Survey 2 only:} We've also answered some frequently asked questions about TEEs. You can read the answers by clicking the "Expand" buttons.
\textit{Buttons for displaying the FAQ in the follow-up survey}

\textit{TEE explanation from Appendix~\ref{app:tee}}

\begin{enumerate}
\setcounter{enumi}{10}

\item Based on what you read above \textit{about TEEs}, please tell us whether the following statements are true or false.
\textit{The True/False questions are shown in Table~\ref{tab:question-text}}

\item How confident are you about your answers to the questions on this page?
  \opt Completely confident
  \opt Somewhat confident
  \opt Not at all confident

\item After answering the questions on this page, how confident are you that you understand what protections a TEE offers?
  \opt Completely confident
  \opt Somewhat confident
  \opt Not at all confident

\item After answering the questions on this page, how safe do you think data would be when it is protected by a TEE?
  \opt Completely safe
  \opt Somewhat safe
  \opt Not at all safe

\end{enumerate}

\subsection{Demographics and Feedback}
In the last part of the survey, we want to learn more about your background. You will also have an opportunity to give us feedback at the end.

\begin{enumerate}
\setcounter{enumi}{13}

\item Before this survey, were you familiar with the concept of TEEs?
  \opt Yes
  \opt No
  \opt I'm not sure/Other \textit{Free-response}

\item Are you employed in a computing field (e.g., IT, software engineer, programmer)?
  \opt Yes
  \opt No
  \opt No, but I have been in the past
  \opt I'm not sure/Other \textit{Free-response}

\item Do you have formal education in a computing field (e.g., degree in computer science or computer engineering)?
  \opt Yes
  \opt No
  \opt I'm not sure/Other \textit{Free-response}

\item Do you have prior experience with medical research?
  \opt Yes, as a participant
  \opt Yes, as a researcher
  \opt No
  \opt I'm not sure/Other \textit{Free-response}

\item Do you work in a medical field (e.g. nurse, doctor, hospital staff)?
  \opt Yes
  \opt No
  \opt No, but I have been in the past
  \opt I'm not sure/Other \textit{Free-response}

\item Do you own any smart devices?
  When we say “smart devices” we are referring to devices that can be controlled remotely or interact with each other over the internet, including: smart TVs, smart voice assistants (e.g., Alexa, Google Home), smart light bulbs, doorbells, cameras, and more.
  \opt Yes, and I use the smart features 
  \opt Yes, but I don't use the smart features
  \opt No
  \opt No, but I have in the past
  \opt I'm not sure/Other \textit{Free-response}

\item Do you have experience with home automation?
  When we say “home automation” we are referring to setting up smart devices to operate automatically, either by using schedules, routines, or scenes; or through another service like IFTTT or SmartThings.
  \opt Yes, I currently use home automation
  \opt Yes, I have used home automation, but not currently
  \opt No, I do not use home automation, but I do own smart devices
  \opt No, I don't own smart devices
  \opt I'm not sure/Other \textit{Free-response}

\end{enumerate}

\section{FAQ Text}
\label{app:faq}
In the follow-up survey, we answer some of the
frequently asked questions from the first survey.

\Paragraph{1. How do TEEs work?}
The details of how different TEEs work can vary. For example, Arm
TrustZone is a feature of modern processors that splits computer
resources between a “Normal World” and a “Trusted World” (a TEE).
Software running in each world has access to different regions of
memory. Software running in the Normal World cannot access or modify
data in the Trusted World. TrustZone is appropriate for protecting
entire trusted applications while Intel SGX, on the other hand, works
well with software that has both trusted and untrusted parts. SGX
allows software to create one or more “enclaves” (TEEs). The data in
the enclave can only be accessed while the trusted part of the
software is running.

\Paragraph{2. How do we know the TEE is working correctly?} TEEs
support hardware-based cryptographic functions that can be used to
guarantee that both the TEE and all the code running in the TEE are
configured properly. This process is called “attestation”. Researchers
are also working on new ways to ensure that software running in a TEE
works as expected.

\Paragraph{3. How are TEEs used in real life?} TEES are used in
computers, smart phones, and other devices. For example,
authentication in modern Android phones is typically handled by code
(called “Gatekeeper”) residing in a TEE based on ARM TrustZone. For
example, when you enter a PIN or scan your fingerprint to unlock your
phone, it is sent to GateKeeper in the secure zone of the CPU to
verify. The response from GateKeeper is encrypted with a secret,
hardware-backed key that is never shared outside the TEE.

\section{Additional Results and Statistics}
\label{app:followup-stats}

\begin{figure*}

\centering
\includegraphics[scale=0.24]{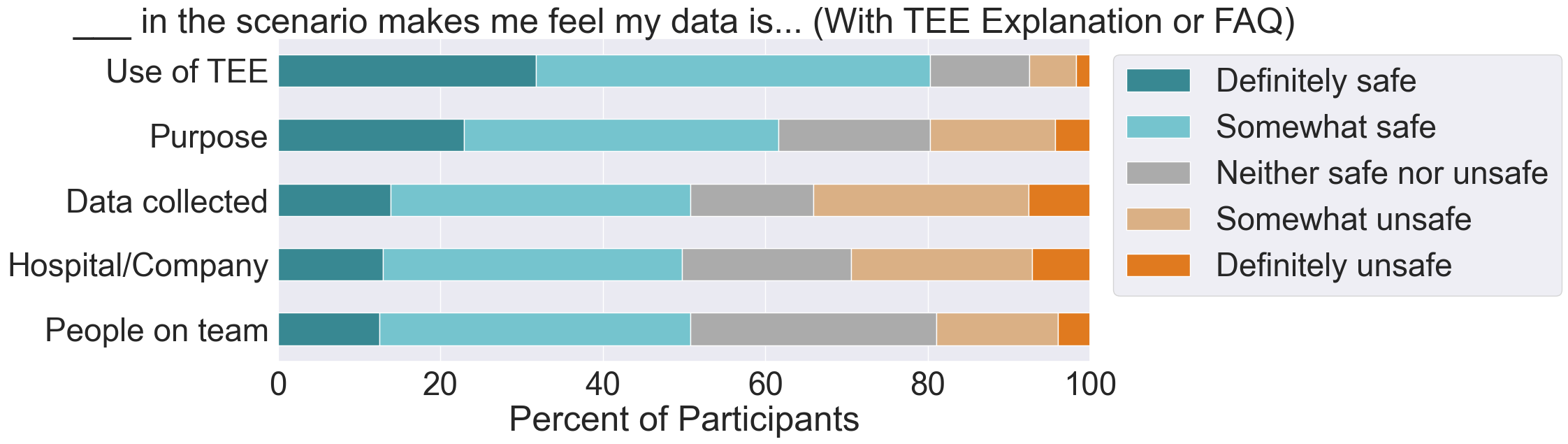}
\caption{Aspects of the scenario contributing to the belief that the data would be safe/unsafe, with a TEE explanation or FAQ}
\label{fig:aspect-expln}
\end{figure*}
\begin{figure*}
\centering
\includegraphics[scale=0.24]{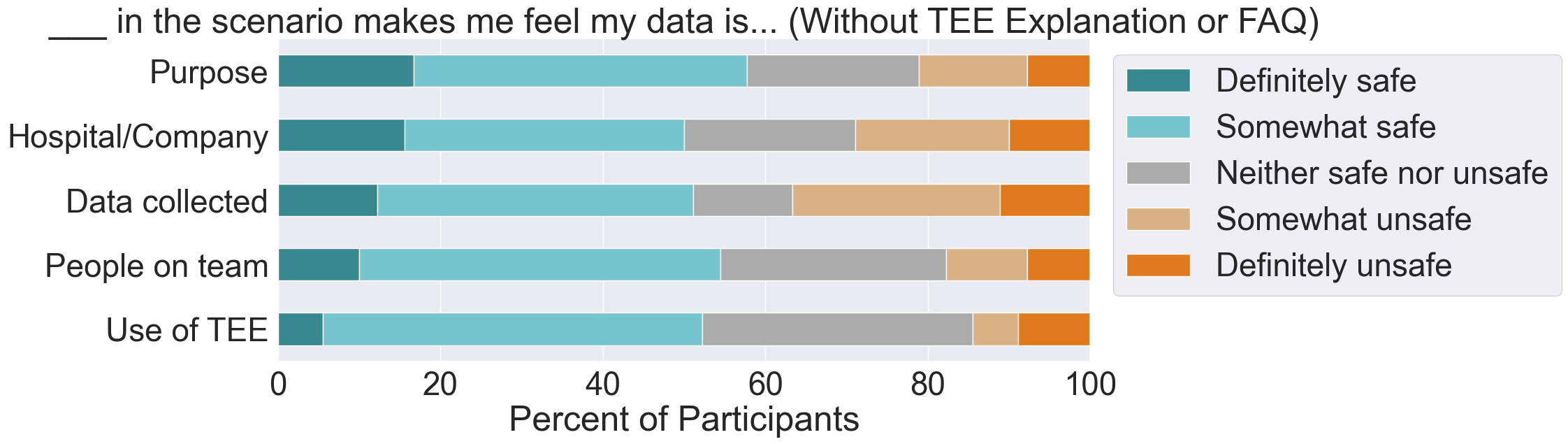}
\caption{Aspects of the scenario contributing to the belief that the data would be safe/unsafe, without a TEE explanation or FAQ}
\label{fig:aspect-none}
\end{figure*}
Results for the True/False comprehension questions in the initial survey are summarized in Tables~\ref{tab:comprehension} and~\ref{tab:comprehension-expln}. The first table includes results overall and split by scenario, while the second table splits the results by TEE explanation. Table~\ref{tab:comp-followup} has similar data for Survey 2, split by TEE explanation and FAQ type.

The regression table for questions about participants' willingness to use the TEE-enhanced technology and their belief that the TEE will keep the data safe.

\begin{table*}
\centering
\begin{tabular}{ccc|cccc}
\toprule
& & & \multicolumn{2}{c}{Medical} & \multicolumn{2}{c}{Smart Home} \\
\textbf{Q\#} & T/F & Overall & With AI & Without AI & With AI & Without AI  \\
\midrule
\multicolumn{6}{l}{\textit{Features of TEEs}} \\
\Qone &     F & 96.7\% & 96.2\% & 97.0\% & 96.6\% & 97.0\%  \\
\Qtwo &     F & 91.9\% & 92.4\% & 93.1\% & 90.6\% & 91.5\%  \\
\Qthree &   F & 79.0\% & 80.9\% & 76.8\% & 78.1\% & 80.1\%  \\
\Qfour &    F & 80.9\% & 83.1\% & 81.5\% & 80.7\% & 78.4\%  \\
\Qnine &    F & 84.2\% & 83.5\% & 84.1\% & 87.1\% & 82.2\%  \\
\midrule
\multicolumn{6}{l}{\textit{Limitations of TEEs}} \\
\Qfive &    T & 87.2\% & 91.1\% & 87.6\% & 83.3\% & 86.9\%  \\
\Qsix &     T & 82.0\% & 86.9\% & 80.3\% & 82.0\% & 78.8\%  \\
\Qseven &   T & 69.8\% & 87.3\% & 63.1\% & 66.1\% & 62.7\%  \\
\Qeight &   F & 61.1\% & 82.2\% & 57.1\% & 54.1\% & 50.8\%  \\
\Qten &     T & 57.2\% & 59.3\% & 55.4\% & 61.4\% & 53.0\%  \\
\bottomrule
\end{tabular}
\caption{Overall scores for each comprehension question, highlighting features and limitations of TEEs and the correct answers. Results are split by scenario.}
\label{tab:comprehension}
\end{table*}

\begin{table*}
\centering
\begin{tabular}{ccc|ccc|cc|cc}
\toprule
&&&\multicolumn{3}{c|}{First sentence}
  &\multicolumn{2}{c|}{Second sentence}
  &\multicolumn{2}{c}{Third sentence} \\
    & T/F & Overall & Hardware & Trust & Unsubstantial & Technical & Non-technical & Prevents & No Prevents \\
\midrule
\multicolumn{6}{l}{\textit{Features of TEEs}} \\
\Qone   & F & 96.7\% & 97.8\% & 94.5\% & 97.8\% & 96.6\% & 96.8\% & 95.7\% & 97.7\% \\
\Qtwo   & F & 91.9\% & 94.0\% & 87.7\% & 93.9\% & 88.7\% & 95.1\% & 91.5\% & 92.3\% \\
\Qthree & F & 79.0\% & 78.6\% & 81.8\% & 76.6\% & 78.4\% & 79.6\% & 82.5\% & 75.5\% \\
\Qfour  & F & 80.9\% & 80.5\% & 80.8\% & 81.4\% & 82.7\% & 79.1\% & 87.8\% & 74.0\% \\
\Qnine  & F & 84.2\% & 84.6\% & 82.5\% & 85.6\% & 82.9\% & 85.5\% & 85.9\% & 82.6\% \\
\midrule
\multicolumn{6}{l}{\textit{Limitations of TEEs}} \\
\Qfive  & T & 87.2\% & 88.1\% & 87.3\% & 86.2\% & 86.1\% & 88.3\% & 87.2\% & 87.2\% \\
\Qsix   & T & 82.0\% & 79.6\% & 82.8\% & 83.7\% & 84.0\% & 80.0\% & 81.6\% & 82.3\% \\
\Qseven & T & 69.8\% & 67.9\% & 73.7\% & 67.9\% & 69.2\% & 70.4\% & 71.2\% & 68.5\% \\
\Qeight & F & 61.1\% & 56.9\% & 60.1\% & 66.3\% & 62.8\% & 59.4\% & 60.5\% & 61.7\% \\
\Qten   & T & 57.2\% & 57.5\% & 58.1\% & 56.1\% & 53.0\% & 61.5\% & 53.8\% & 60.6\% \\
\bottomrule
\end{tabular}
\caption{Overall scores for each comprehension question in Survey 1, highlighting features and limitations of TEEs and the correct answers. Results are split by TEE explanation.}
\label{tab:comprehension-expln}
\end{table*}

\begin{table*}[htp!]
\centering
\begin{tabular}{l|llllllll} 
\toprule
& \multicolumn{4}{c}{Medical} & \multicolumn{4}{c}{Smart Home}  \\ 
& \multicolumn{2}{c}{With AI} & \multicolumn{2}{c}{Without AI} & \multicolumn{2}{c}{With AI} & \multicolumn{2}{c}{Without AI}  \\ 
\midrule          
& W & p-value 
& W & p-value 
& W & p-value 
& W & p-value 
              \\ 
\midrule
\textbf{Average Score}        
& 8226  &  $0.5146$                  
& 11723  &  $3.01e-06^{***}$  
& 10020  &  $0.0058^{**}$ 
& 13756  &  $1.75e-08^{***}$                      \\
\bottomrule
\multicolumn{9}{l}{$^{***}p < 0.001$; $^{**}p < 0.01$; $^{*}p < 0.05$}
\end{tabular}
\vspace{2mm}
\caption{{Table with results of  Wilcoxon signed-rank tests comparing the average score for questions Q1-Q5, to the average score for questions Q6-Q10 in the Study 2. 
Each scenario was treated separately and corrected with the Benjamini-Hochberg procedure.
Statistical significance is noted with asterisks.}}\label{tab:features_limitations_stats_1st}
\end{table*}

\begin{table*}
\centering
\begin{tabular}{ccc|ccc|cccc}
\toprule
\textbf{Q\#} & T/F & Overall & Show & Hide & None 
  & Hardware & Trust & Unsubstantial & None \\
\midrule
\multicolumn{6}{l}{\textit{Features of TEEs}} \\
\Qone &     F & 96.2\% & 97.0\% & 97.6\% & 94.0\% 
  & 97.6\% & 95.5\% & 95.3\% & 96.3\% \\
\Qtwo &     F & 88.3\% & 89.9\% & 89.2\% & 85.9\% 
  & 92.0\% & 85.4\% & 89.4\% & 86.5\% \\
\Qthree &   F & 71.4\% & 74.4\% & 73.2\% & 66.8\% 
  & 77.2\% & 76.4\% & 74.4\% & 57.4\% \\
\Qfour &    F & 73.9\% & 76.2\% & 75.6\% & 70.1\% 
  & 80.8\% & 77.2\% & 78.3\% & 59.0\% \\
\Qnine &    F & 81.6\% & 83.8\% & 83.1\% & 77.8\% 
  & 86.0\% & 79.7\% & 86.2\% & 74.2\% \\
\midrule
\multicolumn{6}{l}{\textit{Limitations of TEEs}} \\
\Qfive &    T & 90.9\% & 91.2\% & 89.8\% & 91.9\% 
  & 90.8\% & 93.5\% & 88.2\% & 91.4\% \\
\Qsix &     T & 86.1\% & 84.5\% & 86.1\% & 87.7\% 
  & 86.0\% & 87.4\% & 87.8\% & 83.2\% \\
\Qseven &   T & 72.0\% & 67.4\% & 72.6\% & 76.0\% 
  & 70.0\% & 72.8\% & 73.6\% & 71.7\% \\
\Qeight &   F & 55.4\% & 54.9\% & 53.6\% & 57.8\% 
  & 56.8\% & 56.9\% & 55.9\% & 52.0\% \\
\Qten &     T & 61.8\% & 60.1\% & 54.8\% & 70.4\% 
  & 56.0\% & 62.2\% & 65.7\% & 63.1\% \\
\midrule
\multicolumn{6}{l}{\textit{FAQ-specific Questions}} \\
\Qeleven &  T & 65.7\% & 89.3\% & 82.2\% & 26.0\% 
  & 66.8\% & 68.3\% & 59.8\% & 68.0\% \\
\Qtwelve &  F & 58.2\% & 71.6\% & 67.8\% & 35.6\% 
  & 62.0\% & 56.5\% & 54.7\% & 59.8\% \\
\bottomrule
\end{tabular}
\caption{Overall scores for each comprehension question in Survey 2, highlighting features and limitations of TEEs and the correct answers. Results are split by FAQ type and TEE explanation.}
\label{tab:comp-followup}
\end{table*}

\begin{figure*}
\centering
\includegraphics[alt={A bar chart showing how frequently participants asked questions about TEEs. Bars are grouped by scenario with one bar showing how frequently participants asked each type of question. The groups are: medical with AI, medical without AI, smart home with AI, smart home without AI. Each group includes the following bars: TEE (pink), scenario (blue), risks (green), guarantees (orange), real-world (purple), other (grey). In each group, questions about TEEs are the most common, while questions about real-world use are the least common. There are more questions overall for the medical scenarios than the smart home scenarios.}, scale=0.47]{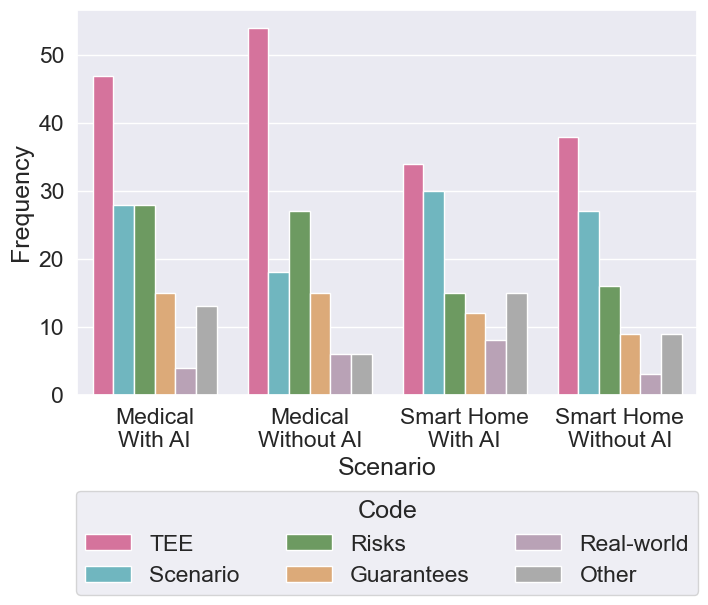}
\caption{Question frequency for Survey 1, split by scenario.}
\label{fig:question-freq}
\end{figure*}
\hfill
\begin{figure*}
\centering
\includegraphics[alt={A bar chart showing how frequently participants asked questions about TEEs. Bars are grouped by FAQ condition with one bar showing how frequently participants asked each type of question. The groups are: show, hide, none. Each group includes the following bars: TEE (pink), scenario (blue), risks (green), guarantees (orange), real-world (purple), other (grey). In each group, questions about TEEs are the most common and questions about the scenario are the second-most common, while questions about real-world use or guarantees are the least common. There are more questions overall for the no FAQ condition than the others.}, scale=0.47]{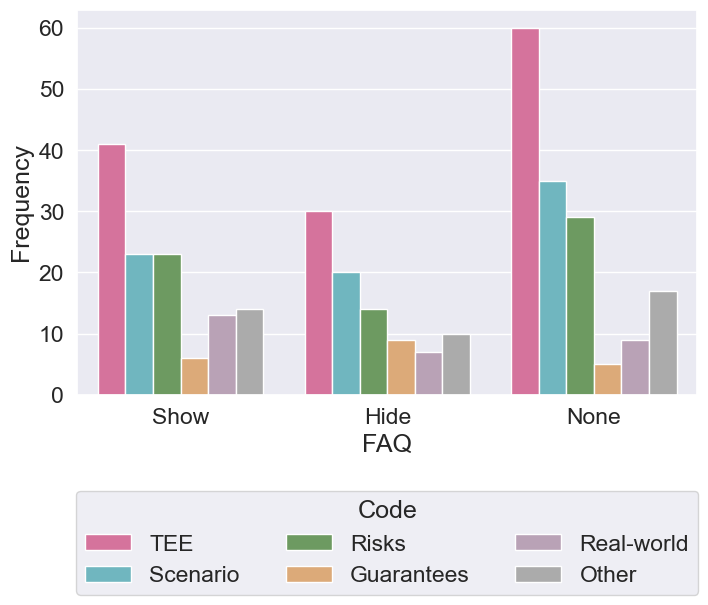}
\caption{Question frequency for Survey 2, split by FAQ condition.}
\label{fig:question-freq-f}
\end{figure*}

\section{Questions About TEEs}
\label{app:questions}
The codebook used to analyze the questions asked by participants is shown in Table~\ref{fig:codebook-qs}. Each response could receive codes from multiple categories and up to two codes from the \textit{TEE}, \textit{Scenario}, and \textit{Risk} categories.

For this first codebook, we identified six main codes: \emph{distrust, risk, general, real, scenario}, and \emph{guarantees} (see~\Cref{fig:codebook-qs} for the full codebook). 
In the \emph{distrust} group, we have sub-codes like \textit{distrust TEE} or \textit{distrust manufacturer} where the participants do not ask questions but instead express a lack of trust in a particular aspect. The second group involves \emph{risk} related codes, capturing instances where participants mention risks such as hacking or data leaks. The third group consists of \emph{general} codes, which are predominantly questions. A frequent sub-code here is \textit{how}, where participants ask broad questions about how the TEE works. This category also includes codes related to questions about encryption, isolation, and other technical aspects. The fourth group consists of questions about the \emph{real} use of TEEs outside of the scope of our study. The fifth group consists of \emph{scenario}-related codes, focusing on the elements not directly tied to the TEE but to the scenario described. For example, when participants ask questions about the hospital in the Medical scenario or the privacy policy of particular manufacturers. Finally, the sixth group deals with \emph{guarantees}, where participants ask for assurances on the TEE functionality.

The frequency of codes for questions asked by participants are shown in Table~\ref{fig:codebook-qs}.

\begin{table*}[htp!]
  \centering
  \resizebox{2\columnwidth}{!}{
\begin{tabular}{@{}llll@{}}
\toprule
\textbf{Code} & \textbf{Description} & \textbf{Frequency} & \textbf{Frequency} \\ 
    & & \textbf{First survey} & \textbf{Follow-up} \\
\midrule
\multicolumn{2}{l}{\textit{Questions about TEEs}} \\   
How?  &             
General questions about how TEEs work  &  
49 & 31 \\
Isolation  &             
Questions about how ``isolation'' works  &  
21 & 7 \\
Implementation  &             
Questions about what hardware is involved/other implementation details  &  
15 & 14 \\
Features  &             
Questions about what else the machine can do  &  
15 & 0 \\
Answered  &             
Questions already answered in the scenario text/TEE explanation received by the participant  &  
13 & 1 \\
Encryption  &             
Questions about the cryptography involved  &  
11 & 2 \\
Tech  &             
Questions comparing TEEs to other technology  &  
9 & 8 \\
Authorization  &             
Questions about the authorization process giving access to the TEE  &  
5 & 1\\
Different  &             
Questions that would have been answered by a different TEE explanation &  
3 & 0 \\
Open-Source$^*$ &             
Questions about whether software supporting TEEs is open-source  &  
- & 3 \\
What?$^*$ &
General questions about what a TEE is &
- & 17 \\
Non-technical$^*$ &
Request for a less technical explanation of TEEs &
- & 13 \\
Visual$^*$ &
Request for a visual description of TEEs &
- & 4 \\
Misc.  &             
Other questions about TEEs, appearing in fewer than 5 responses in the first survey &  
32 & 30 \\
\midrule
\multicolumn{2}{l}{\textit{Questions about the scenario}} \\     
Data  &             
Questions about what is collected, how/when data is anonymized, data retention policies, etc.  &  
41 & 25 \\
People  &             
Questions about the people in the scenario & 
25 & 28 \\
Future  &             
Questions about the future use of data clause in the scenario text  & 
8 & 6 \\
Positive  &             
Mention that they are less worried about this scenario than the other & 
6 & 0 \\
Fail  &             
Questions about the hospital's/company's procedures if there is a data breach  & 
5 & 4 \\
Compensation  &             
Questions about how they will be compensated for giving their data & 
5 & 1 \\
Misc. &
Other questions about the scenario, appearing in fewer than 5 responses in the first survey & 
13 & 14 \\
\midrule
\multicolumn{2}{l}{\textit{Questions about potential risks}} \\  
Hacking  &  
Questions about hacking/hackers & 
27 & 28 \\
People  &             
Questions about how people (on the research/development team, or not) might leak the data & 
23 & 12 \\
Fail  &             
Questions about what happens to their data if the TEE fails  & 
12 & 2 \\
Misc.  &             
Other questions about risks, appearing in fewer than 5 responses in the first survey  & 
24 & 24 \\
\midrule
\multicolumn{2}{l}{\textit{Questions about guarantees}} \\  
General &
Questions about how they know the TEE will work/how good they are &
46 & 15 \\
Testing &
Questions about how TEEs are tested &
4 & 4 \\
Cert &
Questions about whether/how TEEs are certified &
1 & 0 \\
\midrule
\multicolumn{2}{l}{\textit{(Non-)Questions expressing distrust}} \\    
General &             
Express distrust, but the distrust is vague or about data collection in general & 
16 & 23 \\
TEE  &             
Indicates they don't trust the TEE & 
10 & 6 \\
Scenario  &             
Indicates they don't trust the technology in the scenario & 
10 & 9 \\
Manufacturer  &             
Indicates they don't trust the manufacturer of the technology in the scenario or the TEE & 
7 & 3 \\
\midrule
\multicolumn{2}{l}{\textit{Questions about TEE use in real life}} \\    
Breaches  &             
Questions about whether TEEs have been involved in real breaches & 
11 & 5 \\
Request  &             
Request to try one for themselves & 
4 & 1 \\
Real  &             
Questions about whether TEEs are real & 
4 & 7 \\
New  &             
Questions about how new TEEs are & 
2 & 1 \\
\bottomrule
\end{tabular}
}

  \caption{Codebook for questions asked by participants for both surveys. Codes marked with an asterisk$^*$ were used in the follow-up survey only.}
  \label{fig:codebook-qs}
\end{table*}

\end{appendices}

\end{document}